%% file: MSSMPaper.tex
\documentclass[11pt,a4paper]{article} %for JHEP

\usepackage{atlasphysics}
\usepackage{preprintcover}
\usepackage{jheppub}
\usepackage{subfigure}
\usepackage{units}
\usepackage{amsmath,graphicx}
\usepackage{booktabs}
\usepackage{authblk}
\usepackage{lineno}
\usepackage{multirow,xspace}
\usepackage{wasysym}
\usepackage{rotating}
\usepackage[normalem]{ulem}

\def\currentlumirange{19.5--20.3~fb$^{-1}$}
\def\currentlumi{20.3~fb$^{-1}$}
\def\currentlumihh{19.5~fb$^{-1}$}
\def\tautau{\ensuremath{\tau\tau}}
\def\to{\ensuremath{\rightarrow}\,}
\def\MA{\ensuremath{m_{A}}}
\def\tanb{\ensuremath{\tan \beta}}

\def\mt{\ensuremath{m_{\mathrm T}}}

\def\met{\ensuremath{E_{\mathrm T}^{\mathrm{miss}}}}
\def\MET{\ensuremath{E_{\mathrm T}^{\mathrm{miss}}}}
\def\pt{\ensuremath{p_{\mathrm{T}}}}

\def\mumu{\ensuremath{\mu^{+} \mu^{-}}}
\def\ee{\ensuremath{e^{+} e^{-}}}
\def\thad{\ensuremath{\tau_{\mathrm{ had}}}}
\def\tlep{\ensuremath{\tau_{\mathrm{ lep}}}}
\def\telec{\ensuremath{\tau_{e}}}
\def\tmuon{\ensuremath{\tau_{\mu}}}

\def\Htautaulh{\ensuremath{h/H/A \to \tlep\thad\ }}
\def\Htautaull{\ensuremath{h/H/A \to \telec\tmuon\ }}
\def\Htautauhh{\ensuremath{h/H/A \to \thad\thad\ }}
\def\Zmumu{\ensuremath{Z/\gamma^{*}\to\mumu}}
\def\Wjets{\ensuremath{W+\mathrm{jets}}}

\def\Zjets{\ensuremath{Z/\gamma^{*} +\mathrm{jets}}}

\def\Ztautau{\ensuremath{Z/\gamma^{*} \to \tautau}}

\def\ttbar{\ensuremath{t\bar{t}}}
\def\bbbar{\ensuremath{b\bar{b}}}
\def\MTTOT{\ensuremath{m_{\mathrm{T}}^{\mathrm{total}}}}
\def\MMC{\ensuremath{m^{\mathrm{MMC}}_{\tau\tau}}}
\def\Wtaunujets{\ensuremath{W(\to\tau\nu){\rm+jets}}}

\def\mhmax{\ensuremath{m_{h}^{\mathrm{max}}}}
\def\mhmodplus{$m_h^{\text{mod}+}$}
\def\mhmodmin{$m_h^{\text{mod}-}$}

\PreprintCoverPaperTitle{\textbf{
Search for neutral Higgs bosons of the minimal supersymmetric standard model 
in $\boldsymbol{pp}$ collisions at $\boldsymbol{\sqrt{s}=8\TeV}$ with the 
ATLAS detector}}  

\PreprintIdNumber{CERN-PH-EP-2014-210}  

\PreprintCoverAbstract{
A search for the neutral Higgs bosons predicted by the Minimal Supersymmetric
Standard Model (MSSM) is reported. The analysis is performed on data from
proton--proton collisions at a centre-of-mass energy of $8\,\TeV$ collected 
with the ATLAS detector at the Large Hadron Collider. The samples used
for this search were
collected in 2012 and correspond to integrated luminosities
in the range 19.5--20.3~fb$^{-1}$.
The MSSM Higgs bosons are searched for in the $\tau\tau$ final state.
No significant excess over the expected background is observed, and exclusion
limits  are derived for the production cross section times branching fraction
of a  scalar particle as a function of its mass. 
The results are also interpreted in the MSSM
parameter space for various benchmark scenarios.
}

\PreprintJournalName{JHEP}

\mathchardef\mhyphen="2D

\title{\textbf{
Search for neutral Higgs bosons of the minimal supersymmetric standard model 
in $\boldsymbol{pp}$ collisions at $\boldsymbol{\sqrt{s}=8\TeV}$ with the 
ATLAS detector
}}

\author{The ATLAS Collaboration}
\abstract{
A search for the neutral Higgs bosons predicted by the Minimal Supersymmetric
Standard Model (MSSM) is reported. The analysis is performed on data from
proton--proton collisions at a centre-of-mass energy of $8\,\TeV$ collected 
with the ATLAS detector at the Large Hadron Collider. The samples used
for this search were
collected in 2012 and correspond to integrated luminosities
in the range 19.5--20.3~fb$^{-1}$.
The MSSM Higgs bosons are searched for in the $\tau\tau$ final state.
No significant excess over the expected background is observed, and exclusion
limits  are derived for the production cross section times branching fraction
of a  scalar particle as a function of its mass. 
The results are also interpreted in the MSSM
parameter space for various benchmark scenarios.
}

\begin{document}

\maketitle

\flushbottom

\input{introduction}

\input{analysiscategories}

\input{leplep}
\input{lephad}
\input{hadhad}
\input{systematics}

\input{results}

\clearpage

\bibliographystyle{JHEP}
\bibliography{MSSMPaper}

\onecolumn 
\clearpage
\input{atlas_authlist}

\end{document}

%% file: introduction.tex
\section{Introduction} \label{sec:intro}

The discovery of a scalar particle at the Large Hadron 
Collider (LHC) \cite{ATLASHiggsJuly2012, CMSHiggsJuly2012} has provided
important insight into the mechanism of electroweak symmetry breaking.
Experimental studies of the new particle 
\cite{ATLASHiggsProperties1,ATLASHiggsProperties2,
CMSHiggsProperties,CMSHiggs4l,CMSHiggsWW} demonstrate consistency with
the Standard Model (SM) Higgs boson \cite{ENGLERT,HIGGS,HIGGS2,HIGGS3,
Guralnik:1964eu,Kibble:1967sv}. However, it remains possible that the discovered
particle is part of an extended scalar sector, a scenario that is
favoured by a number of theoretical arguments \cite{Djouadi:2005gj,Branco:2011iw}.

The Minimal Supersymmetric Standard Model
(MSSM)~\cite{Fayet:1976et,Fayet:1977yc,Farrar:1978xj,
Fayet:1979sa,Dimopoulos:1981zb} is an
extension of the SM, which provides a framework addressing
naturalness, gauge coupling unification, and the existence of dark matter.
The Higgs sector of the MSSM contains two Higgs doublets, which results in
five physical Higgs bosons after electroweak symmetry breaking.
Of these bosons, two are neutral and CP-even ($h$, $H$),
one is neutral and CP-odd ($A$),
\footnote{By convention the lighter CP-even Higgs boson is
denoted $h$, the heavier CP-even Higgs boson is denoted $H$. The masses of the
three bosons are denoted in the following as $m_h$, $m_H$ and $m_A$
for $h$, $H$ and $A$, respectively.}
and the remaining two are
charged ($H^\pm$).
At tree level, the mass of the light scalar Higgs boson, $m_h$, is
restricted to be smaller than the $Z$ boson mass, $m_Z$. This bound is
weakened due to radiative corrections up to a maximum allowed value of
$m_h \sim 135$~GeV.
Only two additional parameters are needed with respect to the SM
at tree level to describe the MSSM Higgs
sector. These can be chosen to be the mass of the CP-odd Higgs boson,
$\MA$, and the ratio of the vacuum expectation values of the two Higgs
doublets, $\tan\beta$.
Beyond lowest order, the MSSM Higgs sector depends on additional
parameters, which are fixed at specific values in various MSSM benchmark scenarios.
For example, in the $\mhmax$ scenario the radiative corrections are chosen such
that $m_h$ is maximized for a given $\tan\beta$ and 
$M_{\text{SUSY}}$ \cite{Heinemeyer:1999zf,MSSMmhmax}.
\footnote{The supersymmetry scale, $M_{\text{SUSY}}$, is defined here as
the mass of the third generation squarks 
following refs. \cite{Heinemeyer:1999zf,MSSMmhmax,MSSMBenchmarks}.}
This results for $M_{\text{SUSY}}=1$~TeV in $m_h \sim 130$~GeV for 
large $\MA$ and $\tan\beta$.
In addition, in the same region the heavy Higgs bosons, $H$, $A$ and $H^{\pm}$,
are approximately mass degenerate and $h$ has properties very similar to a
SM Higgs boson with the same mass.
This feature is generic in the MSSM Higgs sector: a decoupling limit
exists defined by $\MA \gg m_{Z}$ in which the heavy Higgs bosons have similar
masses and the light CP-even Higgs boson in practice becomes identical to a SM
Higgs boson with the same mass.

The discovery of a SM-like Higgs boson, with mass that is now measured
to be $125.36~\pm~0.37~(\mathrm{stat})~\pm~0.18~(\mathrm{syst})$~GeV \cite{ATLASHiggsMass}, has prompted the definition of
additional MSSM scenarios \cite{MSSMBenchmarks}.  Most notably, the
\mhmodplus{} and \mhmodmin{} scenarios are similar to the $\mhmax$
scenario, apart from the fact that the choice of radiative corrections
is such that the maximum light CP-even Higgs boson mass is $\sim
126$~GeV.  This choice increases the region of the parameter space
that is compatible with the observed Higgs boson being the lightest
CP-even Higgs boson of the MSSM with respect to the $\mhmax$ scenario.
There are many other MSSM parameter choices beyond these scenarios
that are also compatible with the observed SM Higgs boson, for
instance, refs. \cite{Bechtle:2012jw,Arbey:2012dq}.

The couplings of the MSSM Higgs bosons 
to down-type fermions are enhanced with respect to the SM
for large $\tan\beta$ values 
resulting in increased branching fractions to
$\tau$ leptons and $b$-quarks, as well as a higher cross section for
Higgs boson production in association with $b$-quarks.
This has motivated a variety of searches in
$\tau\tau$ and $bb$ final states
at LEP~\cite{LEPLimits}, the
Tevatron~\cite{TevatronLimits1,TevatronLimits2,TevatronLimits3} and the
LHC~\cite{ATLASLimit,CMSLimit,LHCbHtautau}.

\begin{figure}
  \centering
  \subfigure[]{
  \includegraphics[width=0.25\textwidth]{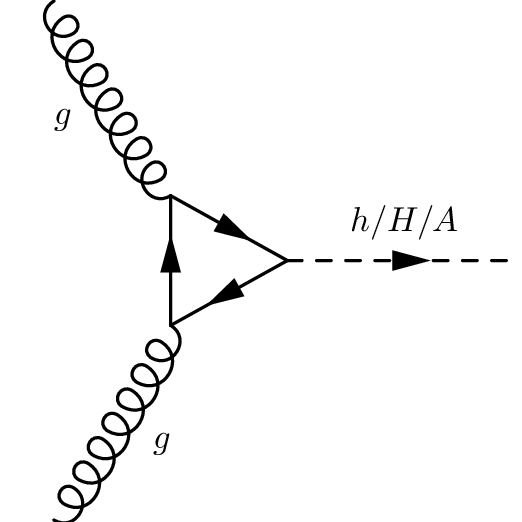} }
  \hspace{0.3cm}
  \subfigure[]{
  \includegraphics[width=0.25\textwidth]{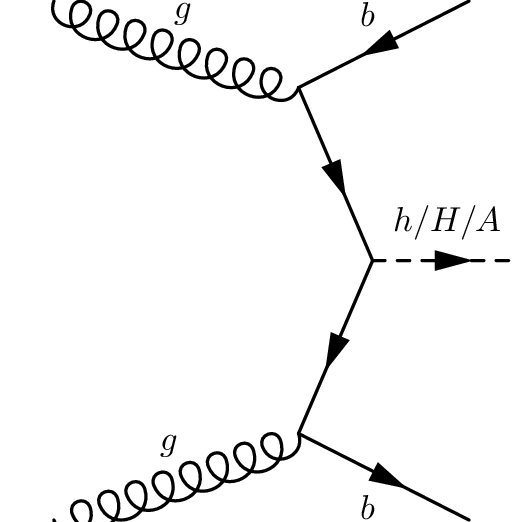} }
  \hspace{1.5cm}
  \subfigure[]{
  \includegraphics[width=0.25\textwidth]{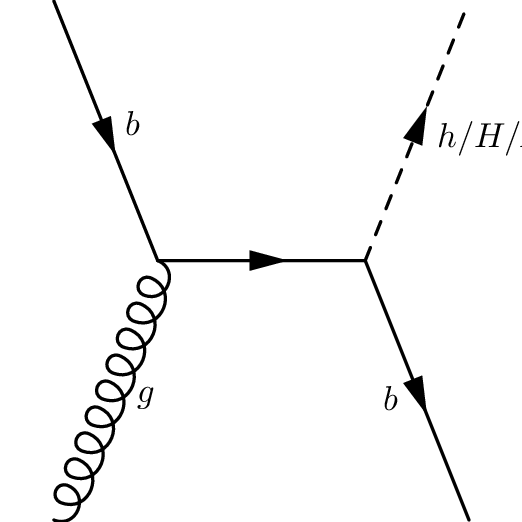} }
  \caption{Example Feynman diagrams for (a) gluon fusion and (b) $b$-associated
production in the four-flavour scheme and (c) five-flavour scheme  of a neutral MSSM Higgs boson. 
}
\label{fig:feynman}
\end{figure}

This paper presents the results of a search for a neutral MSSM Higgs boson in 
the $\tau\tau$ decay mode
using \currentlumirange{} of proton--proton collision data collected with 
the ATLAS detector 
\cite{ATLASDetector} in 2012 at a centre-of-mass energy of 8~TeV.
Higgs boson production through gluon fusion
or in association with $b$-quarks is considered (see figure~\ref{fig:feynman}), with the latter 
mode dominating for high \tanb{} values. The results of the search are
interpreted in various MSSM scenarios.

The ATLAS search for the SM Higgs boson in the $\tau\tau$ channel \cite{SMHtautau2011} 
is similar to that described here. Important differences between the two searches are 
that they are optimized for different production mechanisms and Higgs boson mass ranges. 
Additionally, the three Higgs bosons of the MSSM, which can have different masses, 
are considered in this search. In particular the couplings to $b$-quarks and 
vector bosons are different between the SM and MSSM. The $b$-associated production 
mode is dominant for the $H$ and $A$ bosons and is enhanced for the $h$ boson with 
respect to the SM for large parts of the MSSM parameter space. Furthermore, 
the coupling of the $H$ boson to vector bosons is suppressed with respect to those 
for a SM Higgs boson with the same mass and the coupling of the A boson to vector 
bosons is zero at lowest order, due to the assumption of CP symmetry conservation. 
Hence, vector boson fusion production and production in association with a vector boson, 
which contribute significantly to the SM Higgs boson searches, 
are much less important with respect to the SM. Finally, for high $m_A$ the search 
for the heavy $H$ and $A$ bosons is more sensitive in constraining the MSSM parameter 
space than the search for the $h$ boson. As a consequence, this search has little 
sensitivity to the production of a SM Higgs boson with a mass around 125~GeV. 
For consistency, the SM Higgs signal is not considered part of the SM background, 
as the MSSM contains a SM-like Higgs boson for large parts of the parameter space.

\section{The ATLAS detector} 

The ATLAS experiment~\cite{ATLASDetector}  at the LHC
is a multi-purpose particle detector with a 
forward-backward symmetric cylindrical geometry and a near $4\pi$ coverage in 
solid angle. It consists of an inner tracking detector 
surrounded by a thin superconducting solenoid providing a 2~T axial magnetic 
field, electromagnetic and hadronic calorimeters, and a muon spectrometer.
The inner tracking detector covers the pseudorapidity range\footnote{  
ATLAS uses a right-handed coordinate system with its origin at the nominal 
interaction point (IP) in the centre of the detector
and the $z$-axis along the beam pipe. The $x$-axis points from the IP to the 
centre of the LHC ring, and the $y$-axis points upwards. Cylindrical coordinates
$(r,\phi)$ are used in the transverse plane, $\phi$ being the azimuthal angle 
around the beam pipe. The pseudorapidity is defined in terms of the polar angle
$\theta$ as $\eta=-\ln\tan(\theta/2)$. Angular distance 
is  measured in units of $\Delta R \equiv  \sqrt{(\Delta\eta)^2 + (\Delta\phi)^2} $.} 
$|\eta| < 2.5$.  It consists of silicon pixel, 
silicon micro-strip, and transition radiation tracking detectors. 
Lead/liquid-argon (LAr) sampling calorimeters provide electromagnetic (EM) 
energy measurements with high granularity. A hadronic (iron/scintillator-tile)
calorimeter covers the central pseudorapidity range ($|\eta| < 1.7$).
The end-cap and forward regions are instrumented with LAr calorimeters for 
both the EM and hadronic energy measurements up to $|\eta| = 4.9$. 
The muon spectrometer surrounds the calorimeters and is based on three 
large air-core toroid superconducting magnets with eight coils each. 
Its bending power is in the range  2.0--7.5~Tm.
It includes a system of precision tracking chambers and fast detectors 
for triggering.
A three-level trigger system is used to select events. The first-level 
trigger is implemented in hardware. It is designed to use a subset 
of the detector information to reduce the accepted rate to at most 75~kHz. 
This is followed by two 
software-based trigger levels that together reduce the accepted event rate to 
400~Hz on average, depending on the data-taking conditions, during 2012.

\section{Data and Monte Carlo simulation samples} 
\label{sec:samples}

The data used in this search were recorded by the ATLAS experiment 
during the 2012 LHC run with proton--proton collisions at a centre-of-mass 
energy of 8~TeV. They correspond to an integrated luminosity of
\currentlumirange, depending on the search channel. 

Simulated samples of signal and background events were produced using various
event generators. The presence of  multiple interactions occurring in the same
or neighbouring bunch crossings (pile-up) was accounted for, 
and the ATLAS detector was modelled using GEANT4~\cite{ATLASSIM, Geant4}.

The Higgs boson production mechanisms considered in this analysis are
gluon fusion and $b$-associated production. The cross sections for
these processes were calculated using {\sc Higlu}~\cite{HIGLU}, {\sc
  ggh@nnlo}~\cite{Harlander:2002wh} and {\sc
  SusHi}~\cite{Harlander:2012pb,Harlander:2002wh,Harlander:2003bb,Harlander:2004tp,Harlander:2003kf,Harlander:2003ai,Aglietti:2004nj,Bonciani:2010ms,Degrassi:2010eu,Degrassi:2011vq,Degrassi:2012vt,Heinemeyer:1998yj,Heinemeyer:1998np,Degrassi:2002fi,Frank:2006yh,Harlander:2005rq}.
For $b$-associated production,
four-flavour~\cite{Dittmaier:2003ej,Dawson:2004a} and
five-flavour~\cite{Harlander:2003ai} cross-section calculations are
combined~\cite{SantanderMatching}.  The masses, couplings and
branching fractions of the Higgs bosons are computed with {\sc
  FeynHiggs}~\cite{Frank:2006yh,Heinemeyer:1998yj,Heinemeyer:1998np}.
Gluon fusion production is simulated with {\sc Powheg Box}
1.0~\cite{POWHEG}, while $b$-associated production is simulated with
{\sc Sherpa} 1.4.1~\cite{SHERPA}.  For a mass of $\MA=150$ GeV and
$\tanb=20$, the ratio of the gluon fusion to b associated production
modes is approximately 0.5 for A and H production and three for h
production. For a mass of $\MA=300$ GeV and $\tanb=30$, the ratio of
production modes becomes approximately 0.1 for A and H production and
50 for h production. For both samples the CT10~\cite{Lai:2010vv}
parton distribution function set is used.  Signal samples are
generated using the $A$ boson production mode at discrete values of
$\MA$, with the mass steps chosen by taking the $\tau\tau$ mass
resolution into account.  The signal model is then constructed by
combining three mass samples, one for each of the $h$, $H$ and $A$
bosons, with appropriately scaled cross sections and branching
fractions.  The cross sections and branching fractions, as well as the
masses of the $h$ and $H$ bosons, depend on $\MA$, $\tanb$ and the
MSSM scenario under study. The differences in the kinematic properties
of the decays of CP-odd and CP-even Higgs bosons are expected to be
negligible for this search. Thus the efficiencies and acceptances from
the $A$ boson simulated samples are applicable to all neutral Higgs
bosons.

Background samples of $W$ and $Z$ bosons produced in association with jets are
produced using {\sc Alpgen} 2.14~\cite{Alpgen}, while the high-mass $Z/\gamma^{*}$ tail
is modelled separately using {\sc Pythia8}~\cite{Pythia,Pythia8} 
since in the high-mass range 
the current analysis is rather insensitive to the modelling of $b$-jet production.
$WW$ production is modelled with {\sc Alpgen} 
and $WZ$ and $ZZ$ production is modelled with {\sc Herwig} 6.520~\cite{Herwig}.
The simulation of top pair production uses {\sc Powheg} and {\sc mc@nlo} 4.01~\cite{MCatNLO},
and single-top processes are generated with {\sc AcerMC} 3.8~\cite{AcerMC}. 
All  simulated background samples use the CTEQ6L1~\cite{Pumplin:2002vw} 
parton distribution function set, apart from {\sc mc@nlo}, which uses CT10.

For all the simulated event samples,  the parton shower and 
hadronization are simulated with {\sc Herwig}, {\sc Pythia8} or {\sc Sherpa}.
{\sc Pythia8} is used for {\sc Powheg}-generated samples, 
{\sc Sherpa} for the $b$-associated
signal production and {\sc Herwig} for the remaining samples.
Decays of $\tau$ leptons are generated with 
{\sc Tauola}~\cite{TAUOLA}, {\sc Sherpa} or {\sc Pythia8}. 
{\sc Photos}~\cite{PHOTOS} or {\sc Sherpa} provide additional radiation 
from charged leptons.

\Ztautau{} events form an irreducible background that is particularly 
important when considering low-mass Higgs bosons ($m_A \lesssim 200$~GeV).
It is modelled with  \Zmumu{} events from data, where the muon
tracks and the associated calorimeter cells are replaced by the
corresponding simulated signature of a $\tau$ lepton decay. The two $\tau$ leptons
are simulated by {\sc Tauola}. 
The procedure takes into account the effect of $\tau$ 
polarization and spin correlations \cite{Czyczula:2012ny}. 
In the resulting sample,
the $\tau$ lepton decays and the response of the detector  are
modelled by the simulation, while the underlying event kinematics and all
other properties are obtained from data. 
This $\tau$-embedded \Zmumu{} sample is validated as described in
refs.~\cite{ATLASLimit,SMHtautau2011}.  The $\mu\mu$ event selection
requires two isolated muons in the rapidity range $|\eta|<2.5$, where
the leading muon has $\pt>20$~GeV, the subleading muon $\pt>15$~GeV
and the invariant mass is in the range $m_{\mu\mu}>40$~GeV. This
  results in an almost pure \Zmumu{} sample, which, however, has
  some contribution from $\ttbar$ and diboson production. The contamination
  from these backgrounds that pass the original $\mu\mu$ event selection
  and, after replacement of the muons by tau leptons, enter the final
  event selection are estimated using simulation. Further details can
  be found in section \ref{sec:systematics}. \Ztautau{} events in the invariant mass
  range $m_{\tau\tau}<40$~GeV are modelled using ALPGEN simulated
    events.

\section{Object reconstruction} 
\label{sec:objects}

Electron candidates are formed from energy deposits in the electromagnetic 
calorimeter associated with a charged-particle track measured in the inner 
detector. Electrons are selected if they have a transverse energy 
$\ET > 15$~GeV, lie within $|\eta|<2.47$, but outside  the transition 
region between the barrel and end-cap calorimeters ($1.37<|\eta|<1.52$), 
and meet the ``medium'' identification requirements defined in 
ref.~\cite{ATLAS-CONF-2014-032}. 
Additional isolation criteria, based on tracking and calorimeter
information, are used to suppress backgrounds from misidentified jets
or semileptonic decays of heavy quarks. 
In particular, the sum of the calorimeter deposits in a cone of size
$\Delta R =0.2$ around the electron direction is required to be
less than 6 (8)\% of the electron $\ET$ for the $\tlep\thad$ ($\tlep\tlep$) final state. 
Similarly, the scalar sum of the transverse momentum of tracks with
$\pt > 1$~GeV in a cone of size $\Delta R =0.4$ with respect to the electron direction
is required to be less than 6\% of the electron $\ET$.

Muon candidates are reconstructed by associating an inner detector track
with a muon spectrometer track~\cite{Aad:2014rra}.
For this analysis, the reconstructed muons are required to have a transverse momentum $\pt > 10$~GeV and to lie 
within $|\eta| < 2.5$. Additional track-quality and track-isolation criteria
are required to further suppress backgrounds from cosmic rays, 
hadrons punching through 
the calorimeter, or muons from semileptonic decays of heavy quarks.
The muon calorimetric and track isolation criteria use the same cone
sizes and generally the same threshold values with respect to the muon $\pt$
as in the case of electrons - only for the case of the $\tlep\tlep$ final state is
the muon calorimetric isolation requirement changed to be less than 4\% of the muon momentum.

Jets are reconstructed using the anti-$k_{t}$ algorithm~\cite{AntiKT} with a 
radius parameter $R=0.4$, taking topological clusters~\cite{TopoClusterAlgo}
in the calorimeter as input.
The jet energy is calibrated using a combination of test-beam results, 
simulation and {\it in situ} measurements~\cite{ATLASJETEnergyScale}.
Jets must satisfy $\ET > 20$~GeV and $|\eta| < 4.5$. To reduce the effect of pile-up, it is required that, for jets 
within $|\eta|<2.4$ and $\ET<50$~GeV, at least half of the 
transverse momentum, as measured by the associated charged
particles, be from particles matched to the primary 
vertex.\footnote{The primary vertex is taken 
to be the reconstructed vertex with
the highest $\Sigma \pt^2$ of the associated tracks.}

A multivariate discriminant  is used 
to tag jets, reconstructed within $\left|\eta\right|<2.5$, 
originating from a $b$-quark~\cite{ATLAS-CONF-2014-004}.
The $b$-jet identification has an average efficiency of 70\% 
in simulated $\ttbar$ events, 
whereas the corresponding light-quark jet misidentification probability is 
approximately 0.7\%, but varies as a function of the jet 
$\pt$ and $\eta$ \cite{ATLAS-CONF-2014-046}.

Hadronic decays of $\tau$ leptons (\thad)~\cite{ATLASTauIDNew} are reconstructed starting
from topological clusters in the calorimeter.
A \thad{} candidate must lie within $|\eta| < 2.5$,  have a 
transverse momentum greater than 20~GeV, one or three associated tracks 
and a charge of $\pm 1$. Information on the collimation, isolation, 
and shower profile is combined into a multivariate discriminant against backgrounds from 
jets. Dedicated algorithms that reduce the number of electrons and muons
misreconstructed as hadronic $\tau$ decays are applied.
In this analysis, two \thad{} identification selections are used 
---``loose'' and ``medium''--- with efficiencies 
of about 65\% and 55\%, respectively.

When different objects selected according to the criteria mentioned above
overlap with each other geometrically (within 
$\Delta R = 0.2$) only one of
them is considered. The overlap is resolved by selecting muon, electron,
$\thad$ and jet candidates in this order of priority.

The missing transverse momentum 
is defined as  the negative vectorial sum
of the  muon momenta and energy deposits in the calorimeters~\cite{ATLASmetNEW}.
The magnitude of the missing transverse momentum is denoted by $\met$.
Clusters of calorimeter-cell energy deposits 
belonging to jets, $\thad$ candidates, electrons, and photons, as well as 
cells that are not associated with any object, are treated separately in 
the missing transverse momentum calculation. 
The energy deposits in calorimeter cells that are not matched to any object are weighted by the
fraction of unmatched tracks associated with the primary vertex,
in order to reduce the effect of pile-up on the $\met$ resolution.
The contributions of muons to missing transverse momentum are calculated 
differently for isolated and non-isolated muons, to account for the energy 
deposited by muons in the calorimeters.

%% file: analysiscategories.tex
\section{Search channels} \label{sec:analysiscategories}

The following $\tau\tau$ decay modes are considered in this search:
$\telec\tmuon$~(6\%), $\telec\thad$~(23\%), $\tmuon\thad$~(23\%) 
and $\thad\thad$~(42\%), where  $\telec$ and $\tmuon$ represent the
two leptonic $\tau$ decay modes and the percentages in the parentheses denote 
the corresponding $\tau\tau$  branching fractions.
The selections defined for each of the channels and described
in sections ~\ref{sec:leplep}--\ref{sec:hadhad} are such that there are no
 events common to any two of these channels.

Events are collected using several  single-
and combined-object triggers. The single-electron and single-muon triggers 
require an isolated lepton with a $\pt$ threshold of 24~GeV. 
The single-$\thad$ trigger implements a $\pt$ threshold of 125~GeV. 
The following combined-object triggers are used: 
an electron--muon trigger with lepton $\pt$ thresholds of 
12~GeV and 8~GeV for electrons and muons, respectively, and
a $\thad\thad$ trigger with  $\pt$ thresholds of 38~GeV 
for each hadronically decaying $\tau$ lepton.

With two $\tau$ leptons in the final state, it is not possible to infer the 
neutrino momenta from the reconstructed missing transverse momentum 
vector and, hence, the
$\tau\tau$ invariant mass. Two approaches are used.
The first method used is the Missing Mass Calculator (MMC) \cite{MMCpaper}.
This algorithm assumes that the missing transverse momentum is due
entirely to the neutrinos, and performs a scan over the angles between 
the neutrinos and the visible $\tau$ lepton decay products. 
The MMC mass, \MMC{}, is defined as the  most
likely value chosen by weighting each solution according to 
probability density functions that are derived from simulated $\tau$ lepton decays.
As an example, the MMC resolution,\footnote{The resolution of the mass reconstruction 
is estimated by dividing the root mean square of the mass distribution by its mean.} 
assuming a Higgs boson with mass 
$\MA=150$~GeV, is about 30\% for $\telec\tmuon$ events. The resolution is about 20\% for
$\tlep\thad$ events ($\tlep = \telec$ or $\tmuon$) for Higgs bosons with a 
mass in the range  $150 - 350$~GeV.
The second method uses the $\tau\tau$ total transverse mass, defined as:
\begin{equation*}
 \MTTOT = \sqrt{\mt ^{2}(\tau_{1},\tau_{2}) +\mt ^{2}(\tau_{1},\MET) + \mt ^{2}(\tau_{2},\MET)}~~,
\end{equation*}
where the transverse mass, $\mt$, between two objects with transverse momenta 
$p_{\text{T}1}$ and $p_{\text{T}2}$ and relative angle $\Delta\phi$ is given by
\begin{equation*}
 \mt = \sqrt{2p_{\text{T}1}p_{\text{T}2}(1-\cos\Delta\phi)}~~. 
\end{equation*}
As an example, the $\MTTOT$ mass resolution assuming a Higgs boson with mass $m_A=350$~GeV
for $\thad\thad$ events is approximately 30\%.
While the MMC exhibits a better $\tau\tau$ mass resolution for signal events, 
multi-jet background events tend to be reconstructed at lower masses with $\MTTOT$, 
leading to better overall discrimination between signal and background for topologies 
dominated by multi-jet background.

%% file: leplep.tex
\subsection{The $h/H/A\to\telec\tmuon$ channel} \label{sec:leplep}

Events in the \Htautaull{} channel are selected using either  
single-electron  or  electron--muon triggers. 
The data sample corresponds to an integrated
luminosity of \currentlumi.
Exactly one isolated electron and one isolated 
muon of opposite  charge are required,
with lepton $\pt$ thresholds of
15~GeV for electrons and 10~GeV for muons. 
Electrons with $\pt$ in the range 15--25~GeV are from events 
selected by the electron--muon trigger, whereas electrons 
with $\pt>25$~GeV are from events selected by the single-electron trigger.
Events containing 
hadronically decaying $\tau$ leptons, satisfying the ``loose'' $\thad$
identification criterion, are vetoed.

\begin{figure}[h!]
  \centering

  \subfigure[]{\label{fig:leplep_tag_dphi}
  \includegraphics[width=0.47\textwidth]{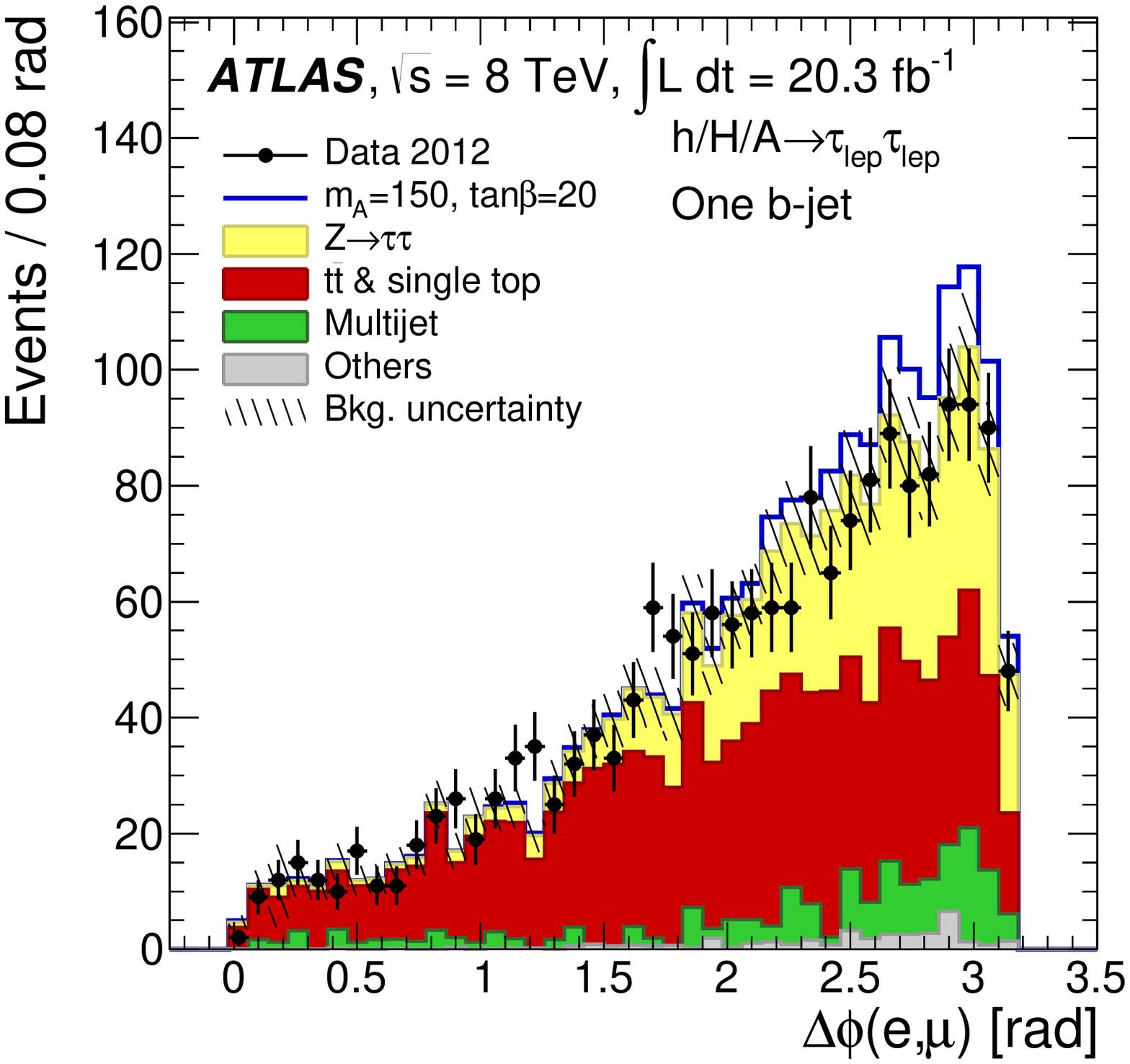} }
  \subfigure[]{\label{fig:leplep_veto_sumcosdphi}
  \includegraphics[width=0.47\textwidth]{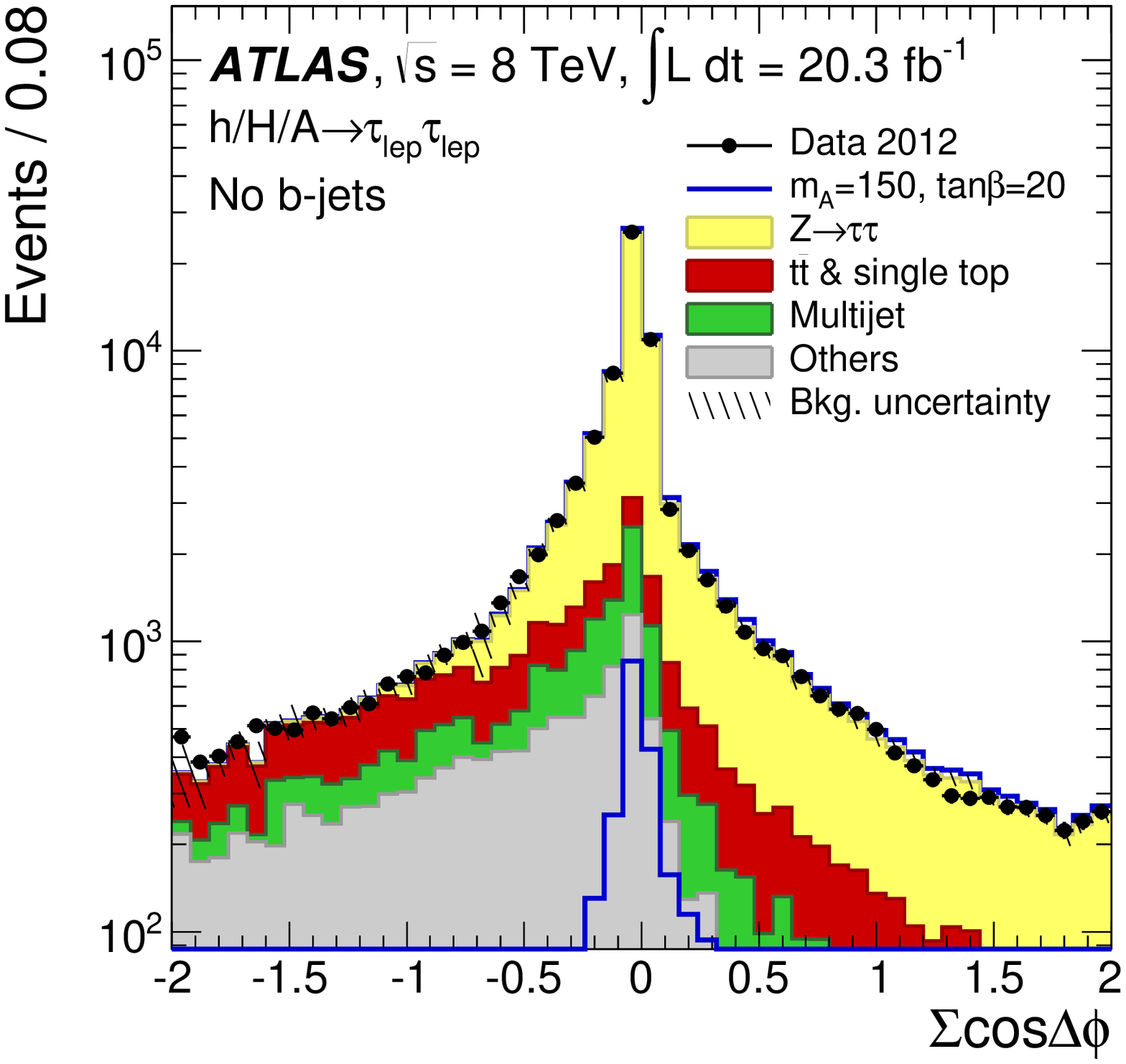} }

  \caption{ Kinematic distributions for the \Htautaull{} channel:
(a) the $\Delta\phi(e,\mu)$ distribution after the tag category 
selection criteria apart from the $\Delta\phi(e,\mu)$ requirement and
(b) the $\Sigma\cos\Delta\phi$ distribution 
after the $b$-jet veto requirement.
The data are compared to the background expectation and a
hypothetical MSSM signal ($\MA=150$~GeV and $\tanb=20$).
In (b) the assumed signal is shown twice:   as a distribution
in the bottom of the plot and on top of the
total background prediction.
The background uncertainty includes statistical and systematic 
uncertainties.}
\label{fig:leplep_distributions}
\end{figure}

To increase the sensitivity of this channel, the events are split
into two categories based on the presence (``tag category'')
or absence (``veto category'') of a $b$-tagged jet. 
The tag category requires
exactly one jet satisfying the $b$-jet identification criterion. 
In addition, a number of kinematic requirements are imposed to reduce the 
background from top quark decays.
The azimuthal angle between the electron
and the muon, $\Delta\phi(e,\mu)$, must be greater than 2.0 
(see figure~\ref{fig:leplep_tag_dphi}). The sum of the
cosines of the azimuthal angles between the leptons and the missing
transverse momentum, 
$\Sigma \cos \Delta\phi \equiv \cos(\phi(e) - \phi(\MET)) + \cos(\phi(\mu) - \phi(\MET))$,
must be greater than $-0.2$. The scalar sum of the $\pt$ of jets with
$\pt>30$~GeV must be less than 100~GeV. Finally, the scalar sum of the $\pt$ 
of the leptons  and the \MET{} must be below 125~GeV. The veto category is 
defined by requiring that no jet satisfies the 
$b$-jet identification criterion.
Because the top quark background is  smaller
in this category, the
imposed kinematic selection requirements, $\Delta\phi(e,\mu)>1.6$ and 
$\Sigma \cos \Delta\phi > -0.4$ (see figure~\ref{fig:leplep_veto_sumcosdphi}), 
are looser than in the tag category.

The most important  background processes 
in this channel are $\Zjets$, $\ttbar$, and multi-jet production.
The \Ztautau{} background is estimated using the $\tau$-embedded \Zmumu{} 
sample outlined in section~\ref{sec:samples}.  
It is normalized using the NNLO $\Zjets$ cross section calculated 
with FEWZ~\cite{Anastasiou:2003ds} and 
a simulation estimate of the efficiency of the trigger, lepton 
$\eta$ and $\pt$, and identification requirements.
The $\ttbar$ background is
estimated from simulation with the normalization taken from a  data control region
enriched in $\ttbar$ events, defined by requiring two $b$-tagged jets. 
The $W$+jet background, where one of the leptons results 
from a misidentified jet, is estimated using simulation.
Smaller backgrounds from single-top and diboson production are also 
estimated from simulation.

The multi-jet background is estimated from data using a two-dimensional 
sideband method. The event sample is split into four regions according 
to the charge product of the 
$e\mu$ pair and the isolation requirements on the electron and muon. 
Region $A$ ($B$) contains events where both leptons pass 
the isolation requirements and are of opposite (same) charge, while 
region $C$ ($D$)  contains events where both leptons 
fail the isolation requirements and are also of opposite (same) charge. 
This way, $A$ is the signal region, while $B$, $C$, and $D$ 
are control regions.
Event contributions to the $B$, $C$ and $D$ control regions from 
processes other than
multi-jet production are estimated using simulation and subtracted. 
The final prediction for the multi-jet contribution to the signal region, $A$, 
is given by the background-subtracted data in region $B$, scaled by the 
opposite-sign to same-sign ratio measured in regions $C$ and $D$, 
$r_{C/D} \equiv n_{C}/n_{D}$. Systematic uncertainties
on the prediction are estimated from the stability of $r_{C/D}$ 
under variations of the lepton isolation requirement.

\begin{figure}
  \centering

\subfigure[]{
  \includegraphics[width=0.44\textwidth]{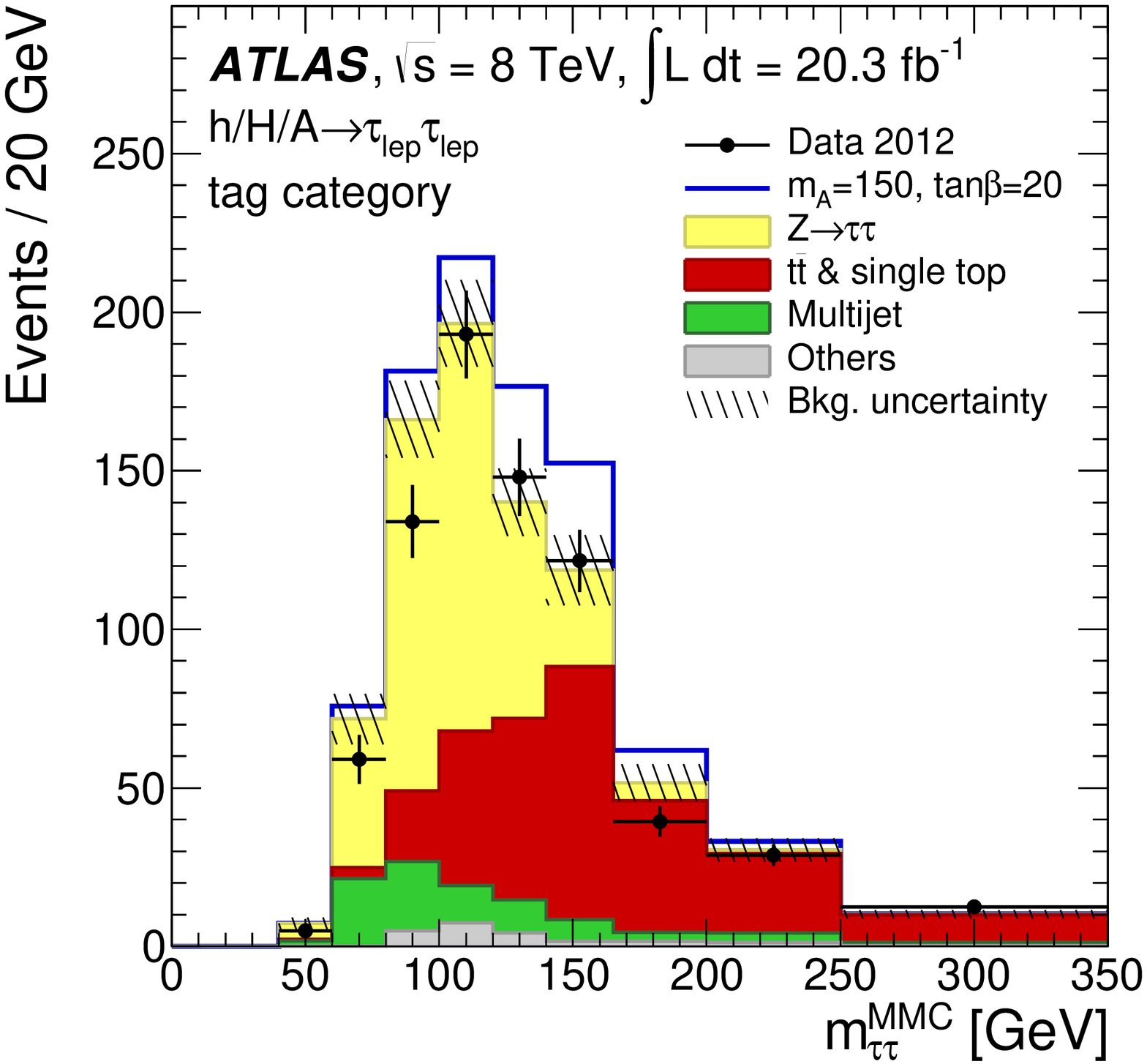}}
\subfigure[]{
  \includegraphics[width=0.44\textwidth]{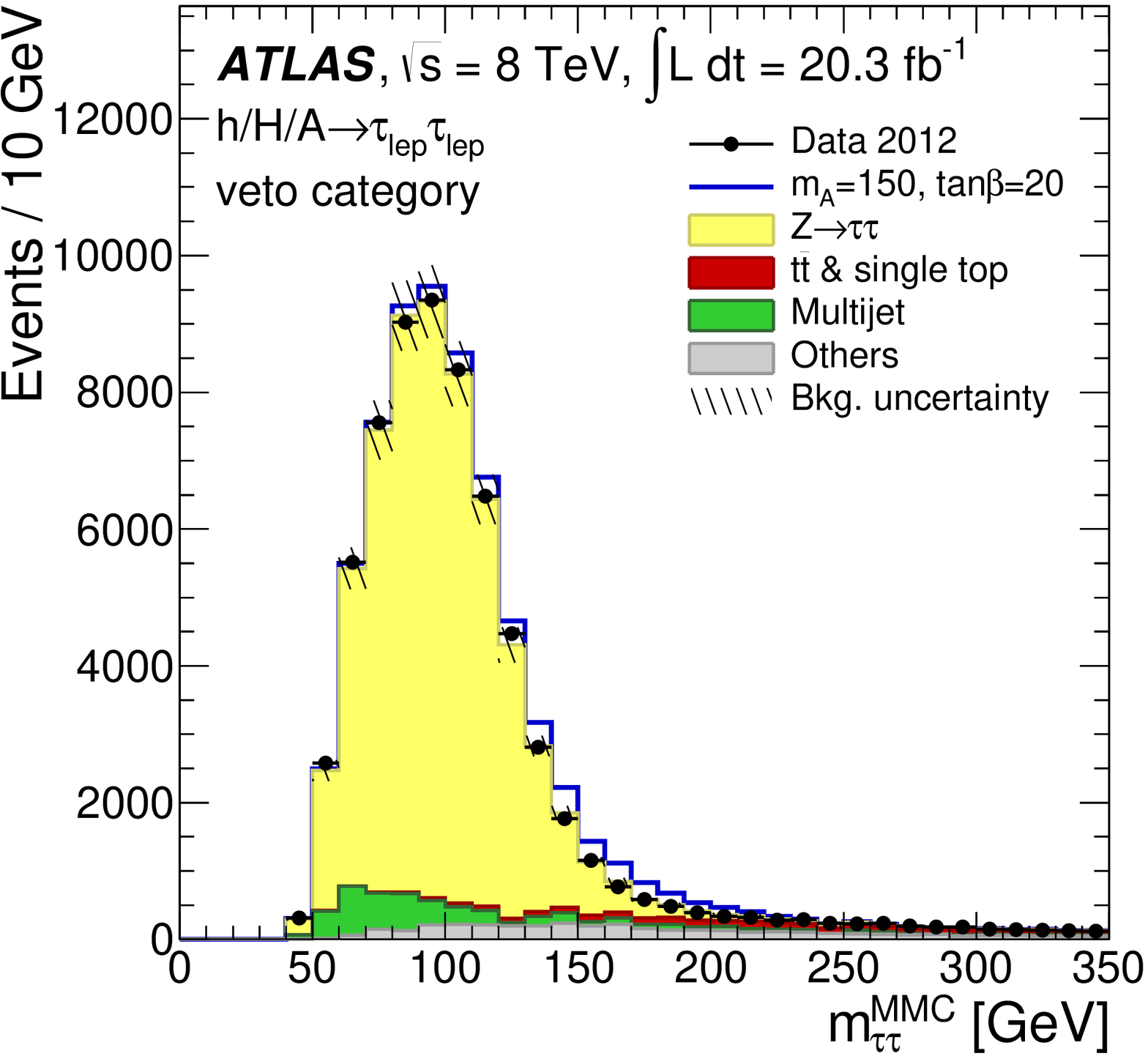}}

  \caption{MMC mass distributions for the \Htautaull{} 
channel.  The MMC mass is shown for (a) the tag   
and (b) the veto  categories.
The data are compared to the background expectation and a
hypothetical MSSM signal ($\MA=150$~GeV and $\tanb=20$). 
The contributions of the diboson, $Z/\gamma^{*}\to ee/\mu\mu$, and \Wjets{} 
background processes are combined and labelled ``Others''.
The background uncertainty includes statistical and systematic uncertainties.
}
\label{fig:leplep_mmc}
\end{figure}

Table~\ref{tab:leplep_eventyield} shows the number of observed
$\telec\tmuon$ events, the predicted  background,
and the signal prediction for the MSSM \mhmax{} scenario 
\cite{Heinemeyer:1999zf,MSSMmhmax} 
parameter choice $m_A = 150$~GeV and $\tan\beta=20$.
The  total combined statistical and systematic uncertainties 
on the  predictions are also quoted on  table~\ref{tab:leplep_eventyield}.
The observed event yields are compatible 
with the expected yields from SM processes.
The MMC mass is used as the discriminating variable in this channel, and is shown in 
figure~\ref{fig:leplep_mmc} for the tag and veto categories separately.

\begin{table} 
\centering
\begin{tabular}{l c c }
& \multicolumn{1}{c}{ Tag category } & \multicolumn{1}{c}{Veto category }  \\ 
\hline \hline
  \multicolumn{3}{l}{Signal  ($m_A=150~\text{GeV},~\tan\beta=20$)} \\
 $h \to\tau\tau$ &  \phantom{0.}8.7  $\pm$  1.9 & 244  $\pm$ 11   \\

 $H \to\tau\tau$ &  \phantom{0.}65 $\pm$ 14    & 882  $\pm$ 45    \\
 $A \to\tau\tau$ &  \phantom{0.}71 $\pm$ 15    & 902  $\pm$ 48    \\

\hline\hline

$\Ztautau$+jets         & 418    $\pm$ 28   & 54700  $\pm$ 3800    \\
Multi-Jet               & 100    $\pm$ 21   &  4180  $\pm$ 670    \\
$\ttbar$ and single top & 421    $\pm$ 46   &  2670  $\pm$ 360    \\
Others                  &  25.8  $\pm$  7.4   &  4010  $\pm$ 280    \\

\hline
Total background        & 965    $\pm$ 59  & 65500  $\pm$ 3900    \\
\hline\hline
Data                    & 904      & 65917  \\
\hline
\hline
\end{tabular}
\caption{Number of events observed in the \Htautaull{} channel
and the predicted background and signal.
The predicted signal event yields correspond to 
the parameter choice $\MA=150$~GeV and $\tanb=20$.
The row labelled ``Others'' includes events from diboson production, 
$Z/\gamma^{*}\to ee/\mu\mu$ and $W$+jets production.
Combined statistical and systematic uncertainties are quoted.
The signal prediction does not include the uncertainty due to the cross-section calculation.
\label{tab:leplep_eventyield}
}

\end{table}

\clearpage

%% file: lephad.tex
\subsection{The $h/H/A\to\tlep\thad$ channel} \label{sec:lephad}

Events in the \Htautaulh{} channel 
are selected using single-electron
or single-muon triggers. The data sample  corresponds to an integrated
luminosity of \currentlumi.
Events are required to contain an electron or a muon with 
$\pt > 26$~GeV and an oppositely charged $\thad$ with $\pt>20$~GeV
satisfying the ``medium'' \thad{} identification criterion.
Events must not contain additional electrons or muons.
The event selection is optimized separately for low-
and high-mass Higgs bosons in order to exploit differences in kinematics
and background composition. 

\begin{figure}[h!]
  \centering

  \subfigure[]{\label{fig:lephad_sdp}
  \includegraphics[width=0.47\textwidth]{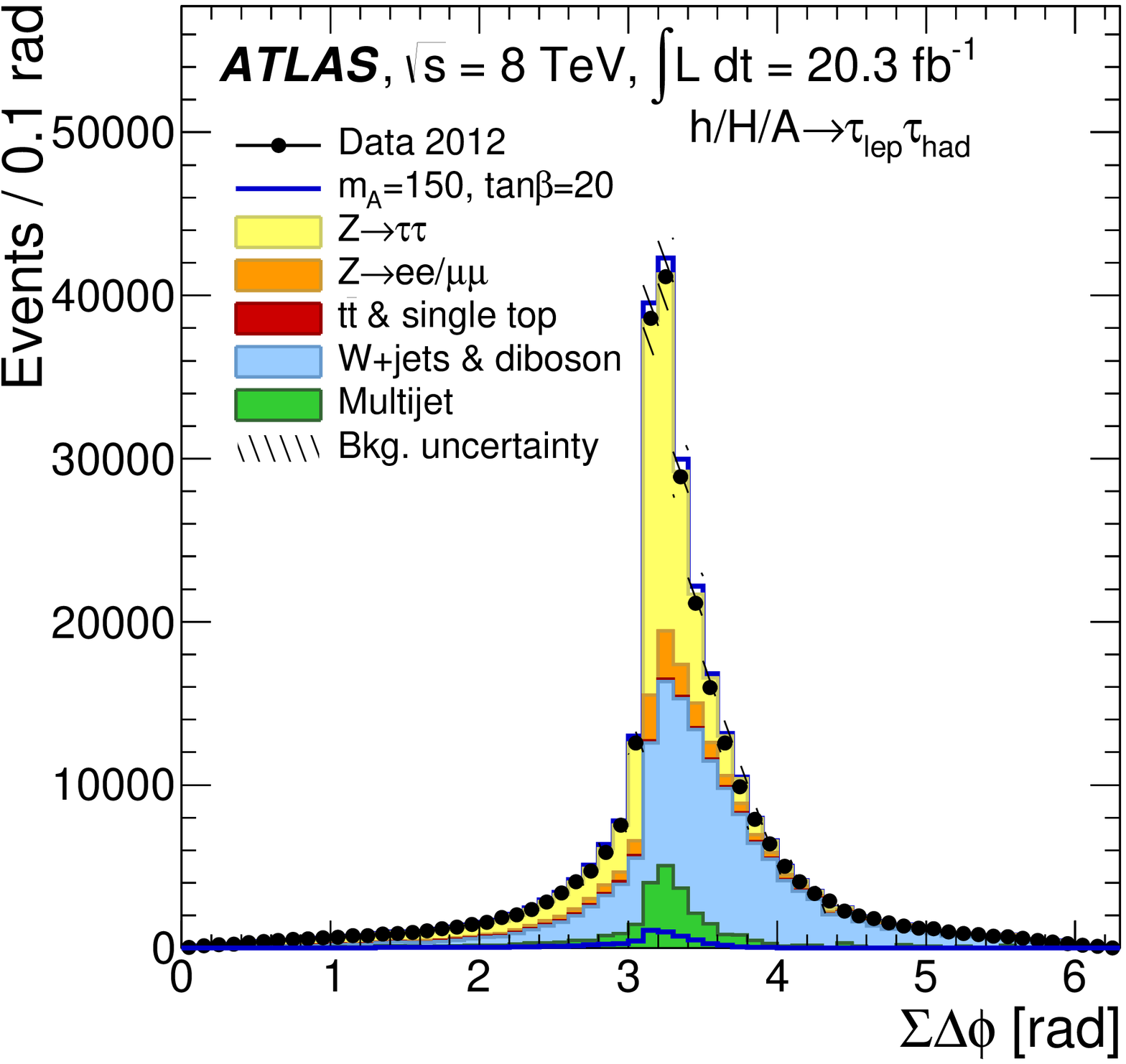} }
  \subfigure[]{\label{fig:lephad_highmass_deltapt}
  \includegraphics[width=0.47\textwidth]{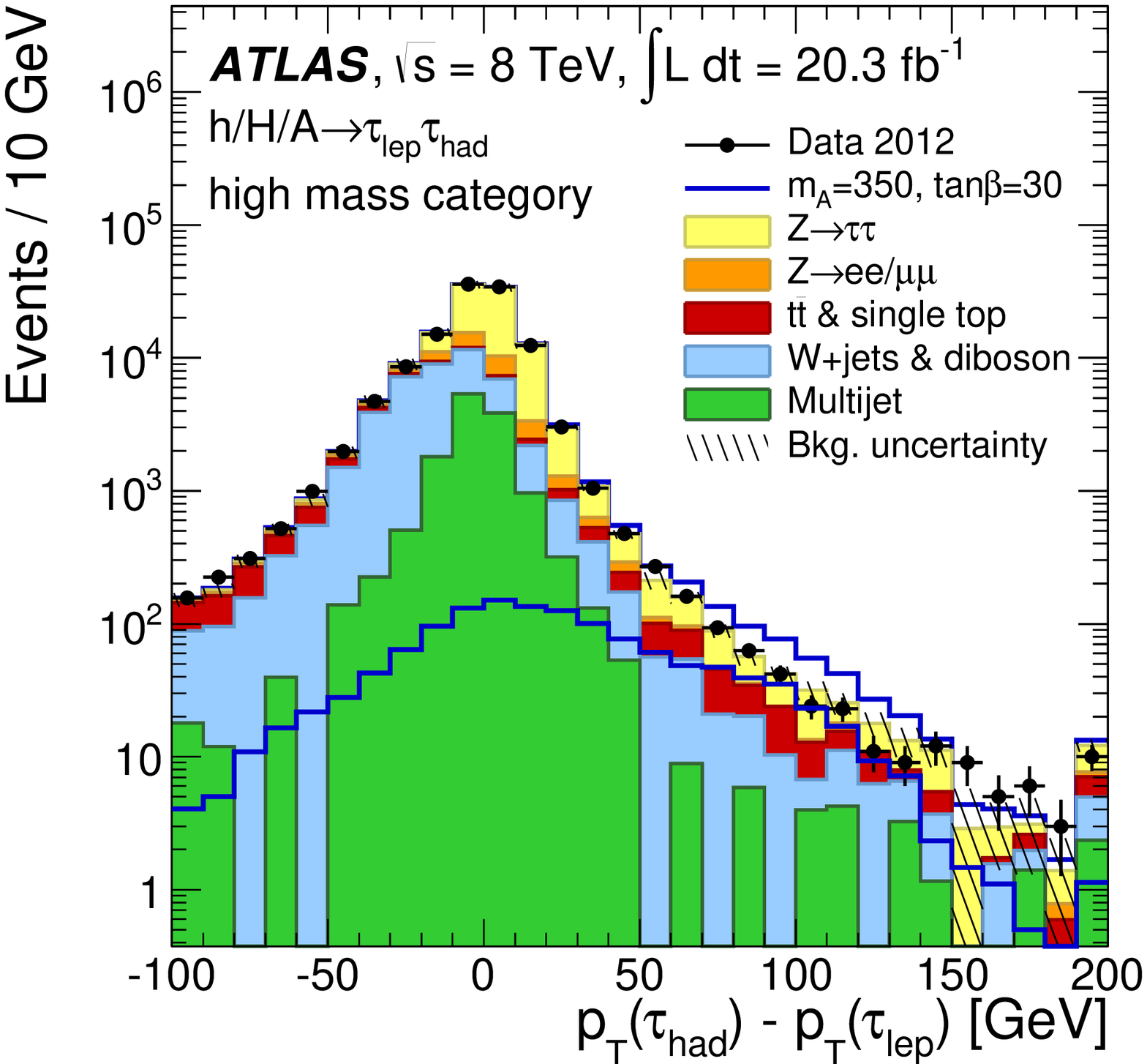} }

  \caption{ Kinematic distributions for the \Htautaulh{} channel:
(a) the $\Sigma\Delta\phi$ distribution after the
kinematic requirements on the $\tlep$ and $\thad$ and
(b) the distribution of 
$\Delta\pt \equiv \pt(\thad)-  \pt(\mathrm{lepton})$
for the high-mass category 
for the combined $\telec\thad$ and $\tmuon\thad$ 
final states. In (b) all the $\tlep\thad$ high-mass selection criteria
are applied apart from the $\Delta\pt > 45$~GeV requirement.
The data are compared to the background expectation and a
hypothetical MSSM signal: $\MA=150$~GeV, $\tanb=20$ for (a)
and $\MA=350$~GeV, $\tanb=30$ for (b). 
The assumed signal is shown twice:   as a distribution
in the bottom of the plot and on top of the
total background prediction.
The background uncertainty includes statistical and systematic 
uncertainties.}
\label{fig:lephad_distributionds}
\end{figure}

The low-mass selection targets the parameter space with $\MA<200$~GeV.
It includes two orthogonal categories: the tag category and the veto category.
In the tag category there must be at least one jet tagged as a $b$-jet.
Events that contain one or more jets with $\pt > 30$~GeV, without taking into account
the leading $b$-jet, are rejected.
In addition, the transverse mass of the lepton and the transverse missing momentum 
is required to not exceed 45~GeV. These requirements 
serve to reduce the otherwise dominant $\ttbar$ background.
In the veto category there must be no jet tagged as a $b$-jet.
Two additional selection requirements are applied to 
reduce the $\Wjets$ background. First, the transverse mass of 
the lepton and the missing transverse momentum must be below 60~GeV. Secondly, 
the sum of the azimuthal angles 
$\Sigma\Delta\phi \equiv \Delta\phi(\thad,\MET)+\Delta\phi(\tlep,\MET)$,
must have a value less than 3.3 (see figure~\ref{fig:lephad_sdp}). 
Finally, in the $\tmuon\thad$ channel of 
the veto category,  dedicated requirements based on kinematic
and shower shape properties of the $\thad$ candidate are applied to reduce the
number of muons faking hadronic $\tau$ lepton decays.

The high-mass selection targets $\MA \ge 200$~GeV. 
It  requires $\Sigma\Delta\phi < 3.3$, 
in order to reduce the $W+$jets background.
The hadronic and leptonic $\tau$ lepton decays are
required to be  back-to-back:
$\Delta\phi(\tlep,\thad) > 2.4$. In addition, the transverse 
momentum difference between the $\thad$ and the lepton, 
$\Delta\pt \equiv \pt(\thad)-  \pt(\mathrm{lepton})$, must be above 45~GeV
(see figure~\ref{fig:lephad_highmass_deltapt}).
This  requirement takes advantage of the fact
that a $\thad$  tends to have a higher visible transverse momentum 
than a $\tlep$ due to the presence of more neutrinos in the latter decay.

In the low-mass categories, the electron and muon channels are treated
separately and combined statistically. For the high-mass category, they 
are treated as a single channel to improve the statistical robustness.

The most important SM background processes 
in this channel are $\Zjets$, $\Wjets$, multi-jet production, top
(including both $\ttbar$ and single top) and diboson production.
%%%
The $\tau$-embedded \Zmumu{} sample is used to estimate the \Ztautau{}
background. It is normalized in the same way as in the $\tlep\tlep$
channel. The rate at which electrons are misidentified as $\thad$,
important mostly for $Z\to ee$ decays, was estimated from data in
ref.~\cite{ATLASTauIDNew}. The contribution of diboson processes is
small and estimated from simulation.  Events originating from
$\Wjets$, $Z(\to \ell\ell)+$~jets ($\ell=e,~\mu$), $\ttbar$ and
single-top production, in which a jet is misreconstructed as $\thad$,
are estimated from simulated samples with normalization estimated by
comparing event yields in background-dominated control regions in
data. Separate regions are defined for each of the background sources
in each of the low-mass tag, low-mass veto, and high-mass categories.
Systematic uncertainties are derived using alternative definitions for
the control regions.  The multi-jet background is estimated with a
two-dimensional sideband method, similar to the one employed for the
$\telec\tmuon$ channel, using the product of the lepton ($e$ or $\mu$)
and \thad{} charges and lepton isolation. The systematic uncertainty
on the predicted event yield is estimated by varying the definitions
of the regions used, and by testing the stability of the $r_{C/D}$
ratio across the \MMC{} range.

\begin{figure}[h!]
  \centering

\subfigure[]{
  \includegraphics[width=0.49\textwidth]{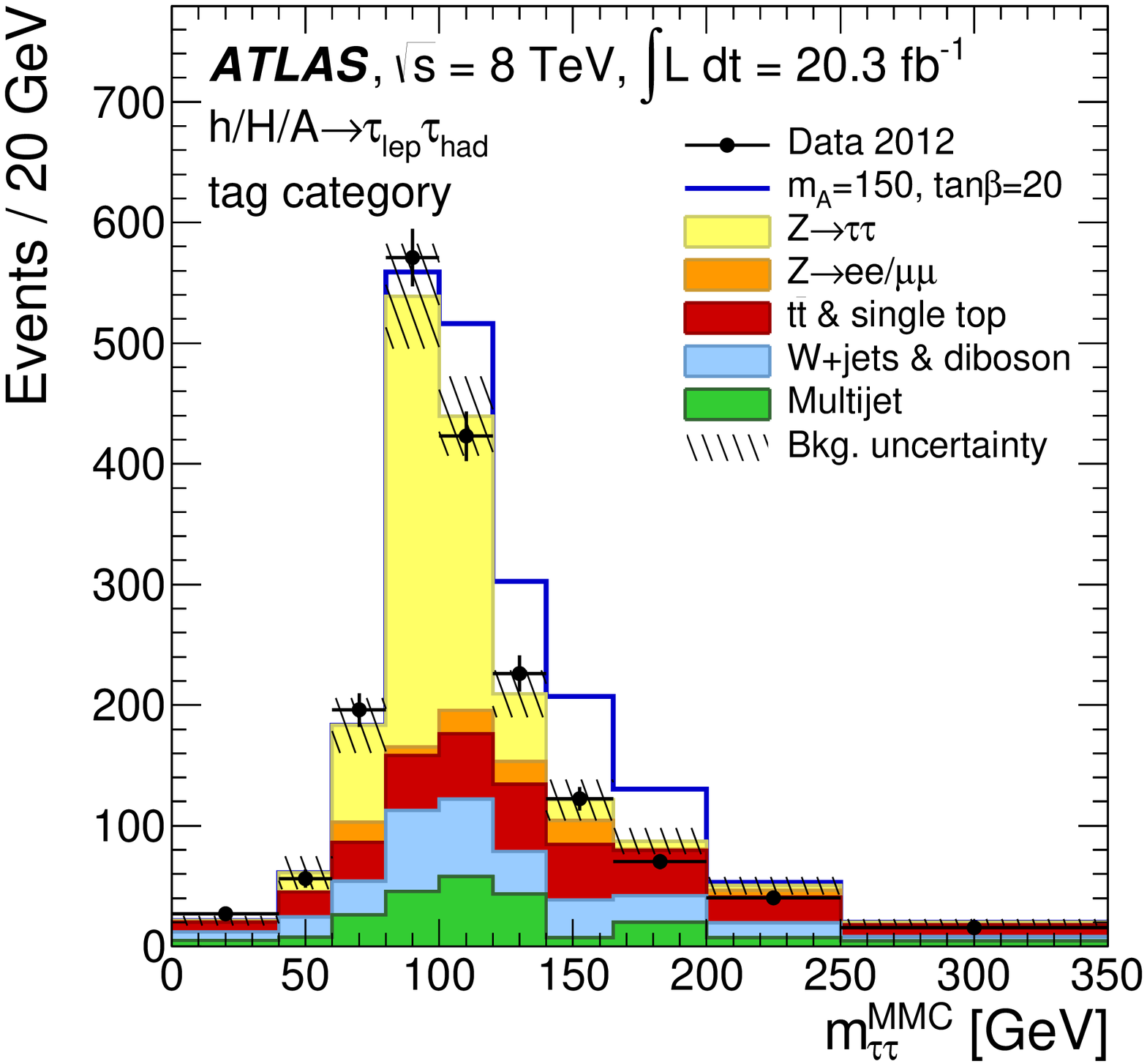}}
\subfigure[]{
  \includegraphics[width=0.49\textwidth]{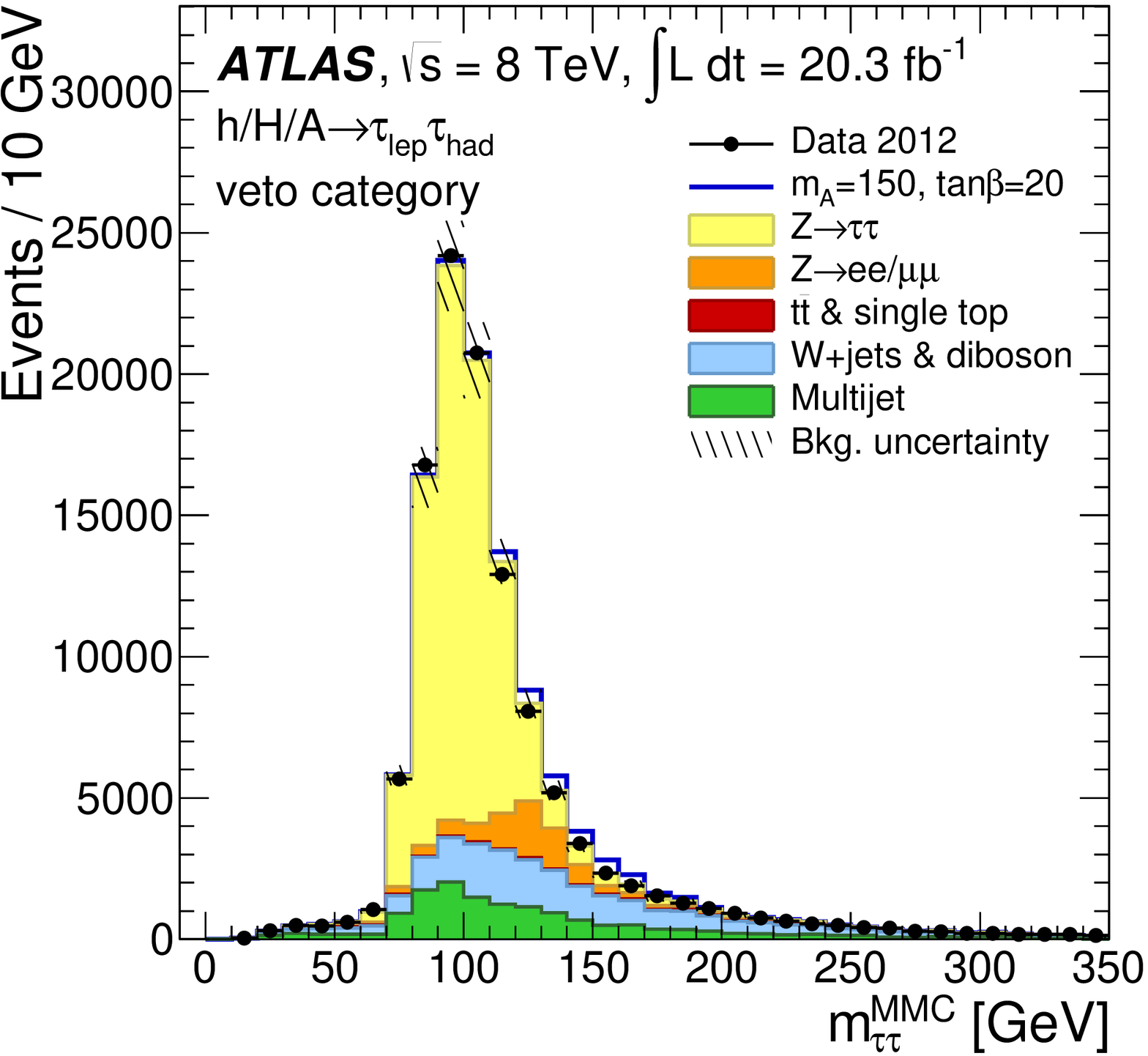}}

  \caption{The MMC mass distributions 
for the low-mass categories of the \Htautaulh{} channel. 
Tag (a) and veto (b) categories are shown 
for the combined $\telec\thad$ and $\tmuon\thad$ final states. 
The data are compared to the background expectation and a
hypothetical MSSM signal ($\MA=150$~GeV and $\tanb=20$). 
The background uncertainty includes statistical and systematic 
uncertainties.}
\label{fig:lephad_lowmass}
\end{figure}

\begin{figure}[h!]
  \centering

  \includegraphics[width=0.49\textwidth]{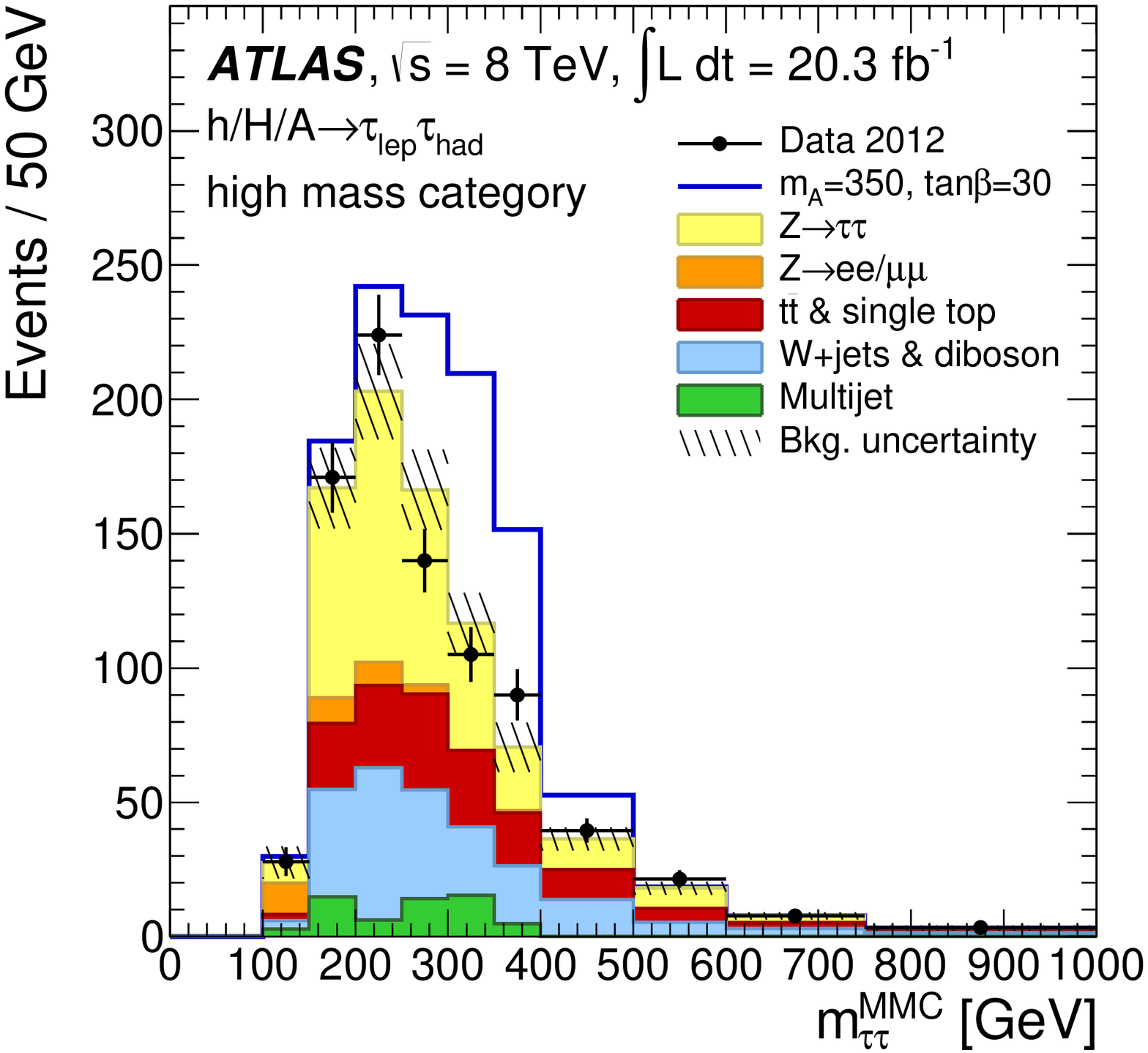}

  \caption{The MMC mass distribution for the 
high-mass category of the \Htautaulh{} channel is shown for the 
combined $\telec\thad$ and $\tmuon\thad$ final states.
The data are compared to the background expectation and a
hypothetical MSSM signal ($\MA=350$~GeV and $\tanb=30$). 
The background uncertainty includes statistical and systematic 
uncertainties.}
\label{fig:lephad_highmass}
\end{figure}

Table~\ref{tab:lephad_eventyield} shows the number of observed
$\tlep\thad$ events, the predicted background, 
and the signal prediction for the MSSM \mhmax{} scenario.
The signal MSSM parameters are  $m_A = 150$~GeV, $\tan\beta=20$ for the 
low-mass categories and $m_A = 350$~GeV, $\tan\beta=30$ for the high mass
category.
The  total combined statistical and systematic uncertainties 
on the  predictions are also quoted in table~\ref{tab:lephad_eventyield}.
 The observed event yields are compatible 
with the expected yields from SM processes within the uncertainties.
The MMC mass is used as the final mass discriminant in this channel
and is shown in figures~\ref{fig:lephad_lowmass} and \ref{fig:lephad_highmass}
for the low- and high-mass categories, respectively.

\begin{table}[h!]
    \centering
\begin{tabular}{l c c c c }

                       \multicolumn{5}{c}{Low-mass categories} \\
  \hline    \hline  
  &   \multicolumn{2}{c}{Tag category}  &   \multicolumn{2}{c}{Veto category}  \\
  &   \multicolumn{1}{c}{$e$ channel}   &  \multicolumn{1}{c}{$\mu$ channel}    &  \multicolumn{1}{c}{$e$ channel}  & \multicolumn{1}{c}{$\mu$ channel} \\
  \hline  \hline
     \multicolumn{5}{l}{Signal ($m_A=150~\text{GeV},~\tan\beta=20$) } \\ 
  $h\to \tau\tau$                     &   \phantom{0}10.5 $\pm$ 2.8  & 10.5 $\pm$ 2.6          & \phantom{0}194 $\pm$ 13 &   \phantom{00}192  $\pm$  14\phantom{00}     \\
  $H\to \tau\tau$                     &   \phantom{00}86 $\pm$ 26    & \phantom{0}86 $\pm$ 24  & \phantom{0}836 $\pm$ 60 &   \phantom{00}822  $\pm$  61\phantom{00}     \\
  $A\to \tau\tau$                     &   \phantom{00}94 $\pm$ 29    & \phantom{0}94 $\pm$ 27  & \phantom{0}840 $\pm$ 64 &   \phantom{00}825  $\pm$  62\phantom{00}     \\

  \hline  \hline

 $Z\to\tau\tau$+jets                  &   \phantom{0}403 $\pm$ 39 & 425 $\pm$ 42           & 31700 $\pm$ 2800                       &38400 $\pm$ 3300  \\    
 $Z\to \ell\ell$+jets ($\ell=e,~\mu$) &   \phantom{00}72 $\pm$ 24 & \phantom{0}33 $\pm$ 14 & \phantom{0}5960  $\pm$  920\phantom{0} &\phantom{0}2860 $\pm$ 510\phantom{0}  \\
 $W$+jets                             &   \phantom{0}158 $\pm$ 44 & 185 $\pm$ 58           & \phantom{0}9100  $\pm$ 1300            & \phantom{0}9800 $\pm$ 1400   \\   
 Multi-jet                            &   \phantom{0}185 $\pm$ 35 & \phantom{0}66 $\pm$ 31 & 11700 $\pm$ 490\phantom{0}             & \phantom{0}3140 $\pm$  430\phantom{0}   \\   
 $\ttbar$ and single top              &   \phantom{0}232 $\pm$ 36 & 236 $\pm$ 34           & \phantom{00}533 $\pm$ 91\phantom{00}   & \phantom{00}535  $\pm$ 98\phantom{00}   \\   
 Diboson                              &   \phantom{00}9.1 $\pm$ 2.3   & 10.0 $\pm$ 2.5     &  \phantom{00}466 $\pm$ 40\phantom{00}        & \phantom{00}468 $\pm$ 42\phantom{00}    \\  
  \hline                                                                                                         
                                                                                                   
 Total background                     &    1059 $\pm$ 81              &    955 $\pm$ 86   &   59500 $\pm$ 3300     &    55200 $\pm$ 3600   \\
  \hline    \hline                                                                                                        
 
 Data                                 &    1067     &    947  &    60351  &     54776  \\
 
 \hline    \hline  
\vspace{0.3cm}
 \end{tabular}
 
 \centering
 
 \begin{tabular}{l c }

   \multicolumn{2}{c}{High-mass category}    \\
  \hline  \hline 
 
  \multicolumn{2}{l}{Signal  ($m_A=350~\text{GeV},~\tan\beta=30$)} \\
  $h\to\tau\tau$            &  \phantom{0}5.60\phantom{0}$\pm$\phantom{0}0.68     \\ 
  $H\to\tau\tau$          &   157\phantom{0}$\pm$\phantom{0}13     \\ 
  $A\to\tau\tau$          &   152\phantom{0}$\pm$\phantom{0}13     \\ 

  \hline  \hline

 $Z\to\tau\tau$+jets     &  \phantom{.}380\phantom{0}$\pm$\phantom{0}50\phantom{.}          \\
 $Z\to \ell\ell$+jets ($\ell=e,~\mu$)     &   34.9\phantom{0}$\pm$\phantom{0}7.3   \\
 $W$+jets                &  \phantom{.}213\phantom{0}$\pm$\phantom{0}40\phantom{.}          \\
 Multi-jet               &  \phantom{.0}57\phantom{0}$\pm$\phantom{0}20\phantom{.}   \\
 $\ttbar$ and single top &  \phantom{.}184\phantom{0}$\pm$\phantom{0}26\phantom{.}    \\
 Diboson                 &  30.1\phantom{0}$\pm$\phantom{0}4.8   \\
  \hline 
 Total background        &  \phantom{.}900\phantom{0}$\pm$\phantom{0}72\phantom{.}      \\   
 
  \hline \hline                                                                                             

 Data        &  920        \\   
 \hline    \hline   
\end{tabular}

\caption{Numbers of events observed in the \Htautaulh{} channel
and the predicted background and signal.
The predicted signal event yields correspond to 
the parameter choice $\MA=150$~GeV, $\tanb=20$ for the low-mass categories
and $\MA=350$~GeV, $\tanb=30$ for the high-mass category. 
Combined statistical and systematic uncertainties are quoted.
The signal prediction does not include the uncertainty due to the cross-section calculation.
\label{tab:lephad_eventyield}
}

\end{table}

\clearpage

%% file: hadhad.tex
\subsection{The $h/H/A\to\thad\thad$ channel} \label{sec:hadhad}

\begin{figure}[h!]
  \centering

  \subfigure[]{\label{fig:hadhad_distributions-1}
  \includegraphics[width=0.44\textwidth]{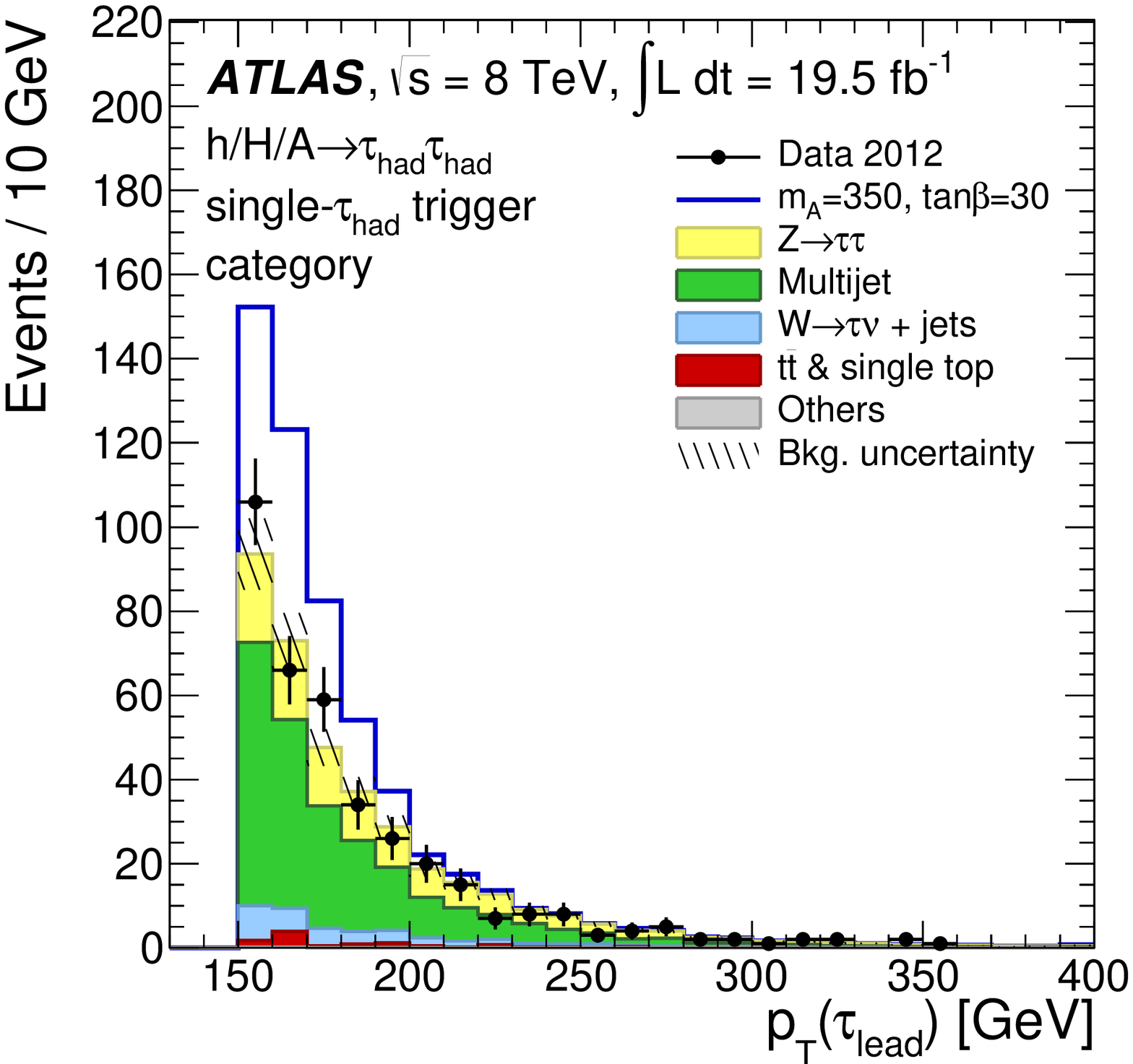} }
  \subfigure[]{\label{fig:hadhad_distributions-2}
  \includegraphics[width=0.44\textwidth]{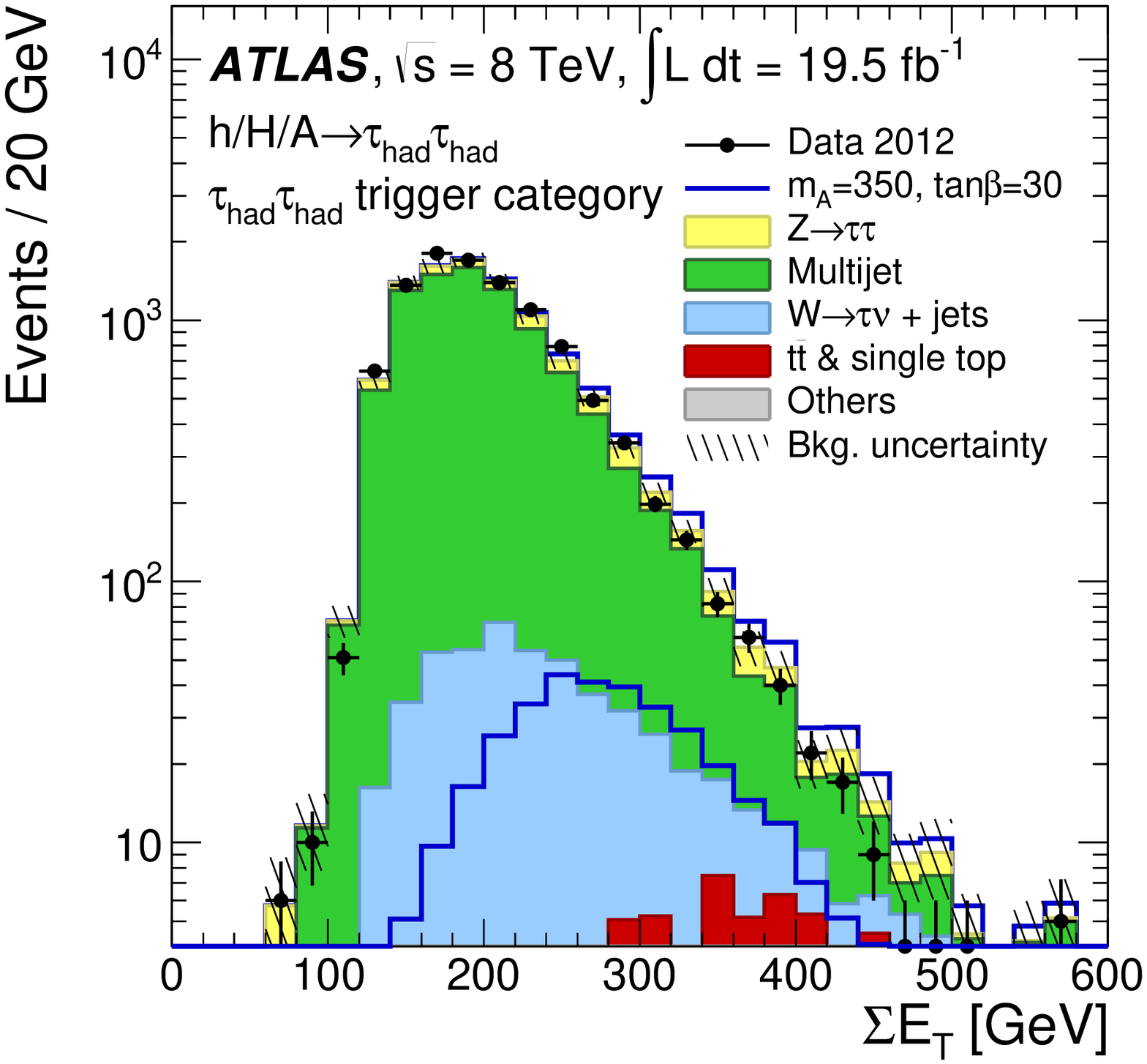} }

  \caption{Kinematic distributions for the \Htautauhh{} channel:
(a) the transverse momentum of the highest-$\pt$ $\thad$
for the STT category and (b) the scalar 
sum of transverse energy of all deposits, $\Sigma E_{\text{T}}$,
in the DTT category, before the application of this requirement. 
The data are compared to the background expectation and a
hypothetical MSSM signal ($\MA=350$~GeV and $\tanb=30$). 
The background labelled ``Others'' includes 
events from diboson production, $Z\to \ell\ell$ 
and $W\to \ell\nu$ with $\ell=e,~\mu$.
In (b) the assumed signal is shown twice:   as a distribution
in the bottom of the plot and on top of the
total background prediction.
The background uncertainty includes statistical and systematic 
uncertainties.
}
\label{fig:hadhad_distributions}
\end{figure}

Events in the \Htautauhh{} channel are selected using either a single-$\thad$ 
trigger or a $\thad\thad$ trigger. The data sample  corresponds to an 
integrated luminosity of \currentlumihh.
Events are required to contain at least two \thad{}, identified using
the ``loose''  identification criterion. If more than two \thad{} are present, 
the two with the highest \pt{} values are considered. 
Events containing an electron 
or muon are rejected to ensure orthogonality with the other channels. 
The two \thad{} are required to have $\pt>50$~GeV,  
have opposite electric charges, and to be back-to-back in the azimuthal plane 
($\Delta\phi > 2.7$). 
Two event categories are defined as follows.
The single-\thad{} trigger category (STT category)
includes the events selected by the single-\thad{} trigger 
which contain at least one \thad{} with $\pt > 150$~GeV 
(see figure~\ref{fig:hadhad_distributions-1}).
The  \thad\thad{} trigger category (DTT category)
includes the events selected by the $\thad\thad$ trigger,
with the leading \thad{} required to have $\pt$ less than 150~GeV, to ensure orthogonality with the STT category, 
and with both $\tau$ leptons satisfying the ``medium'' identification criterion.
In addition, events in the DTT category are required to have $\MET>10$~GeV, 
and the scalar sum of transverse energy of all deposits 
in the calorimeter to be greater than 160~GeV (see figure~\ref{fig:hadhad_distributions-2}). 

The dominant background in this channel is multi-jet production
and for this reason  $\MTTOT$ is used as the final discriminant.
Other background samples include $\Zjets$, $\Wjets$, $\ttbar$ and diboson.

The  multi-jet background is estimated separately for the STT and DTT categories. 
In the STT category, a control region is obtained by requiring 
the next-to-highest-\pt{} \thad{} to fail the ``loose''   
\thad{} identification requirement, thus obtaining a
 high-purity sample of multi-jet events. 
The probability of a jet to be misidentified as a  \thad{} is measured 
in a high purity sample of dijet events in data, as a function of 
the number of associated tracks with the jet and the jet \pt. 
These efficiencies are used to obtain the shape and the 
normalization of the multi-jet background from the 
control region with the next-to-highest-\pt{} \thad{} that fails the 
\thad{} identification requirement. 
The systematic uncertainty on the method is obtained by repeating the
multijet estimation, but requiring either a same-sign or opposite-sign
between the two jets. The difference between the calculated
efficiencies for the two measurements is then taken as the systematic
uncertainty. This procedure has some sensitivity to
differences related to whether the jets in the dijet sample are quark- or gluon-initiated.
The resulting uncertainty is on average 11\%.
A two-dimensional sideband method is used in the DTT category  by defining
four regions based on the charge product of the two $\thad$ and the
 $\MET>10$~GeV requirement. 
A systematic uncertainty is derived by measuring the
variation of the ratio of opposite-sign to same-sign $\thad\thad$ pairs
for different sideband region definitions, as well as across the \MTTOT{} range,
and amounts to 5\%.

The remaining backgrounds are modelled using simulation.
Non-multi-jet processes with jets misidentified as $\thad$ are
dominated by $\Wtaunujets$.  In such events the $\thad$ identification
requirements are only applied to the $\thad$ from the $W$ decay and
not the jet that may be misidentified as the second $\thad$. Instead the event is
weighted using misidentification probabilities, measured in a
control region in data, to estimate the background yield.
$\Zjets$ background is also estimated using simulation. 
Due to the small number of remaining events after the $\pt$ thresholds 
of the $\thad$ trigger requirements, the $\tau$-embedded $Z\to\mu\mu$ 
sample is not used.

\begin{figure}
  \centering

  \subfigure[]{\includegraphics[width=0.44\textwidth]{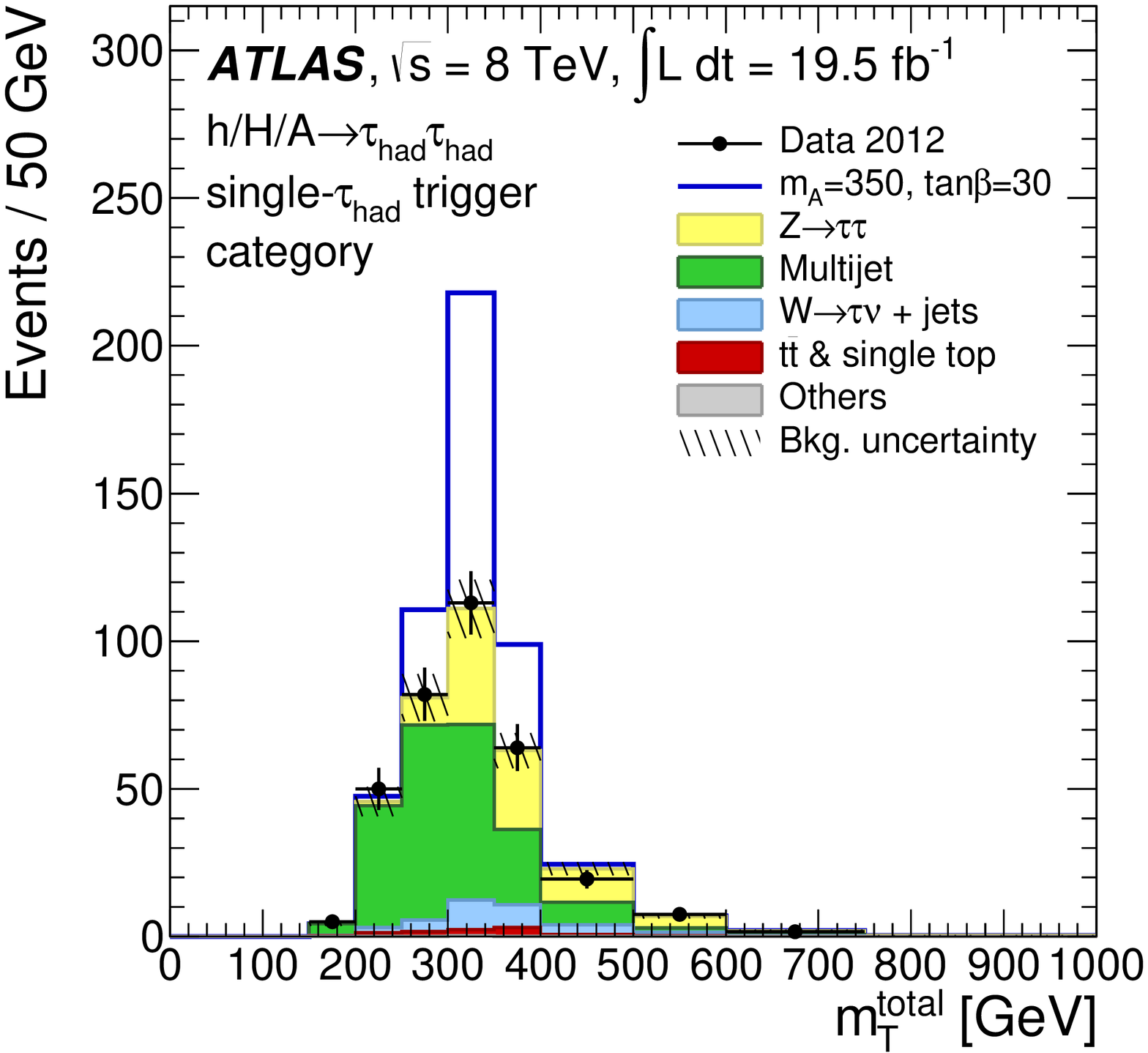}}
  \subfigure[]{\includegraphics[width=0.44\textwidth]{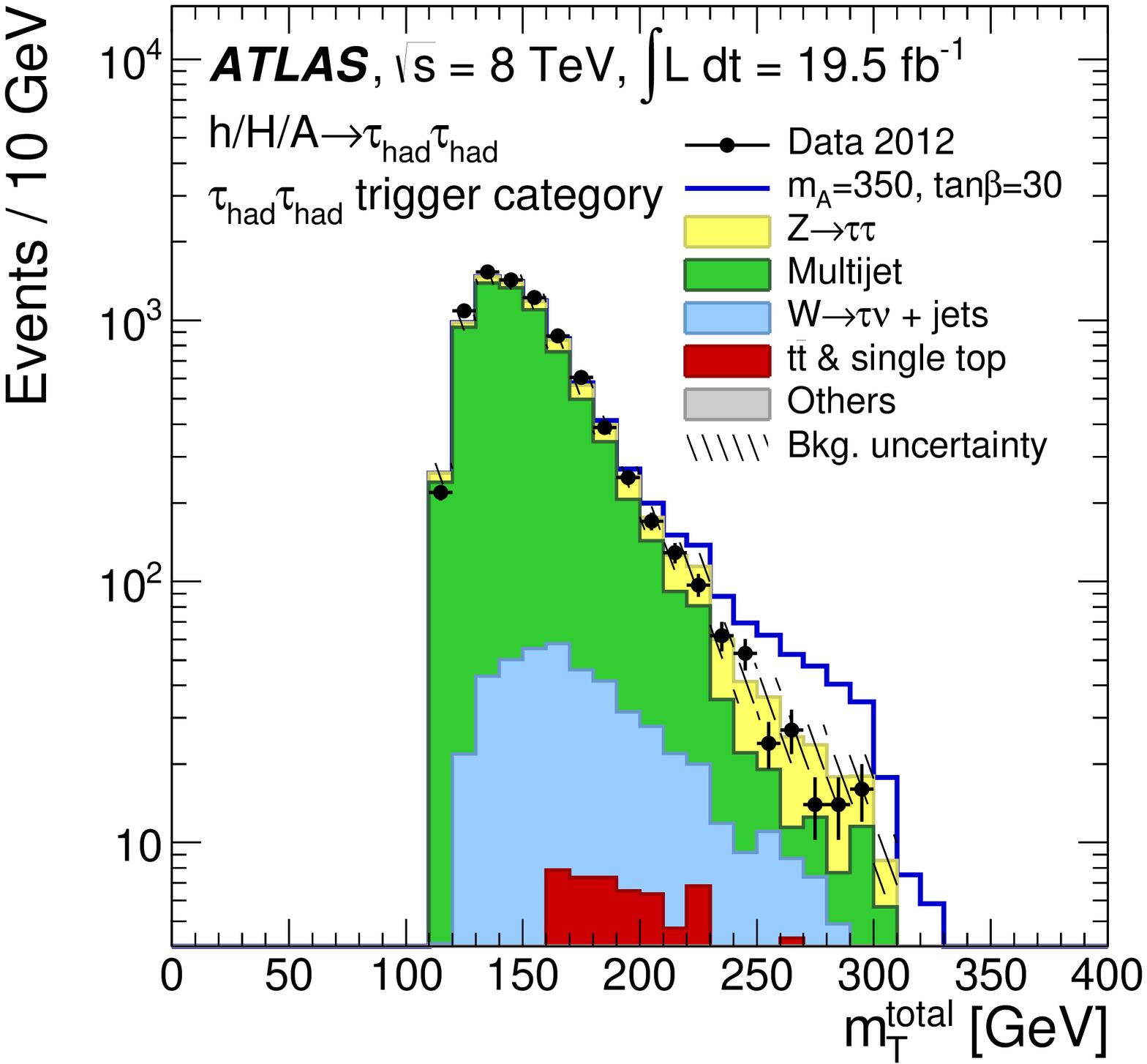}}

  \caption{Total transverse mass distributions for (a)
STT  and (b) DTT  categories of the \Htautauhh{} channel. 
The data are compared to the background expectation and a
hypothetical MSSM signal ($\MA=350$~GeV and $\tanb=30$). 
The background labelled ``Others'' includes 
events from diboson production, $Z\to \ell\ell$ and $W\to \ell\nu$ with $\ell=e,~\mu$.
The background uncertainty includes statistical and systematic 
uncertainties.
}
\label{fig:hadhad_mttot}
\end{figure}

Table~\ref{tab:hadhad_eventyield} shows the number of observed
$\thad\thad$ events, the predicted background, 
and the signal prediction for the MSSM \mhmax{} scenario 
parameter choice $m_A = 350$~GeV, $\tan\beta=30$.
The  total combined statistical and systematic uncertainties 
on the  predictions are also quoted in table~\ref{tab:hadhad_eventyield}.
The observed event yields are compatible 
with the expected yields from SM processes within the uncertainties.
The distributions of  the total transverse mass are shown in 
figure~\ref{fig:hadhad_mttot} for the STT and the DTT categories separately.

\begin{table}% [h]
\centering
\begin{tabular}{l c c}
 & \multicolumn{1}{c}{Single-\thad{} trigger} & \multicolumn{1}{c}{ \thad\thad{} trigger}  \\
 & \multicolumn{1}{c}{(STT) category}  & \multicolumn{1}{c}{(DTT) category}   \\
\hline\hline
  \multicolumn{3}{l}{Signal  ($m_A=350~\text{GeV},~\tan\beta=30$)} \\
 $h \to\tau\tau$ &  0.042 $\pm$ 0.039 & \phantom{0}11.2 $\pm$ 4.5\phantom{0}   \\
 $H \to\tau\tau$ &  \phantom{.0}95 $\pm$ 18\phantom{0.} & \phantom{0}182 $\pm$ 27\phantom{0}   \\
 $A \to\tau\tau$ &  \phantom{.0}82 $\pm$ 16\phantom{0.} & \phantom{0}158 $\pm$ 24\phantom{0}   \\
\hline\hline
Multi-jet                &  \phantom{.}216 $\pm$ 25\phantom{0.} & \phantom{.}6770 $\pm$  430\phantom{.}   \\
\Ztautau                 &  \phantom{.}113 $\pm$ 18\phantom{0.} & \phantom{0.}750  $\pm$ 210\phantom{.}    \\
\Wtaunujets              &  \phantom{.0}34 $\pm$ 8.1\phantom{0} & \phantom{0.}410  $\pm$ 100\phantom{.}    \\
$\ttbar$ and single top  &           10.2 $\pm$ 4.4\phantom{0}  & \phantom{00.}76   $\pm$ 26\phantom{0.}    \\
Others                   &  0.50 $\pm$ 0.20                     & \phantom{0}3.40  $\pm$ 0.80    \\
\hline
Total background         &  \phantom{.}374 $\pm$ 32\phantom{0.} & \phantom{.}8010  $\pm$ 490\phantom{.}   \\
\hline\hline
Data                     &  373            & 8225            \\

\hline\hline
\end{tabular}
\caption{\label{tab:hadhad_eventyield}
Number of events observed in the \Htautauhh{} channel
and the predicted background and signal.
The predicted signal event yields correspond to 
the parameter choice $\MA=350$~GeV, $\tanb=30$.
The row labelled ``Others'' includes events from diboson production, $Z\to \ell\ell$
and $W\to \ell\nu$ with $\ell=e,~\mu$.
Combined statistical and systematic uncertainties are quoted.
The signal prediction does not include the uncertainty due to the cross-section calculation.
}
\end{table}

\clearpage

%% file: systematics.tex
\section{Systematic uncertainties}
\label{sec:systematics}

The event yields for several of the backgrounds in this search are
estimated using control samples in data as described in
section~\ref{sec:analysiscategories} and their associated
uncertainties are discussed there.  In this section, the remaining
uncertainties are discussed and the overall effect of the systematic
uncertainties is presented. Many of the systematic uncertainties
affect both the signal and background estimates based on MC. These
correlations are used in the limit calculation described in section \ref{sec:results}.

Signal cross-section uncertainties are taken from the study in
ref.~\cite{LHCHiggsCrossSectionWorkingGroup:2011}. Typical uncertainty
values are in the range 10--15\% for gluon fusion and 15--20\% for
$b$-associated production.

The uncertainty on the signal acceptance from the parameters used in the event
generation of signal and background samples is also considered.
This is done by evaluating the change in acceptance after varying
the factorisation and renormalisation scale parameters, parton distribution function choices,
and if applicable, conditions for the matching of the partons used in the
fixed-order calculation and the parton shower.
The uncertainty on the signal acceptance is largest in the tag 
category for $b$-associated production, where it is about 13\%.

Uncertainties for single-boson and diboson production cross sections
are estimated for  missing higher-order corrections,
parton distribution functions and the value of the strong coupling constant,
and are considered wherever applicable. Acceptance uncertainties
for these background processes are estimated in the same way as for
signal.
The most important theoretical uncertainties on the background are the
$Z$+jets cross section and acceptance, which affect 
the normalization by about 7\%.

The uncertainty on the integrated luminosity is 2.8\%. It is derived, 
following the same methodology as that detailed in ref.~\cite{lumi}, 
from a preliminary calibration of the luminosity scale derived from beam-separation 
scans performed in November 2012.

The single-$\thad$ and $\thad\thad$ trigger efficiencies are studied in 
$Z\rightarrow\tau\tau$ events. 
Their uncertainties are in the range 3--25\% depending on the 
number of the tracks matched to the $\thad$, the $\thad$
pseudorapidity and $\pt$, 
as well as the data-taking period. They are estimated with a method
similar to the one in ref.~\cite{ATLAS-CONF-2013-006} and updated for the 2012 
data-taking conditions.

The $\thad$ identification efficiency is measured using $Z\rightarrow\tau\tau$
events. The uncertainty is in the range 3--10\%, depending on
the $\thad$   pseudorapidity and the number of tracks matched 
to the $\tau$ lepton \cite{ATLASTauIDNew}. 
Extrapolated uncertainties are used for $\thad$ candidates with
transverse momenta above those accessible in
$Z\rightarrow\tau\tau$ events.

The $\thad$ energy scale uncertainty is estimated by propagating the 
single-particle response to the individual $\thad$ decay products (neutral and
charged pions). This uncertainty is in the range 2--4\% \cite{ATLASTauES} 
depending on $\pt$, pseudorapidity and the number of associated
tracks.

The jet energy scale (JES) and resolution uncertainties are described in 
refs.~\cite{ATLASJETEnergyScale,ATLASJETEnergyResolution}. The JES 
is established by exploiting the \pt{} balance between a jet and a 
reference object such as a $Z$ boson or a photon. The uncertainty 
range is between $3$\% and $7$\%, depending on the \pt{} and 
pseudorapidity.
 
The $b$-jet identification efficiency uncertainty range is from
$2$\% to $8$\%, depending on the jet $\pt$.
The estimation of this uncertainty
is based on a study that uses $\ttbar$ events in data~\cite{ATLAS-CONF-2014-004}.

The \MET{} uncertainties are derived by propagating all 
energy scale uncertainties of reconstructed objects.
Additionally, the uncertainty on the scale for energy deposits outside
reconstructed objects and the resolution uncertainties are considered
\cite{ATLASMET}.

Electron and muon reconstruction, identification, isolation and trigger 
efficiency uncertainties are estimated from data in refs. 
\cite{egammapaper,Aad:2014rra}. 
Uncertainties related to the electron energy scale and resolution 
and to  the muon momentum scale and resolution 
are also estimated from data 
\cite{ATLAS_egamma_scale,Aad:2014rra} and  taken into account.

Systematic uncertainties associated with the $\tau$-embedded $\Zmumu$+jets 
data event sample are examined in refs.~\cite{ATLASLimit,SMHtautau2011}. Two
are found to be the most significant: the uncertainty due to the muon selection,
which is estimated by varying the muon isolation requirement used in
selecting the  $\Zmumu$+jets events, and the uncertainty from the
subtraction of the calorimeter cell energy associated with the muon.
The embedded sample contains a small contamination of $\ttbar$ events at
high MMC values. This is found to have a non-negligible influence in
the $\tlep\thad$ tag and high-mass categories only. The effect on the search
result is found to be very small in the tag category since other
background contributions are dominant in the relevant MMC region. Its
effect is taken into account by adding an additional uncertainty of 50\% to the
$Z\to \tau\tau$ background for MMC values exceeding 135~GeV. For the 
high-mass category, the estimated background level is subtracted from the
data and an uncertainty contribution of the same size is applied.

The relative effect of each of the systematic uncertainties
can be seen by their influence 
on the signal strength parameter, $\mu$, defined as 
the ratio of the fitted to the assumed signal cross section times 
branching fraction (see also section~\ref{sec:results}).
The effects of the most important sources of systematic uncertainty 
are shown for two signal assumptions: table~\ref{tab:systematics-1}
shows a low-mass pseudoscalar boson hypothesis ($m_A=150$~GeV, $\tan\beta=5.7$)
and table~\ref{tab:systematics-2} a high-mass pseudoscalar boson 
hypothesis ($m_A=350$~GeV, $\tan\beta=14$).
The $\tan\beta$ values chosen  correspond
to the observed limits for the respective $m_A$ assumptions (see section ~\ref{sec:results}).
The size of the systematic uncertainty on $\mu$ varies strongly 
with $\tan\beta$. In these tables,  ``Multi-jet background'' entries refer 
to uncertainties inherent to the methods used in estimation of the 
multi-jet background in the various channels of this search.
The largest contribution comes from the stability of the 
ratio of opposite-sign to same-sign events used in the 
two-dimensional sideband extrapolation method for the multi-jet background
estimation.

\begin{table}
\centering
\begin{tabular}{lc}
\hline
Source of uncertainty & Uncertainty on $\mu$ (\%)  \\
\hline

  Lepton-to-$\thad$ fake rate                              & 14 \\
  \thad{} energy scale                                     & 12 \\
  Jet energy scale and resolution                          & 11 \\
  Electron reconstruction \& identification                & 8.1 \\
  Simulated backgrounds cross section and acceptance       & 7.5 \\
  Luminosity                                               & 7.4 \\
  Muon reconstruction \& identification                    & 7.2 \\
  $b$-jet identification                                   & 6.6 \\
  Jet-to-$\thad$ fake rate for electroweak processes  ($\tlep\thad$)  & 6.2 \\
  Multi-jet background ($\tlep\tlep$, $\tlep\thad$)        & 6.1 \\
  Associated with the $\tau$-embedded $Z\to\mu\mu$  sample & 5.3 \\
  Signal acceptance                                        & 2.0 \\
  $e\mu$ trigger                                           & 1.5 \\
  \thad{} identification                                   & 0.8 \\

\hline%\hline
\end{tabular}
\caption{\label{tab:systematics-1}
The effect of the most important sources of uncertainty on the signal
strength parameter, $\mu$, for the signal hypothesis of $m_A=150$~GeV, $\tan\beta=5.7$.
For this signal hypothesis only the \Htautaulh{} and \Htautaull{} channels are used.
}
\end{table}

\begin{table}[t]
\centering
\begin{tabular}{lc}
\hline
Source of uncertainty & Uncertainty on $\mu$ (\%)  \\
\hline

 \thad{} energy scale                                           &  15  \\
 Multi-jet background ($\thad\thad$, $\tlep\thad$)              &  9.8  \\
 \thad{} identification                                         &  7.9 \\
 Jet-to-$\thad$ fake rate for electroweak processes                 &  7.6 \\
 \thad{} trigger                                                &  7.4 \\
 Simulated backgrounds cross section and acceptance             &  6.6 \\
 Signal acceptance                                              &  4.7 \\
 Luminosity                                                     &  4.1 \\
Associated with the $\tau$-embedded $Z\to\mu\mu$  sample        &  1.2 \\
 Lepton identification                                          &  0.7 \\

\hline
\end{tabular}
\caption{\label{tab:systematics-2}
The effect of the most important sources of uncertainty on the signal
strength parameter, $\mu$, for the signal hypothesis of $m_A=350$~GeV, 
$\tan\beta=14$. For this signal hypothesis only the \Htautaulh{} and 
\Htautauhh{} channels are used. 
}
\end{table}

\clearpage

%% file: results.tex
\section{Results}
\label{sec:results}

The results from the  channels studied in this search
are combined to improve the sensitivity to  MSSM Higgs boson production.
Each of the  channels used here is optimized for a specific
Higgs boson mass regime.
In particular, the $\telec\tmuon$ channel, the $\tlep\thad$ tag category,
and the $\tlep\thad$ veto category 
are used for the range $90 \leq m_A < 200$~GeV. 
The $\tlep\thad$ high mass category and the $\thad\thad$ channel
are used for $m_A \geq 200$~GeV.
The event selection in these categories is such that  the low
mass categories, i.e. those that target $90 \leq m_A < 200$~GeV, are sensitive
to the production of all three MSSM Higgs bosons, $h$, $H$ and $A$.
In contrast, the categories that target $m_A \geq 200$~GeV are sensitive only
to $H$ and $A$ production.

The parameter of interest in this search is the signal strength, 
$\mu$,  defined as the ratio of the fitted 
signal cross section times branching fraction to the  signal
cross section times branching fraction predicted by the particular MSSM 
signal assumption. The value $\mu=0$ corresponds to the absence of
signal, whereas the value $\mu=1$ suggests signal presence 
as predicted by the theoretical model under study.
The statistical analysis of the data employs a binned likelihood
function constructed as the product of Poisson probability terms
as an estimator of $\mu$. Signal and background predictions depend
on systematic uncertainties, which are parameterized as nuisance parameters 
and are constrained using Gaussian functions.
The binned likelihood function is constructed in bins of the MMC mass
for the $\telec\tmuon$ and the $\tlep\thad$ channels
and in bins of total transverse mass for the
$\thad\thad$ channel.

Since the data are in good agreement with the predicted
background yields, exclusion limits are calculated.
The significance of any small observed excess in data is
evaluated by quoting $p$-values to quantify the level of
consistency of the data with the mu=0 hypothesis.
Exclusion limits use the modified frequentist method known 
as CL$_{s}$~\cite{CLs_2002}. Both the exclusion limits and
$p$-values are calculated using the asymptotic approximation \cite{CCGV}.
The test statistic used for the exclusion limits derivation is
the $\tilde{q}_\mu$ test statistic 
and for the $p$-values the $q_{0}$ test statistic\footnote{The 
definition of the test
statistics used in this search is the following: 

\[ \tilde q_\mu = \left\{
  \begin{array}{l l}
    -2 \ln (\mathcal{L}(\mu,\hat{\hat{\theta}})/\mathcal{L}(0,\hat{\hat\theta})) & \quad \text{if $\hat\mu < 0$}\\
    -2 \ln (\mathcal{L}(\mu,\hat{\hat{\theta}})/\mathcal{L}(\hat \mu,\hat\theta)) & \quad \text{if $0 \leq \hat\mu \leq \mu$}\\
    0 & \quad \text{if $\hat\mu > \mu$}
  \end{array} \right.\]
and 
\[ q_0 = \left\{
  \begin{array}{l l}
    -2 \ln (\mathcal{L}(0,\hat{\hat{\theta}})/\mathcal{L}(\hat\mu,\hat\theta)) &\quad \text{if $\hat\mu \geq 0$}\\
    0 & \quad \text{if $\hat\mu <0$}
  \end{array} \right.\]
where $\mathcal L(\mu,\theta)$ denotes the binned likelihood function, $\mu$ is the parameter of interest (i.e. 
the signal strength parameter), and $\theta$ denotes the nuisance parameters. The pair $(\hat\mu, \hat\theta)$
corresponds to the global maximum of the likelihood, whereas $(x, \hat{\hat\theta})$ corresponds to a conditional
maximum in which $\mu$ is fixed to a given value $x$.
} 
\cite{CCGV}.

The lowest local $p$-values are calculated assuming a single
scalar boson $\phi$ with narrow natural width with respect to the
experimental mass resolution.
The lowest local $p$-value for the 
combination of all channels corresponds to 0.20, or $0.8~\sigma$
in terms of Gaussian standard deviations, at $m_{\phi} = 200$~GeV.
For the individual channels, the lowest local $p$-value in $\thad\thad$
is 0.10 (or $1.3~\sigma$) at $m_{\phi} = 250$~GeV and
for the  $\tlep\thad$   0.10 (or $1.3~\sigma$) at $m_{\phi} = 90$~GeV.
In the  $\tlep\tlep$ channel there is no excess in the mass region used
for the combination ($90 \le m_{\phi} < 200$~GeV).

\begin{figure}
  \centering
  \subfigure[]{ \label{fig:limits-tb-1}
    \includegraphics[width=.45\textwidth]{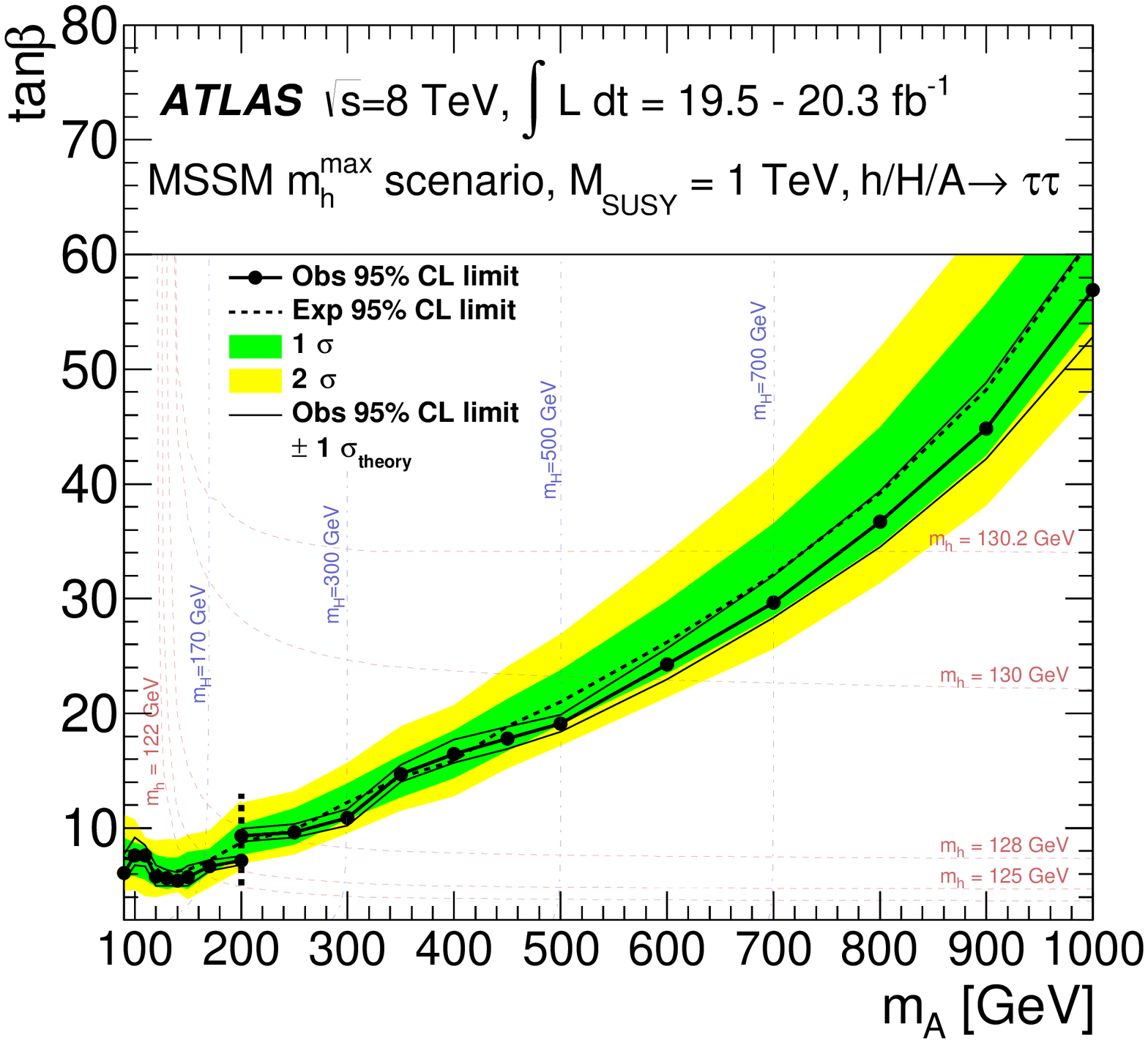}
  }
  \subfigure[]{ \label{fig:limits-tb-2}
    \includegraphics[width=.45\textwidth]{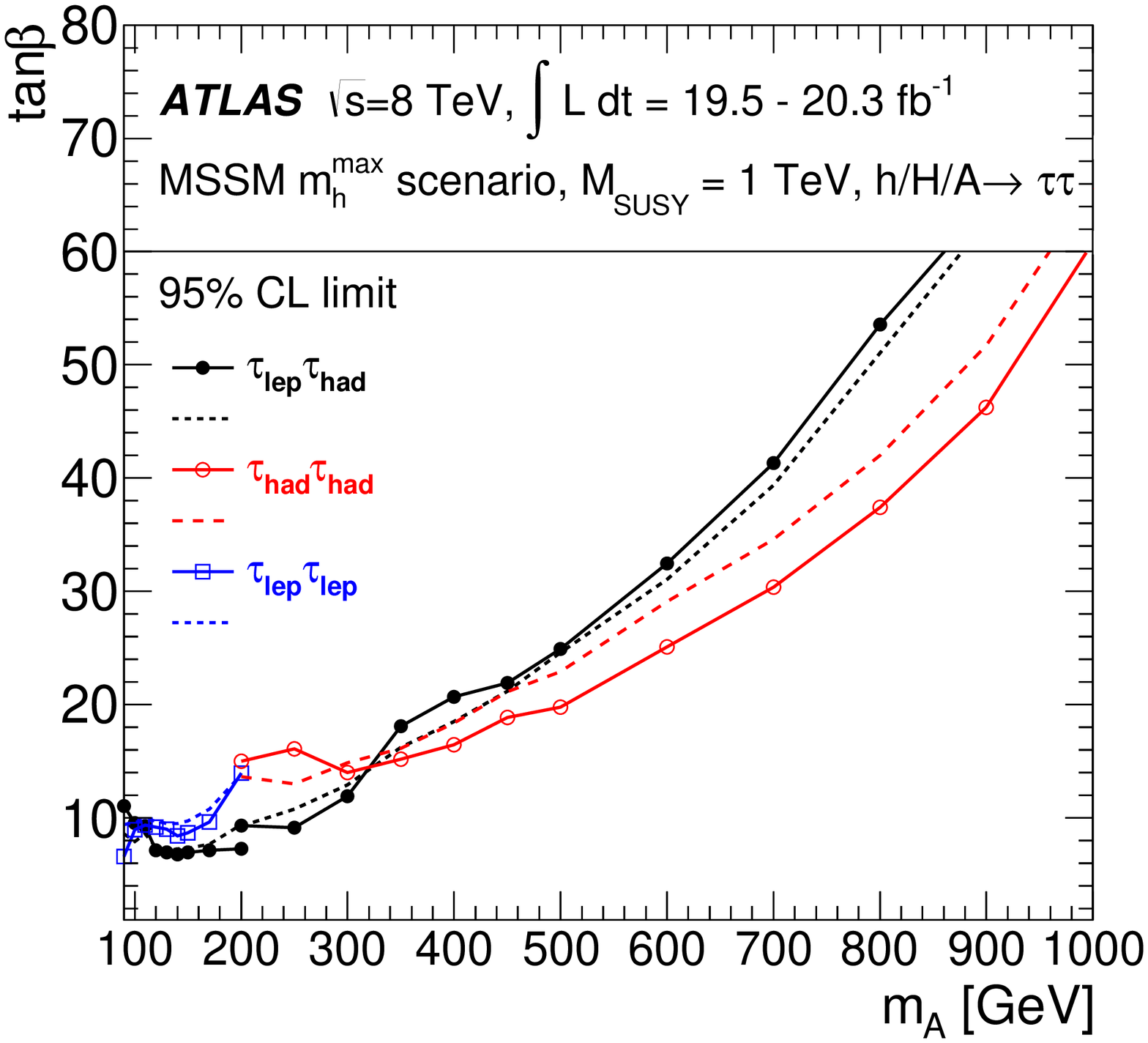}
  }
  \caption{
Expected (dashed line) and observed (solid line with markers) 95\% CL 
upper limits on \tanb{} as a function of \MA{} for the \mhmax{} scenario 
of the MSSM (a) for the combination of all channels and (b) for each channel separately.
Values of \tanb{} above the  lines are excluded. 
The vertical dashed line at 200~GeV in (a) indicates the transition point 
between low- and high-mass categories.  
Lines of constant $m_h$ and $m_H$ are also shown in (a) in 
red and blue colour, respectively.
For more information, see text.
}
  \label{fig:limits-tb}
\end{figure}

\begin{figure}[h]
  \centering
  \subfigure[]{ % [$m_h^{\text{mod}+}$]{
    \includegraphics[width=.45\textwidth]{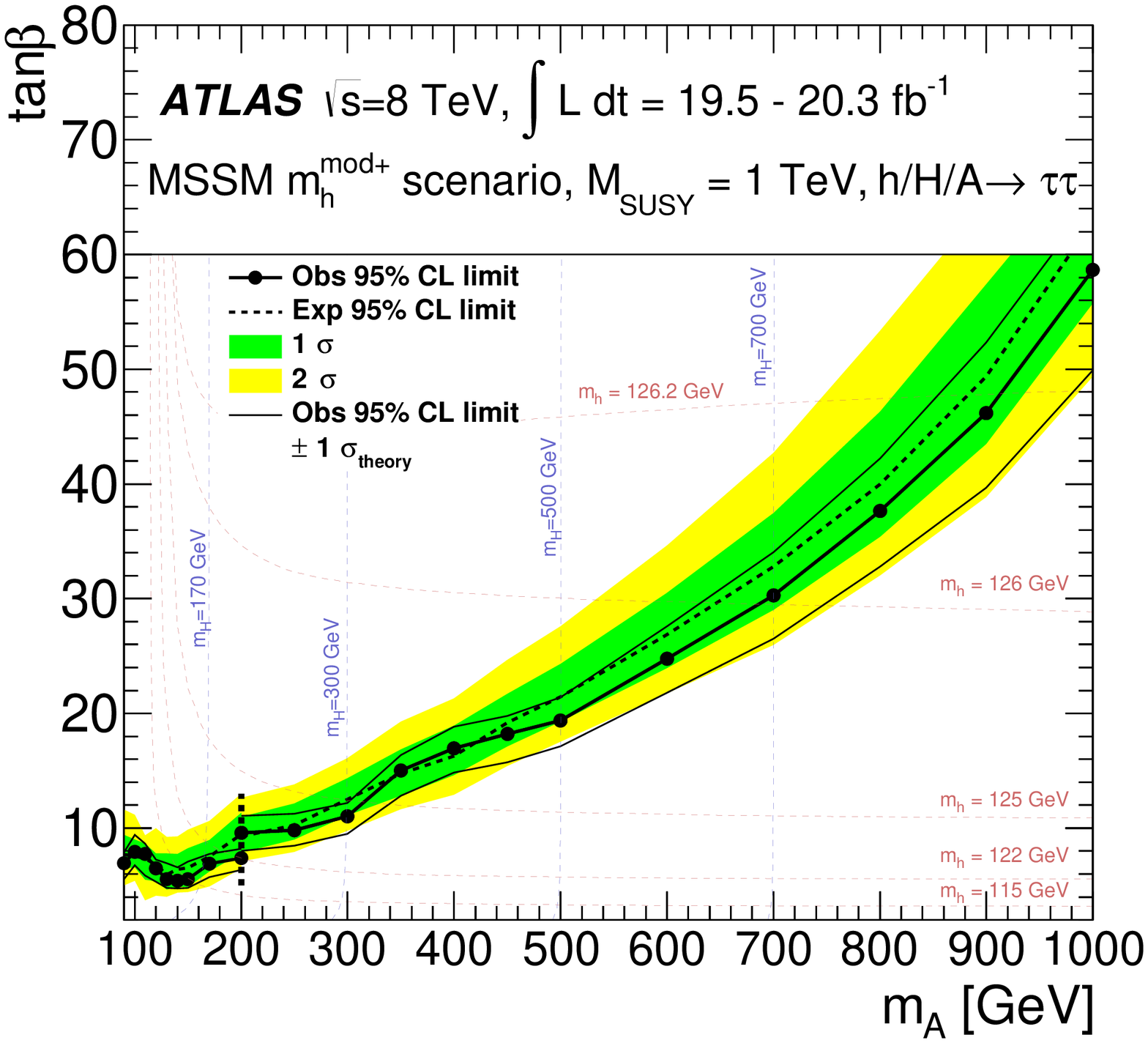}
  }
  \subfigure[]{ %[$m_h^{\text{mod}-}$]{
    \includegraphics[width=.45\textwidth]{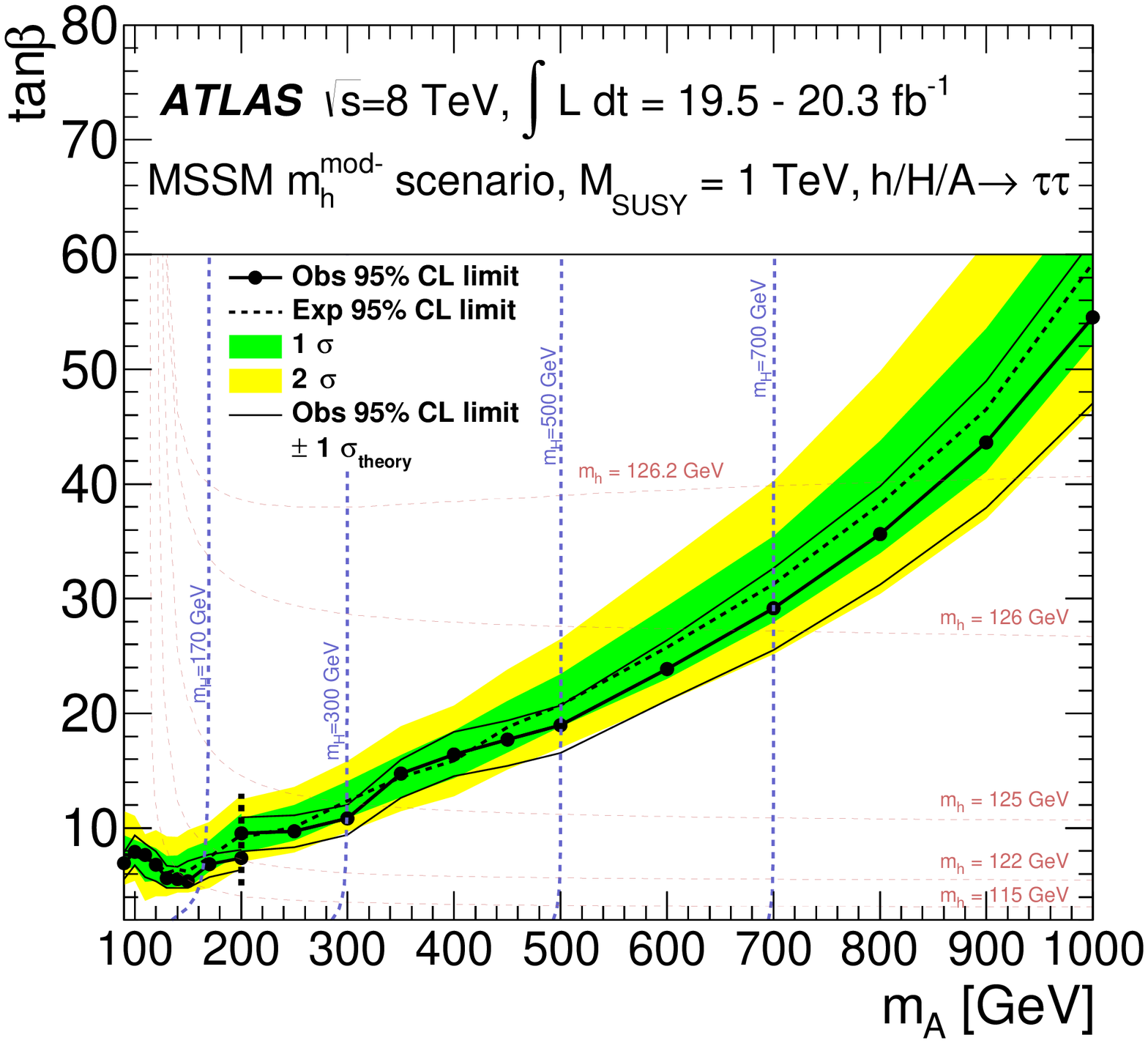}
  }

  \caption{
Expected (dashed  line) and observed (solid  line with markers) 95\% CL 
upper limits on \tanb{} as a function of \MA{}
for (a) the \mhmodplus{}  and (b) the \mhmodmin{}  benchmark scenarios of the MSSM.
The same notation as in figure~\ref{fig:limits-tb-1} is used.
}
  \label{fig:limits-altscenarios}
\end{figure}

Expected and observed 95\% confidence level (CL) upper limits for the  combination of all
channels are shown in figure~\ref{fig:limits-tb-1} for the MSSM \mhmax{}
scenario with $M_{\text{SUSY}}=1$~TeV \cite{Heinemeyer:1999zf,MSSMmhmax}. In this figure, the
theoretical MSSM Higgs cross-section uncertainties are not included in the
reported result, but their impact is shown separately, by recalculating 
the upper limits again after considering the relevant  
$\pm 1 \sigma$ variations.
Figure~\ref{fig:limits-tb-2} shows the upper limits for each channel
separately for comparison.
The best $\tan\beta$ constraint for the combined search excludes 
$\tan\beta > 5.4$ for $m_A = 140$~GeV, whereas, as an example,
$\tan\beta > 37$ is excluded for $m_A = 800$~GeV. 
Figure~\ref{fig:limits-tb-1} shows also contours of constant
$m_{h}$ and $m_{H}$ for the  MSSM \mhmax{} scenario.
Assuming that the light CP-even Higgs boson of the MSSM has a mass
of about 125~GeV and taking into consideration the 3~GeV 
uncertainty in the $m_{h}$ calculation in the MSSM~\cite{MSSMBenchmarks}, 
only the parameter space that is compatible with $122 < m_{h} < 128$~GeV
is allowed.
From this consideration it is concluded that  if
the light CP-even Higgs boson of the MSSM is identified with the particle
discovered at the LHC, then for this particular MSSM scenario
$m_A < 160$~GeV is excluded for all $\tan\beta$ values. Similarly,
$\tan\beta > 10$ and $\tan\beta < 4$ are excluded for all $m_A$
values. Figure~\ref{fig:limits-altscenarios} shows the expected and
observed upper limits for the MSSM \mhmodplus{} and \mhmodmin{}
scenarios \cite{MSSMBenchmarks}. 
Again, the  combination of all channels is
shown and the impact of signal cross-section uncertainties is shown
separately. Under the assumption that
the light CP-even Higgs boson of the MSSM is identified with the particle
discovered at the LHC and taking into account the
same considerations as in the $\mhmax$ scenario case, 
the region with  $m_A < 200$~GeV is excluded for
all $\tan\beta$ values, whereas $\tan\beta < 5.5$ is excluded
for all values of $m_A$ for both the MSSM \mhmodplus{} and \mhmodmin{} scenarios.

\begin{figure}
  \centering
  \subfigure[]{ \label{fig:limits-xs-1}
    \includegraphics[width=.45\textwidth]{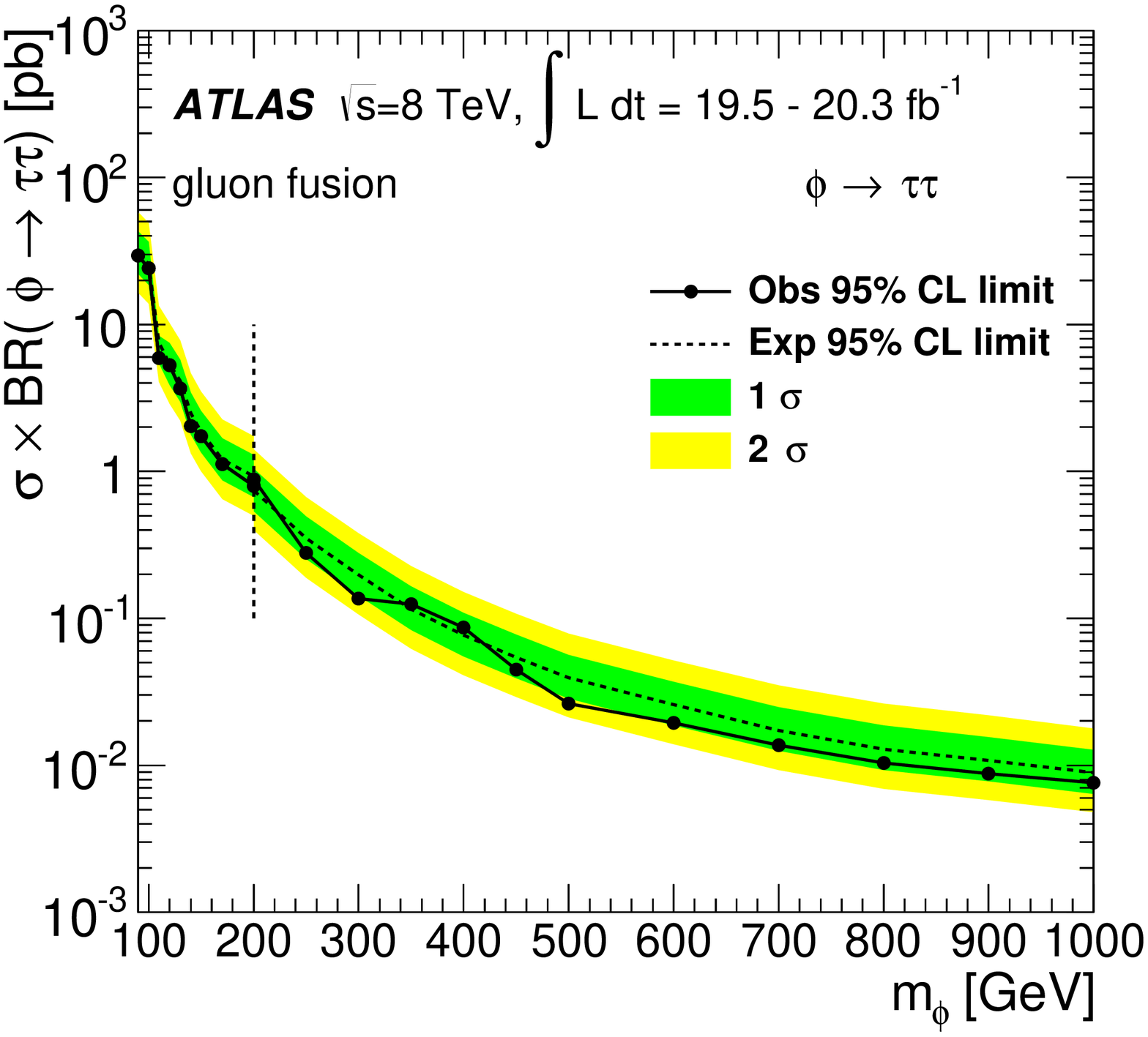}
  }
  \subfigure[]{ \label{fig:limits-xs-2}
    \includegraphics[width=.45\textwidth]{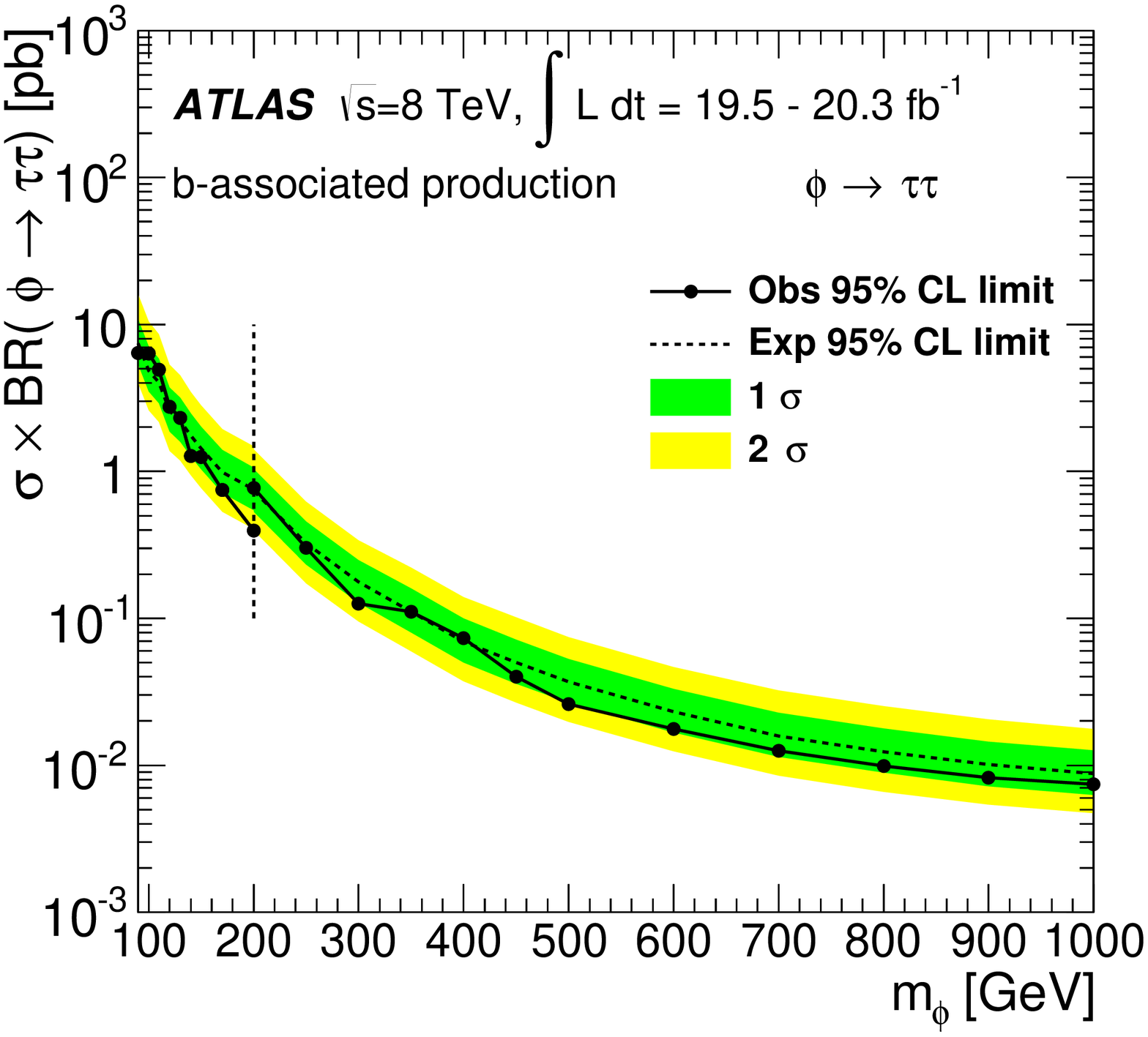}
  }
  \caption{
Expected (dashed bold line) and observed (solid bold line) 95\% CL upper limits
on the cross section of a scalar boson $\phi$ produced via 
(a) gluon fusion and (b) in association with $b$-quarks
times the branching fraction into $\tau$ pairs. 
The vertical dashed line at 200~GeV indicates the transition point 
between low- and high-mass categories. 
}
  \label{fig:limits-xs}
\end{figure}

The outcome of the search is further interpreted in the 
case of a single scalar boson $\phi$, with narrow width relative 
to the experimental mass resolution, produced in either the 
gluon fusion or $b$-associated production mode and decaying to \tautau.
Figure~\ref{fig:limits-xs} shows 95\% CL upper limits on the cross section
times the $\tau\tau$ branching fraction based on this interpretation.
The exclusion limits for the production cross section times the branching 
fraction for a scalar boson decaying to \tautau{} are shown as a function of 
the scalar boson mass. The excluded cross section times branching fraction
values range from $\sigma\times BR > 29$~pb
at $m_{\phi}=90$~GeV to $\sigma\times BR > 7.4$~fb at $m_{\phi}=1000$~GeV
for a scalar boson produced via gluon fusion.
The exclusion range for the $b$-associated production mechanism ranges from
 $\sigma\times BR > 6.4$~pb at $m_{\phi}=90$~GeV 
to $\sigma\times BR > 7.2$~fb at $m_{\phi}=1000$~GeV.

\clearpage

\section{Summary}

A search is presented for the neutral Higgs bosons of the Minimal 
Supersymmetric Standard Model in proton--proton collisions at the 
centre-of-mass energy of 8~TeV with the ATLAS experiment at the LHC.
The integrated luminosity used in the search is \currentlumirange.
The search  uses the \tautau{} final state. In particular, the following
cases are considered:
one $\tau$ lepton decays to an electron and the other to a muon (\telec\tmuon),
one $\tau$ lepton decays to an electron or muon and the other hadronically (\tlep\thad)
and finally both $\tau$ leptons decay hadronically (\thad\thad). 
The sensitivity is improved by performing a categorisation based on expected 
Higgs boson mass and production mechanisms.
The search finds no indication of an excess over the expected background
in the channels considered and 95\% CL limits are set, which provide
tight constraints in the MSSM parameter space.
In particular, in the context of the MSSM \mhmax{} scenario
the lowest $\tan\beta$ constraint excludes
$\tan\beta > 5.4$ for $m_A = 140$~GeV. 
Upper limits for the production cross section times 
$\tau\tau$ branching fraction of a scalar boson versus its mass, 
depending on the production mode, are also presented. 
The excluded cross section times $\tau\tau$ branching fraction
ranges from about 30~pb to about 7~fb depending on the Higgs boson
mass and the production mechanism.

% Acknowledgements for papers with collision data
% Version 01-Mar-2014

\section*{Acknowledgements}

% Standard acknowledgements start here
%----------------------------------------------
We thank CERN for the very successful operation of the LHC, as well as the
support staff from our institutions without whom ATLAS could not be
operated efficiently.

We acknowledge the support of ANPCyT, Argentina; YerPhI, Armenia; ARC,
Australia; BMWFW and FWF, Austria; ANAS, Azerbaijan; SSTC, Belarus; CNPq and FAPESP,
Brazil; NSERC, NRC and CFI, Canada; CERN; CONICYT, Chile; CAS, MOST and NSFC,
China; COLCIENCIAS, Colombia; MSMT CR, MPO CR and VSC CR, Czech Republic;
DNRF, DNSRC and Lundbeck Foundation, Denmark; EPLANET, ERC and NSRF, European Union;
IN2P3-CNRS, CEA-DSM/IRFU, France; GNSF, Georgia; BMBF, DFG, HGF, MPG and AvH
Foundation, Germany; GSRT and NSRF, Greece; ISF, MINERVA, GIF, I-CORE and Benoziyo Center,
Israel; INFN, Italy; MEXT and JSPS, Japan; CNRST, Morocco; FOM and NWO,
Netherlands; BRF and RCN, Norway; MNiSW and NCN, Poland; GRICES and FCT, Portugal; MNE/IFA, Romania; MES of Russia and ROSATOM, Russian Federation; JINR; MSTD,
Serbia; MSSR, Slovakia; ARRS and MIZ\v{S}, Slovenia; DST/NRF, South Africa;
MINECO, Spain; SRC and Wallenberg Foundation, Sweden; SER, SNSF and Cantons of
Bern and Geneva, Switzerland; NSC, Taiwan; TAEK, Turkey; STFC, the Royal
Society and Leverhulme Trust, United Kingdom; DOE and NSF, United States of
America.

The crucial computing support from all WLCG partners is acknowledged
gratefully, in particular from CERN and the ATLAS Tier-1 facilities at
TRIUMF (Canada), NDGF (Denmark, Norway, Sweden), CC-IN2P3 (France),
KIT/GridKA (Germany), INFN-CNAF (Italy), NL-T1 (Netherlands), PIC (Spain),
ASGC (Taiwan), RAL (UK) and BNL (USA) and in the Tier-2 facilities
worldwide.
%----------------------------------------------

%% file: atlas_authlist.tex
% ATLAS Collaboration author list
% Data extracted on 14-Oct-2014 for paper reference HIGG-2013-31
%%\documentclass[11pt]{article}
%%\usepackage{a4wide}\begin{document}
\begin{flushleft}
{\Large The ATLAS Collaboration}

\bigskip

G.~Aad$^{\rm 85}$,
B.~Abbott$^{\rm 113}$,
J.~Abdallah$^{\rm 152}$,
S.~Abdel~Khalek$^{\rm 117}$,
O.~Abdinov$^{\rm 11}$,
R.~Aben$^{\rm 107}$,
B.~Abi$^{\rm 114}$,
M.~Abolins$^{\rm 90}$,
O.S.~AbouZeid$^{\rm 159}$,
H.~Abramowicz$^{\rm 154}$
% ...
\bigskip
\\
% ...
$^{11}$ Institute of Physics, Azerbaijan Academy of Sciences, Baku, Azerbaijan\\
% ...
$^{85}$ CPPM, Aix-Marseille Universit{\'e} and CNRS/IN2P3, Marseille, France\\
% ...
$^{90}$ Department of Physics and Astronomy, Michigan State University, East Lansing MI, United States of America\\
% ...
$^{107}$ Nikhef National Institute for Subatomic Physics and University of Amsterdam, Amsterdam, Netherlands\\
% ...
$^{113}$ Homer L. Dodge Department of Physics and Astronomy, University of Oklahoma, Norman OK, United States of America\\
$^{114}$ Department of Physics, Oklahoma State University, Stillwater OK, United States of America\\
% ...
$^{117}$ LAL, Universit{\'e} Paris-Sud and CNRS/IN2P3, Orsay, France\\
% ...
$^{152}$ Institute of Physics, Academia Sinica, Taipei, Taiwan\\
% ...
$^{154}$ Raymond and Beverly Sackler School of Physics and Astronomy, Tel Aviv University, Tel Aviv, Israel\\
% ...
$^{159}$ Department of Physics, University of Toronto, Toronto ON, Canada\\
% ...
\end{flushleft}

%%\end{document}
% Created with ./xml2latex.py

%% file: MSSMPaper.bbl
\providecommand{\href}[2]{#2}\begingroup\raggedright\begin{thebibliography}{10}

\bibitem{ATLASHiggsJuly2012}
{ATLAS Collaboration}, {\it {Observation of a new particle in the search for
  the Standard Model Higgs boson with the ATLAS detector at the LHC}},
  \href{http://dx.doi.org/10.1016/j.physletb.2012.08.020}{{\em Phys. Lett.}}
  \href{http://dx.doi.org/10.1016/j.physletb.2012.08.020}{{\bf B 716}}
  \href{http://dx.doi.org/10.1016/j.physletb.2012.08.020}{(2012)}
  \href{http://dx.doi.org/10.1016/j.physletb.2012.08.020}{1--29},
  [\href{http://xxx.lanl.gov/abs/1207.7214}{{\tt arXiv:1207.7214}}].
%%CITATION = ARXIV:1207.7214;%%

\bibitem{CMSHiggsJuly2012}
{CMS Collaboration}, {\it {Observation of a new boson at a mass of 125 GeV with
  the CMS experiment at the LHC}},
  \href{http://dx.doi.org/10.1016/j.physletb.2012.08.021}{{\em Phys. Lett.}}
  \href{http://dx.doi.org/10.1016/j.physletb.2012.08.021}{{\bf B 716}}
  \href{http://dx.doi.org/10.1016/j.physletb.2012.08.021}{(2012)}
  \href{http://dx.doi.org/10.1016/j.physletb.2012.08.021}{30--61},
  [\href{http://xxx.lanl.gov/abs/1207.7235}{{\tt arXiv:1207.7235}}].
%%CITATION = ARXIV:1207.7235;%%

\bibitem{ATLASHiggsProperties1}
{ATLAS Collaboration}, {\it {Measurements of Higgs boson production and
  couplings in diboson final states with the ATLAS detector at the LHC}},
  \href{http://dx.doi.org/10.1016/j.physletb.2013.08.010}{{\em Phys. Lett.}}
  \href{http://dx.doi.org/10.1016/j.physletb.2013.08.010}{{\bf B 726}}
  \href{http://dx.doi.org/10.1016/j.physletb.2013.08.010}{(2013)}
  \href{http://dx.doi.org/10.1016/j.physletb.2013.08.010}{88--119},
  [\href{http://xxx.lanl.gov/abs/1307.1427}{{\tt arXiv:1307.1427}}].
%%CITATION = ARXIV:1307.1427;%%

\bibitem{ATLASHiggsProperties2}
{ATLAS Collaboration}, {\it {Evidence for the spin-0 nature of the Higgs boson
  using ATLAS data}},
  \href{http://dx.doi.org/10.1016/j.physletb.2013.08.026}{{\em Phys. Lett.}}
  \href{http://dx.doi.org/10.1016/j.physletb.2013.08.026}{{\bf B 726}}
  \href{http://dx.doi.org/10.1016/j.physletb.2013.08.026}{(2013)}
  \href{http://dx.doi.org/10.1016/j.physletb.2013.08.026}{120--144},
  [\href{http://xxx.lanl.gov/abs/1307.1432}{{\tt arXiv:1307.1432}}].
%%CITATION = ARXIV:1307.1432;%%

\bibitem{CMSHiggsProperties}
{CMS Collaboration}, {\it {Evidence for the direct decay of the 125 GeV Higgs
  boson to fermions}},  \href{http://dx.doi.org/10.1038/nphys3005}{{\em Nature
  Phys.}} \href{http://dx.doi.org/10.1038/nphys3005}{{\bf 10}}
  \href{http://dx.doi.org/10.1038/nphys3005}{(2014)}
  [\href{http://xxx.lanl.gov/abs/1401.6527}{{\tt arXiv:1401.6527}}].
%%CITATION = ARXIV:1401.6527;%%

\bibitem{CMSHiggs4l}
{CMS Collaboration}, {\it {Measurement of the properties of a Higgs boson in
  the four-lepton final state}},
  \href{http://dx.doi.org/10.1103/PhysRevD.89.092007}{{\em Phys.Rev.}}
  \href{http://dx.doi.org/10.1103/PhysRevD.89.092007}{{\bf D 89}}
  \href{http://dx.doi.org/10.1103/PhysRevD.89.092007}{(2014)}
  \href{http://dx.doi.org/10.1103/PhysRevD.89.092007}{092007},
  [\href{http://xxx.lanl.gov/abs/1312.5353}{{\tt arXiv:1312.5353}}].
%%CITATION = ARXIV:1312.5353;%%

\bibitem{CMSHiggsWW}
{CMS Collaboration}, {\it {Measurement of Higgs boson production and properties
  in the WW decay channel with leptonic final states}},
  \href{http://dx.doi.org/10.1007/JHEP01(2014)096}{{\em JHEP}}
  \href{http://dx.doi.org/10.1007/JHEP01(2014)096}{{\bf 01}}
  \href{http://dx.doi.org/10.1007/JHEP01(2014)096}{(2014)}
  \href{http://dx.doi.org/10.1007/JHEP01(2014)096}{096},
  [\href{http://xxx.lanl.gov/abs/1312.1129}{{\tt arXiv:1312.1129}}].
%%CITATION = ARXIV:1312.1129;%%

\bibitem{ENGLERT}
F.~Englert and R.~Brout, {\it {Broken symmetry and the mass of gauge vector
  mesons}},  \href{http://dx.doi.org/10.1103/PhysRevLett.13.321}{{\em Phys.
  Rev. Lett.}} \href{http://dx.doi.org/10.1103/PhysRevLett.13.321}{{\bf 13}}
  \href{http://dx.doi.org/10.1103/PhysRevLett.13.321}{(1964)}
  \href{http://dx.doi.org/10.1103/PhysRevLett.13.321}{321--323}.
%%CITATION = PRLTA,13,321;%%

\bibitem{HIGGS}
P.~W. Higgs, {\it {Broken symmetries, massless particles and gauge fields}},
  \href{http://dx.doi.org/10.1016/0031-9163(64)91136-9}{{\em Phys. Lett.}}
  \href{http://dx.doi.org/10.1016/0031-9163(64)91136-9}{{\bf 12}}
  \href{http://dx.doi.org/10.1016/0031-9163(64)91136-9}{(1964)}
  \href{http://dx.doi.org/10.1016/0031-9163(64)91136-9}{132--133}.
%%CITATION = PHLTA,12,132;%%

\bibitem{HIGGS2}
P.~W. Higgs, {\it {Broken symmetries and the masses of gauge bosons}},
  \href{http://dx.doi.org/10.1103/PhysRevLett.13.508}{{\em Phys. Rev. Lett.}}
  \href{http://dx.doi.org/10.1103/PhysRevLett.13.508}{{\bf 13}}
  \href{http://dx.doi.org/10.1103/PhysRevLett.13.508}{(1964)}
  \href{http://dx.doi.org/10.1103/PhysRevLett.13.508}{508--509}.
%%CITATION = PRLTA,13,508;%%

\bibitem{HIGGS3}
P.~W. Higgs, {\it {Spontaneous symmetry breakdown without massless bosons}},
  \href{http://dx.doi.org/10.1103/PhysRev.145.1156}{{\em Phys. Rev.}}
  \href{http://dx.doi.org/10.1103/PhysRev.145.1156}{{\bf 145}}
  \href{http://dx.doi.org/10.1103/PhysRev.145.1156}{(1966)}
  \href{http://dx.doi.org/10.1103/PhysRev.145.1156}{1156--1163}.
%%CITATION = PHRVA,145,1156;%%

\bibitem{Guralnik:1964eu}
G.~Guralnik, C.~Hagen, and T.~Kibble, {\it {Global conservation laws and
  massless particles}},
  \href{http://dx.doi.org/10.1103/PhysRevLett.13.585}{{\em Phys. Rev. Lett.}}
  \href{http://dx.doi.org/10.1103/PhysRevLett.13.585}{{\bf 13}}
  \href{http://dx.doi.org/10.1103/PhysRevLett.13.585}{(1964)}
  \href{http://dx.doi.org/10.1103/PhysRevLett.13.585}{585--587}.
%%CITATION = PRLTA,13,585;%%

\bibitem{Kibble:1967sv}
T.~Kibble, {\it {Symmetry breaking in non-Abelian gauge theories}},
  \href{http://dx.doi.org/10.1103/PhysRev.155.1554}{{\em Phys. Rev.}}
  \href{http://dx.doi.org/10.1103/PhysRev.155.1554}{{\bf 155}}
  \href{http://dx.doi.org/10.1103/PhysRev.155.1554}{(1967)}
  \href{http://dx.doi.org/10.1103/PhysRev.155.1554}{1554--1561}.
%%CITATION = PHRVA,155,1554;%%

\bibitem{Djouadi:2005gj}
A.~Djouadi, {\it {The Anatomy of electro-weak symmetry breaking. II. The Higgs
  bosons in the minimal supersymmetric model}},
  \href{http://dx.doi.org/10.1016/j.physrep.2007.10.005}{{\em Phys. Rept.}}
  \href{http://dx.doi.org/10.1016/j.physrep.2007.10.005}{{\bf 459}}
  \href{http://dx.doi.org/10.1016/j.physrep.2007.10.005}{(2008)}
  \href{http://dx.doi.org/10.1016/j.physrep.2007.10.005}{1--241},
  [\href{http://xxx.lanl.gov/abs/hep-ph/0503173}{{\tt hep-ph/0503173}}].
%%CITATION = HEP-PH/0503173;%%

\bibitem{Branco:2011iw}
{\it {Theory and phenomenology of two-Higgs-doublet models}},
  \href{http://dx.doi.org/10.1016/j.physrep.2012.02.002}{{\em Phys. Rept.}}
  \href{http://dx.doi.org/10.1016/j.physrep.2012.02.002}{{\bf 516}}
  \href{http://dx.doi.org/10.1016/j.physrep.2012.02.002}{(2012)}
  \href{http://dx.doi.org/10.1016/j.physrep.2012.02.002}{1--102},
  [\href{http://xxx.lanl.gov/abs/1106.0034}{{\tt arXiv:1106.0034}}].
%%CITATION = ARXIV:1106.0034;%%

\bibitem{Fayet:1976et}
P.~Fayet, {\it {Supersymmetry and Weak, Electromagnetic and Strong
  Interactions}},  \href{http://dx.doi.org/10.1016/0370-2693(76)90319-1}{{\em
  Phys. Lett.}} \href{http://dx.doi.org/10.1016/0370-2693(76)90319-1}{{\bf B
  64}} \href{http://dx.doi.org/10.1016/0370-2693(76)90319-1}{(1976)}
  \href{http://dx.doi.org/10.1016/0370-2693(76)90319-1}{159}.
%%CITATION = PHLTA,B64,159;%%

\bibitem{Fayet:1977yc}
P.~Fayet, {\it {Spontaneously Broken Supersymmetric Theories of Weak,
  Electromagnetic and Strong Interactions}},
  \href{http://dx.doi.org/10.1016/0370-2693(77)90852-8}{{\em Phys. Lett.}}
  \href{http://dx.doi.org/10.1016/0370-2693(77)90852-8}{{\bf B 69}}
  \href{http://dx.doi.org/10.1016/0370-2693(77)90852-8}{(1977)}
  \href{http://dx.doi.org/10.1016/0370-2693(77)90852-8}{489}.
%%CITATION = PHLTA,B69,489;%%

\bibitem{Farrar:1978xj}
G.~R. Farrar and P.~Fayet, {\it {Phenomenology of the Production, Decay, and
  Detection of New Hadronic States Associated with Supersymmetry}},
  \href{http://dx.doi.org/10.1016/0370-2693(78)90858-4}{{\em Phys. Lett.}}
  \href{http://dx.doi.org/10.1016/0370-2693(78)90858-4}{{\bf B 76}}
  \href{http://dx.doi.org/10.1016/0370-2693(78)90858-4}{(1978)}
  \href{http://dx.doi.org/10.1016/0370-2693(78)90858-4}{575--579}.
%%CITATION = PHLTA,B76,575;%%

\bibitem{Fayet:1979sa}
P.~Fayet, {\it {Relations Between the Masses of the Superpartners of Leptons
  and Quarks, the Goldstino Couplings and the Neutral Currents}},
  \href{http://dx.doi.org/10.1016/0370-2693(79)91229-2}{{\em Phys. Lett.}}
  \href{http://dx.doi.org/10.1016/0370-2693(79)91229-2}{{\bf B 84}}
  \href{http://dx.doi.org/10.1016/0370-2693(79)91229-2}{(1979)}
  \href{http://dx.doi.org/10.1016/0370-2693(79)91229-2}{416}.
%%CITATION = PHLTA,B84,416;%%

\bibitem{Dimopoulos:1981zb}
S.~Dimopoulos and H.~Georgi, {\it {Softly Broken Supersymmetry and SU(5)}},
  \href{http://dx.doi.org/10.1016/0550-3213(81)90522-8}{{\em Nucl. Phys.}}
  \href{http://dx.doi.org/10.1016/0550-3213(81)90522-8}{{\bf B 193}}
  \href{http://dx.doi.org/10.1016/0550-3213(81)90522-8}{(1981)}
  \href{http://dx.doi.org/10.1016/0550-3213(81)90522-8}{150}.
%%CITATION = NUPHA,B193,150;%%

\bibitem{Heinemeyer:1999zf}
S.~Heinemeyer, W.~Hollik, and G.~Weiglein, {\it {Constraints on tan Beta in the
  MSSM from the upper bound on the mass of the lightest Higgs boson}},
  \href{http://dx.doi.org/10.1088/1126-6708/2000/06/009}{{\em JHEP}}
  \href{http://dx.doi.org/10.1088/1126-6708/2000/06/009}{{\bf 06}}
  \href{http://dx.doi.org/10.1088/1126-6708/2000/06/009}{(2000)}
  \href{http://dx.doi.org/10.1088/1126-6708/2000/06/009}{009},
  [\href{http://xxx.lanl.gov/abs/hep-ph/9909540}{{\tt hep-ph/9909540}}].
%%CITATION = HEP-PH/9909540;%%

\bibitem{MSSMmhmax}
{M. Carena, S. Heinemeyer, C. E. M. Wagner and G. Weiglein}, {\it {Suggestions
  for benchmark scenarios for MSSM Higgs boson searches at hadron colliders}},
  \href{http://dx.doi.org/10.1140/epjc/s2002-01084-3}{{\em Eur. Phys. J.}}
  \href{http://dx.doi.org/10.1140/epjc/s2002-01084-3}{{\bf C 26}}
  \href{http://dx.doi.org/10.1140/epjc/s2002-01084-3}{(2003)}
  \href{http://dx.doi.org/10.1140/epjc/s2002-01084-3}{601},
  [\href{http://xxx.lanl.gov/abs/hep-ph/0202167}{{\tt hep-ph/0202167}}].
%%CITATION = HEP-PH/0202167;%%

\bibitem{MSSMBenchmarks}
M.~Carena, S.~Heinemeyer, O.~St\r{a}l, C.~Wagner, and G.~Weiglein, {\it {MSSM
  Higgs Boson Searches at the LHC: Benchmark Scenarios after the Discovery of a
  Higgs-like Particle}},
  \href{http://dx.doi.org/10.1140/epjc/s10052-013-2552-1}{{\em Eur. Phys. J.}}
  \href{http://dx.doi.org/10.1140/epjc/s10052-013-2552-1}{{\bf C 73}}
  \href{http://dx.doi.org/10.1140/epjc/s10052-013-2552-1}{(2013)}
  \href{http://dx.doi.org/10.1140/epjc/s10052-013-2552-1}{2552},
  [\href{http://xxx.lanl.gov/abs/1302.7033}{{\tt arXiv:1302.7033}}].
%%CITATION = ARXIV:1302.7033;%%

\bibitem{ATLASHiggsMass}
{\it {Measurement of the Higgs boson mass from the $H\rightarrow \gamma\gamma$
  and $H \rightarrow ZZ^{*} \rightarrow 4\ell$ channels with the ATLAS detector
  using 25 fb$^{-1}$ of $pp$ collision data}},
  \href{http://xxx.lanl.gov/abs/1406.3827}{{\tt arXiv:1406.3827}}.
%%CITATION = ARXIV:1406.3827;%%

\bibitem{Bechtle:2012jw}
P.~Bechtle et~al., {\it {MSSM Interpretations of the LHC Discovery: Light or
  Heavy Higgs?}},  \href{http://dx.doi.org/10.1140/epjc/s10052-013-2354-5}{{\em
  Eur. Phys. J.}} \href{http://dx.doi.org/10.1140/epjc/s10052-013-2354-5}{{\bf
  C 73}} \href{http://dx.doi.org/10.1140/epjc/s10052-013-2354-5}{(2013)}
  \href{http://dx.doi.org/10.1140/epjc/s10052-013-2354-5}{2354},
  [\href{http://xxx.lanl.gov/abs/1211.1955}{{\tt arXiv:1211.1955}}].
%%CITATION = ARXIV:1211.1955;%%

\bibitem{Arbey:2012dq}
A.~Arbey, M.~Battaglia, A.~Djouadi, and F.~Mahmoudi, {\it {The Higgs sector of
  the phenomenological MSSM in the light of the Higgs boson discovery}},
  \href{http://dx.doi.org/10.1007/JHEP09(2012)107}{{\em JHEP}}
  \href{http://dx.doi.org/10.1007/JHEP09(2012)107}{{\bf 09}}
  \href{http://dx.doi.org/10.1007/JHEP09(2012)107}{(2012)}
  \href{http://dx.doi.org/10.1007/JHEP09(2012)107}{107},
  [\href{http://xxx.lanl.gov/abs/1207.1348}{{\tt arXiv:1207.1348}}].
%%CITATION = ARXIV:1207.1348;%%

\bibitem{LEPLimits}
{ALEPH, DELPHI, L3, and OPAL Collaborations, S. Schael et al.}, {\it {Search
  for neutral MSSM Higgs bosons at LEP}},
  \href{http://dx.doi.org/10.1140/epjc/s2006-02569-7}{{\em Eur. Phys. J.}}
  \href{http://dx.doi.org/10.1140/epjc/s2006-02569-7}{{\bf C 47}}
  \href{http://dx.doi.org/10.1140/epjc/s2006-02569-7}{(2006)}
  \href{http://dx.doi.org/10.1140/epjc/s2006-02569-7}{547--587},
  [\href{http://xxx.lanl.gov/abs/hep-ex/0602042}{{\tt hep-ex/0602042}}].
%%CITATION = HEP-EX/0602042;%%

\bibitem{TevatronLimits1}
{Tevatron New Phenomena \& Higgs Working Group Collaboration, B. Doug et al.},
  {\it {Combined CDF and D0 upper limits on MSSM Higgs boson production in
  $\tau\tau$ final states with up to $2.2$~fb$^{-1}$}},
  \href{http://xxx.lanl.gov/abs/1003.3363}{{\tt arXiv:1003.3363}}.
%%CITATION = ARXIV:1003.3363;%%

\bibitem{TevatronLimits2}
{CDF Collaboration, T. Aaltonen et al.}, {\it {Search for Higgs bosons
  predicted in two-Higgs-doublet models via decays to $\tau$ lepton pairs in
  1.96 TeV proton--antiproton collisions}},
  \href{http://dx.doi.org/10.1103/PhysRevLett.103.201801}{{\em Phys. Rev.
  Lett.}} \href{http://dx.doi.org/10.1103/PhysRevLett.103.201801}{{\bf 103}}
  \href{http://dx.doi.org/10.1103/PhysRevLett.103.201801}{(2009)}
  \href{http://dx.doi.org/10.1103/PhysRevLett.103.201801}{201801},
  [\href{http://xxx.lanl.gov/abs/0906.1014}{{\tt arXiv:0906.1014}}].
%%CITATION = ARXIV:0906.1014;%%

\bibitem{TevatronLimits3}
{D0 Collaboration, V. M. Abazov et al.}, {\it {Search for Higgs bosons decaying
  to $\tau$ pairs in $p\bar{p}$ collisions with the D0 detector}},
  \href{http://dx.doi.org/10.1103/PhysRevLett.101.071804}{{\em Phys. Rev.
  Lett.}} \href{http://dx.doi.org/10.1103/PhysRevLett.101.071804}{{\bf 101}}
  \href{http://dx.doi.org/10.1103/PhysRevLett.101.071804}{(2008)}
  \href{http://dx.doi.org/10.1103/PhysRevLett.101.071804}{071804},
  [\href{http://xxx.lanl.gov/abs/0805.2491}{{\tt arXiv:0805.2491}}].
%%CITATION = ARXIV:0805.2491;%%

\bibitem{ATLASLimit}
{ATLAS Collaboration}, {\it {Search for the neutral Higgs bosons of the minimal
  supersymmetric standard model in $pp$ collisions at $\sqrt{s}=7$ TeV with the
  ATLAS detector}},  \href{http://dx.doi.org/10.1007/JHEP02(2013)095}{{\em
  JHEP}} \href{http://dx.doi.org/10.1007/JHEP02(2013)095}{{\bf 02}}
  \href{http://dx.doi.org/10.1007/JHEP02(2013)095}{(2013)}
  \href{http://dx.doi.org/10.1007/JHEP02(2013)095}{095},
  [\href{http://xxx.lanl.gov/abs/1211.6956}{{\tt arXiv:1211.6956}}].
%%CITATION = ARXIV:1211.6956;%%

\bibitem{CMSLimit}
{CMS Collaboration}, {\it {Search for neutral MSSM Higgs bosons decaying to a
  pair of tau leptons in pp collisions}},
  \href{http://xxx.lanl.gov/abs/1408.3316}{{\tt arXiv:1408.3316}}.
%%CITATION = ARXIV:1408.3316;%%

\bibitem{LHCbHtautau}
{LHCb Collaboration, R. Aaij et al.}, {\it {Limits on neutral Higgs boson
  production in the forward region in $pp$ collisions at $\sqrt{s} = 7$ TeV}},
  \href{http://dx.doi.org/10.1007/JHEP05(2013)132}{{\em JHEP}}
  \href{http://dx.doi.org/10.1007/JHEP05(2013)132}{{\bf 05}}
  \href{http://dx.doi.org/10.1007/JHEP05(2013)132}{(2013)}
  \href{http://dx.doi.org/10.1007/JHEP05(2013)132}{132},
  [\href{http://xxx.lanl.gov/abs/1304.2591}{{\tt arXiv:1304.2591}}].
%%CITATION = ARXIV:1304.2591;%%

\bibitem{ATLASDetector}
{ATLAS Collaboration}, {\it {The ATLAS experiment at the CERN Large Hadron
  Collider}},  \href{http://dx.doi.org/10.1088/1748-0221/3/08/S08003}{{\em
  JINST}} \href{http://dx.doi.org/10.1088/1748-0221/3/08/S08003}{{\bf 3}}
  \href{http://dx.doi.org/10.1088/1748-0221/3/08/S08003}{(2008)}
  \href{http://dx.doi.org/10.1088/1748-0221/3/08/S08003}{S08003}.
%%CITATION = JINST,3,S08003;%%

\bibitem{SMHtautau2011}
{ATLAS Collaboration}, {\it {Search for the Standard Model Higgs boson in the
  $H \to \tau \tau$ decay mode in $\sqrt{s}$=7~TeV pp collisions with ATLAS}},
  \href{http://dx.doi.org/10.1007/JHEP09(2012)070}{{\em JHEP}}
  \href{http://dx.doi.org/10.1007/JHEP09(2012)070}{{\bf 09}}
  \href{http://dx.doi.org/10.1007/JHEP09(2012)070}{(2012)}
  \href{http://dx.doi.org/10.1007/JHEP09(2012)070}{070},
  [\href{http://xxx.lanl.gov/abs/1206.5971}{{\tt arXiv:1206.5971}}].
%%CITATION = ARXIV:1206.5971;%%

\bibitem{ATLASSIM}
{ATLAS Collaboration}, {\it {The ATLAS simulation infrastructure}},
  \href{http://dx.doi.org/10.1140/epjc/s10052-010-1429-9}{{\em Eur. Phys. J.}}
  \href{http://dx.doi.org/10.1140/epjc/s10052-010-1429-9}{{\bf C 70}}
  \href{http://dx.doi.org/10.1140/epjc/s10052-010-1429-9}{(2010)}
  \href{http://dx.doi.org/10.1140/epjc/s10052-010-1429-9}{823--874},
  [\href{http://xxx.lanl.gov/abs/1005.4568}{{\tt arXiv:1005.4568}}].
%%CITATION = ARXIV:1005.4568;%%

\bibitem{Geant4}
{GEANT4 Collaboration, S. Agostinelli et al.}, {\it {GEANT4 - a simulation
  toolkit}},  \href{http://dx.doi.org/10.1016/S0168-9002(03)01368-8}{{\em Nucl.
  Instrum. Meth.}} \href{http://dx.doi.org/10.1016/S0168-9002(03)01368-8}{{\bf
  A 506}} \href{http://dx.doi.org/10.1016/S0168-9002(03)01368-8}{(2003)}
  \href{http://dx.doi.org/10.1016/S0168-9002(03)01368-8}{250--303}.
%%CITATION = NUIMA,A506,250;%%

\bibitem{HIGLU}
{M.~Spira}, {\it {HIGLU: A program for the calculation of the total Higgs
  production cross section at hadron colliders via gluon fusion including QCD
  corrections}},  \href{http://xxx.lanl.gov/abs/hep-ph/9510347}{{\tt
  hep-ph/9510347}}.
%%CITATION = HEP-PH/9510347;%%

\bibitem{Harlander:2002wh}
{R. V. Harlander and W. B. Kilgore}, {\it {Next-to-Next-to-Leading Order Higgs
  Production at Hadron Colliders}},
  \href{http://dx.doi.org/10.1103/PhysRevLett.88.201801}{{\em Phys. Rev.
  Lett.}} \href{http://dx.doi.org/10.1103/PhysRevLett.88.201801}{{\bf 88}}
  \href{http://dx.doi.org/10.1103/PhysRevLett.88.201801}{(2002)}
  \href{http://dx.doi.org/10.1103/PhysRevLett.88.201801}{201801},
  [\href{http://xxx.lanl.gov/abs/hep-ph/0201206}{{\tt hep-ph/0201206}}].
%%CITATION = HEP-PH/0201206;%%

\bibitem{Harlander:2012pb}
R.~V. Harlander, S.~Liebler, and H.~Mantler, {\it {SusHi: A program for the
  calculation of Higgs production in gluon fusion and bottom-quark annihilation
  in the Standard Model and the MSSM}},
  \href{http://dx.doi.org/10.1016/j.cpc.2013.02.006}{{\em Comp. Phys. Commun.}}
  \href{http://dx.doi.org/10.1016/j.cpc.2013.02.006}{{\bf 184}}
  \href{http://dx.doi.org/10.1016/j.cpc.2013.02.006}{(2013)}
  \href{http://dx.doi.org/10.1016/j.cpc.2013.02.006}{1605--1617},
  [\href{http://xxx.lanl.gov/abs/1212.3249}{{\tt arXiv:1212.3249}}].
%%CITATION = ARXIV:1212.3249;%%

\bibitem{Harlander:2003bb}
R.~V. Harlander and M.~Steinhauser, {\it {Hadronic Higgs production and decay
  in supersymmetry at next-to-leading order}},
  \href{http://dx.doi.org/10.1016/j.physletb.2003.09.013}{{\em Phys. Lett.}}
  \href{http://dx.doi.org/10.1016/j.physletb.2003.09.013}{{\bf B 574}}
  \href{http://dx.doi.org/10.1016/j.physletb.2003.09.013}{(2003)}
  \href{http://dx.doi.org/10.1016/j.physletb.2003.09.013}{258--268},
  [\href{http://xxx.lanl.gov/abs/hep-ph/0307346}{{\tt hep-ph/0307346}}].
%%CITATION = HEP-PH/0307346;%%

\bibitem{Harlander:2004tp}
R.~V. Harlander and M.~Steinhauser, {\it {Supersymmetric Higgs production in
  gluon fusion at next-to-leading order}},
  \href{http://dx.doi.org/10.1088/1126-6708/2004/09/066}{{\em JHEP}}
  \href{http://dx.doi.org/10.1088/1126-6708/2004/09/066}{{\bf 09}}
  \href{http://dx.doi.org/10.1088/1126-6708/2004/09/066}{(2004)}
  \href{http://dx.doi.org/10.1088/1126-6708/2004/09/066}{066},
  [\href{http://xxx.lanl.gov/abs/hep-ph/0409010}{{\tt hep-ph/0409010}}].
%%CITATION = HEP-PH/0409010;%%

\bibitem{Harlander:2003kf}
R.~Harlander and M.~Steinhauser, {\it {Effects of SUSY QCD in hadronic Higgs
  production at next-to-next-to-leading order}},
  \href{http://dx.doi.org/10.1103/PhysRevD.68.111701}{{\em Phys.Rev.}}
  \href{http://dx.doi.org/10.1103/PhysRevD.68.111701}{{\bf D68}}
  \href{http://dx.doi.org/10.1103/PhysRevD.68.111701}{(2003)}
  \href{http://dx.doi.org/10.1103/PhysRevD.68.111701}{111701},
  [\href{http://xxx.lanl.gov/abs/hep-ph/0308210}{{\tt hep-ph/0308210}}].
%%CITATION = HEP-PH/0308210;%%

\bibitem{Harlander:2003ai}
{R. Harlander and W. B. Kilgore}, {\it Higgs boson production in bottom quark
  fusion at next-to-next-to-leading order},
  \href{http://dx.doi.org/10.1103/PhysRevD.68.013001}{{\em Phys. Rev.}}
  \href{http://dx.doi.org/10.1103/PhysRevD.68.013001}{{\bf D 68}}
  \href{http://dx.doi.org/10.1103/PhysRevD.68.013001}{(2003)}
  \href{http://dx.doi.org/10.1103/PhysRevD.68.013001}{013001},
  [\href{http://xxx.lanl.gov/abs/hep-ph/0304035}{{\tt hep-ph/0304035}}].
%%CITATION = HEP-PH/0304035;%%

\bibitem{Aglietti:2004nj}
U.~Aglietti, R.~Bonciani, G.~Degrassi, and A.~Vicini, {\it {Two loop light
  fermion contribution to Higgs production and decays}},
  \href{http://dx.doi.org/10.1016/j.physletb.2004.06.063}{{\em Phys.Lett.}}
  \href{http://dx.doi.org/10.1016/j.physletb.2004.06.063}{{\bf B595}}
  \href{http://dx.doi.org/10.1016/j.physletb.2004.06.063}{(2004)}
  \href{http://dx.doi.org/10.1016/j.physletb.2004.06.063}{432--441},
  [\href{http://xxx.lanl.gov/abs/hep-ph/0404071}{{\tt hep-ph/0404071}}].
%%CITATION = HEP-PH/0404071;%%

\bibitem{Bonciani:2010ms}
R.~Bonciani, G.~Degrassi, and A.~Vicini, {\it {On the Generalized Harmonic
  Polylogarithms of One Complex Variable}},
  \href{http://dx.doi.org/10.1016/j.cpc.2011.02.011}{{\em Comput.Phys.Commun.}}
  \href{http://dx.doi.org/10.1016/j.cpc.2011.02.011}{{\bf 182}}
  \href{http://dx.doi.org/10.1016/j.cpc.2011.02.011}{(2011)}
  \href{http://dx.doi.org/10.1016/j.cpc.2011.02.011}{1253--1264},
  [\href{http://xxx.lanl.gov/abs/1007.1891}{{\tt arXiv:1007.1891}}].
%%CITATION = ARXIV:1007.1891;%%

\bibitem{Degrassi:2010eu}
G.~Degrassi and P.~Slavich, {\it {NLO QCD bottom corrections to Higgs boson
  production in the MSSM}},
  \href{http://dx.doi.org/10.1007/JHEP11(2010)044}{{\em JHEP}}
  \href{http://dx.doi.org/10.1007/JHEP11(2010)044}{{\bf 11}}
  \href{http://dx.doi.org/10.1007/JHEP11(2010)044}{(2010)}
  \href{http://dx.doi.org/10.1007/JHEP11(2010)044}{044},
  [\href{http://xxx.lanl.gov/abs/1007.3465}{{\tt arXiv:1007.3465}}].
%%CITATION = ARXIV:1007.3465;%%

\bibitem{Degrassi:2011vq}
G.~Degrassi, S.~Di~Vita, and P.~Slavich, {\it {NLO QCD corrections to
  pseudoscalar Higgs production in the MSSM}},
  \href{http://dx.doi.org/10.1007/JHEP08(2011)128}{{\em JHEP}}
  \href{http://dx.doi.org/10.1007/JHEP08(2011)128}{{\bf 08}}
  \href{http://dx.doi.org/10.1007/JHEP08(2011)128}{(2011)}
  \href{http://dx.doi.org/10.1007/JHEP08(2011)128}{128},
  [\href{http://xxx.lanl.gov/abs/1107.0914}{{\tt arXiv:1107.0914}}].
%%CITATION = ARXIV:1107.0914;%%

\bibitem{Degrassi:2012vt}
G.~Degrassi, S.~Di~Vita, and P.~Slavich, {\it {On the NLO QCD Corrections to
  the Production of the Heaviest Neutral Higgs Scalar in the MSSM}},
  \href{http://dx.doi.org/10.1140/epjc/s10052-012-2032-z}{{\em Eur. Phys. J.}}
  \href{http://dx.doi.org/10.1140/epjc/s10052-012-2032-z}{{\bf C 72}}
  \href{http://dx.doi.org/10.1140/epjc/s10052-012-2032-z}{(2012)}
  \href{http://dx.doi.org/10.1140/epjc/s10052-012-2032-z}{2032},
  [\href{http://xxx.lanl.gov/abs/1204.1016}{{\tt arXiv:1204.1016}}].
%%CITATION = ARXIV:1204.1016;%%

\bibitem{Heinemeyer:1998yj}
S.~Heinemeyer, W.~Hollik, and G.~Weiglein, {\it {FeynHiggs: A Program for the
  calculation of the masses of the neutral CP even Higgs bosons in the MSSM}},
  \href{http://dx.doi.org/10.1016/S0010-4655(99)00364-1}{{\em Comput. Phys.
  Commun.}} \href{http://dx.doi.org/10.1016/S0010-4655(99)00364-1}{{\bf 124}}
  \href{http://dx.doi.org/10.1016/S0010-4655(99)00364-1}{(2000)}
  \href{http://dx.doi.org/10.1016/S0010-4655(99)00364-1}{76--89},
  [\href{http://xxx.lanl.gov/abs/hep-ph/9812320}{{\tt hep-ph/9812320}}].
%%CITATION = HEP-PH/9812320;%%

\bibitem{Heinemeyer:1998np}
S.~Heinemeyer, W.~Hollik, and G.~Weiglein, {\it {The Masses of the neutral CP -
  even Higgs bosons in the MSSM: Accurate analysis at the two loop level}},
  \href{http://dx.doi.org/10.1007/s100529900006}{{\em Eur. Phys. J.}}
  \href{http://dx.doi.org/10.1007/s100529900006}{{\bf C 9}}
  \href{http://dx.doi.org/10.1007/s100529900006}{(1999)}
  \href{http://dx.doi.org/10.1007/s100529900006}{343--366},
  [\href{http://xxx.lanl.gov/abs/hep-ph/9812472}{{\tt hep-ph/9812472}}].
%%CITATION = HEP-PH/9812472;%%

\bibitem{Degrassi:2002fi}
G.~Degrassi, S.~Heinemeyer, W.~Hollik, P.~Slavich, and G.~Weiglein, {\it
  {Towards high precision predictions for the MSSM Higgs sector}},
  \href{http://dx.doi.org/10.1140/epjc/s2003-01152-2}{{\em Eur. Phys. J.}}
  \href{http://dx.doi.org/10.1140/epjc/s2003-01152-2}{{\bf C 28}}
  \href{http://dx.doi.org/10.1140/epjc/s2003-01152-2}{(2003)}
  \href{http://dx.doi.org/10.1140/epjc/s2003-01152-2}{133--143},
  [\href{http://xxx.lanl.gov/abs/hep-ph/0212020}{{\tt hep-ph/0212020}}].
%%CITATION = HEP-PH/0212020;%%

\bibitem{Frank:2006yh}
{M. Frank et al.}, {\it {The Higgs boson masses and mixings of the complex MSSM
  in the Feynman-diagrammatic approach}},
  \href{http://dx.doi.org/10.1088/1126-6708/2007/02/047}{{\em JHEP}}
  \href{http://dx.doi.org/10.1088/1126-6708/2007/02/047}{{\bf 02}}
  \href{http://dx.doi.org/10.1088/1126-6708/2007/02/047}{(2007)}
  \href{http://dx.doi.org/10.1088/1126-6708/2007/02/047}{047},
  [\href{http://xxx.lanl.gov/abs/hep-ph/0611326}{{\tt hep-ph/0611326}}].
%%CITATION = HEP-PH/0611326;%%

\bibitem{Harlander:2005rq}
R.~Harlander and P.~Kant, {\it {Higgs production and decay: Analytic results at
  next-to-leading order QCD}},
  \href{http://dx.doi.org/10.1088/1126-6708/2005/12/015}{{\em JHEP}}
  \href{http://dx.doi.org/10.1088/1126-6708/2005/12/015}{{\bf 12}}
  \href{http://dx.doi.org/10.1088/1126-6708/2005/12/015}{(2005)}
  \href{http://dx.doi.org/10.1088/1126-6708/2005/12/015}{015},
  [\href{http://xxx.lanl.gov/abs/hep-ph/0509189}{{\tt hep-ph/0509189}}].
%%CITATION = HEP-PH/0509189;%%

\bibitem{Dittmaier:2003ej}
{S.~Dittmaier, M. Kr\"amer and M. Spira}, {\it {Higgs radiation off bottom
  quarks at the Tevatron and the LHC}},
  \href{http://dx.doi.org/10.1103/PhysRevD.70.074010}{{\em Phys. Rev.}}
  \href{http://dx.doi.org/10.1103/PhysRevD.70.074010}{{\bf D 70}}
  \href{http://dx.doi.org/10.1103/PhysRevD.70.074010}{(2004)}
  \href{http://dx.doi.org/10.1103/PhysRevD.70.074010}{074010},
  [\href{http://xxx.lanl.gov/abs/hep-ph/0309204}{{\tt hep-ph/0309204}}].
%%CITATION = HEP-PH/0309204;%%

\bibitem{Dawson:2004a}
{S. Dawson, C. B. Jackson, L. Reina and D. Wackeroth}, {\it {Exclusive Higgs
  boson production with bottom quarks at hadron colliders}},
  \href{http://dx.doi.org/10.1103/PhysRevD.69.074027}{{\em Phys. Rev.}}
  \href{http://dx.doi.org/10.1103/PhysRevD.69.074027}{{\bf D 69}}
  \href{http://dx.doi.org/10.1103/PhysRevD.69.074027}{(2004)}
  \href{http://dx.doi.org/10.1103/PhysRevD.69.074027}{074027},
  [\href{http://xxx.lanl.gov/abs/hep-ph/0311067}{{\tt hep-ph/0311067}}].
%%CITATION = HEP-PH/0311067;%%

\bibitem{SantanderMatching}
R.~Harlander, M.~Kr{\"a}mer, and M.~Schumacher, {\it {Bottom-quark associated
  Higgs-boson production: reconciling the four- and five-flavour scheme
  approach}},  \href{http://xxx.lanl.gov/abs/1112.3478}{{\tt arXiv:1112.3478}}.
%%CITATION = ARXIV:1112.3478;%%

\bibitem{POWHEG}
S.~Alioli, P.~Nason, C.~Oleari, and E.~Re, {\it {NLO Higgs boson production via
  gluon fusion matched with shower in POWHEG}},
  \href{http://dx.doi.org/10.1088/1126-6708/2009/04/002}{{\em JHEP}}
  \href{http://dx.doi.org/10.1088/1126-6708/2009/04/002}{{\bf 04}}
  \href{http://dx.doi.org/10.1088/1126-6708/2009/04/002}{(2009)}
  \href{http://dx.doi.org/10.1088/1126-6708/2009/04/002}{002},
  [\href{http://xxx.lanl.gov/abs/0812.0578}{{\tt arXiv:0812.0578}}].
%%CITATION = ARXIV:0812.0578;%%

\bibitem{SHERPA}
{T. Gleisberg et al.}, {\it {Event generation with SHERPA 1.1}},
  \href{http://dx.doi.org/10.1088/1126-6708/2009/02/007}{{\em JHEP}}
  \href{http://dx.doi.org/10.1088/1126-6708/2009/02/007}{{\bf 02}}
  \href{http://dx.doi.org/10.1088/1126-6708/2009/02/007}{(2009)}
  \href{http://dx.doi.org/10.1088/1126-6708/2009/02/007}{007},
  [\href{http://xxx.lanl.gov/abs/0811.4622}{{\tt arXiv:0811.4622}}].
%%CITATION = ARXIV:0811.4622;%%

\bibitem{Lai:2010vv}
H.-L. Lai et~al., {\it {New parton distributions for collider physics}},
  \href{http://dx.doi.org/10.1103/PhysRevD.82.074024}{{\em Phys. Rev.}}
  \href{http://dx.doi.org/10.1103/PhysRevD.82.074024}{{\bf D 82}}
  \href{http://dx.doi.org/10.1103/PhysRevD.82.074024}{(2010)}
  \href{http://dx.doi.org/10.1103/PhysRevD.82.074024}{074024},
  [\href{http://xxx.lanl.gov/abs/1007.2241}{{\tt arXiv:1007.2241}}].
%%CITATION = ARXIV:1007.2241;%%

\bibitem{Alpgen}
{M. L. Mangano, M. Moretti, F. Piccinini, R. Pittau and A. D. Polosa}, {\it
  {ALPGEN, a generator for hard multiparton processes in hadronic collisions}},
   \href{http://dx.doi.org/10.1088/1126-6708/2003/07/001}{{\em JHEP}}
  \href{http://dx.doi.org/10.1088/1126-6708/2003/07/001}{{\bf 07}}
  \href{http://dx.doi.org/10.1088/1126-6708/2003/07/001}{(2003)}
  \href{http://dx.doi.org/10.1088/1126-6708/2003/07/001}{001},
  [\href{http://xxx.lanl.gov/abs/hep-ph/0206293}{{\tt hep-ph/0206293}}].
%%CITATION = HEP-PH/0206293;%%

\bibitem{Pythia}
{T. Sj\"ostrand, S. Mrenna and P. Skands}, {\it {PYTHIA 6.4 physics and
  manual}},  \href{http://dx.doi.org/10.1088/1126-6708/2006/05/026}{{\em JHEP}}
  \href{http://dx.doi.org/10.1088/1126-6708/2006/05/026}{{\bf 05}}
  \href{http://dx.doi.org/10.1088/1126-6708/2006/05/026}{(2006)}
  \href{http://dx.doi.org/10.1088/1126-6708/2006/05/026}{026},
  [\href{http://xxx.lanl.gov/abs/hep-ph/0603175}{{\tt hep-ph/0603175}}].
%%CITATION = HEP-PH/0603175;%%

\bibitem{Pythia8}
{T. Sj\"ostrand, S. Mrenna and P. Skands}, {\it {A Brief Introduction to PYTHIA
  8.1}},  \href{http://dx.doi.org/10.1016/j.cpc.2008.01.036}{{\em Comput. Phys.
  Commun.}} \href{http://dx.doi.org/10.1016/j.cpc.2008.01.036}{{\bf 178}}
  \href{http://dx.doi.org/10.1016/j.cpc.2008.01.036}{(2008)}
  \href{http://dx.doi.org/10.1016/j.cpc.2008.01.036}{852--867},
  [\href{http://xxx.lanl.gov/abs/0710.3820}{{\tt arXiv:0710.3820}}].
%%CITATION = ARXIV:0710.3820;%%

\bibitem{Herwig}
{G. Corcella et al.}, {\it {HERWIG 6.5: an event generator for hadron emission
  reactions with interfering gluons (including supersymmetric processes)}},
  \href{http://dx.doi.org/10.1088/1126-6708/2001/01/010}{{\em JHEP}}
  \href{http://dx.doi.org/10.1088/1126-6708/2001/01/010}{{\bf 01}}
  \href{http://dx.doi.org/10.1088/1126-6708/2001/01/010}{(2001)}
  \href{http://dx.doi.org/10.1088/1126-6708/2001/01/010}{010},
  [\href{http://xxx.lanl.gov/abs/hep-ph/0011363}{{\tt hep-ph/0011363}}].
%%CITATION = HEP-PH/0011363;%%

\bibitem{MCatNLO}
S.~Frixione and B.~R. Webber, {\it {Matching NLO QCD computations and parton
  shower simulations}},
  \href{http://dx.doi.org/10.1088/1126-6708/2002/06/029}{{\em JHEP}}
  \href{http://dx.doi.org/10.1088/1126-6708/2002/06/029}{{\bf 06}}
  \href{http://dx.doi.org/10.1088/1126-6708/2002/06/029}{(2002)}
  \href{http://dx.doi.org/10.1088/1126-6708/2002/06/029}{029},
  [\href{http://xxx.lanl.gov/abs/hep-ph/0204244}{{\tt hep-ph/0204244}}].
%%CITATION = HEP-PH/0204244;%%

\bibitem{AcerMC}
{B. P. Kersevan and E. Richter-W\c{a}s}, {\it {The Monte Carlo event generator
  AcerMC version 2.0 with interfaces to PYTHIA 6.2 and HERWIG 6.5}},
  \href{http://xxx.lanl.gov/abs/hep-ph/0405247}{{\tt hep-ph/0405247}}.
%%CITATION = HEP-PH/0405247;%%

\bibitem{Pumplin:2002vw}
J.~Pumplin, D.~Stump, J.~Huston, H.~Lai, P.~M. Nadolsky, et~al., {\it {New
  generation of parton distributions with uncertainties from global QCD
  analysis}},  \href{http://dx.doi.org/10.1088/1126-6708/2002/07/012}{{\em
  JHEP}} \href{http://dx.doi.org/10.1088/1126-6708/2002/07/012}{{\bf 07}}
  \href{http://dx.doi.org/10.1088/1126-6708/2002/07/012}{(2002)}
  \href{http://dx.doi.org/10.1088/1126-6708/2002/07/012}{012},
  [\href{http://xxx.lanl.gov/abs/hep-ph/0201195}{{\tt hep-ph/0201195}}].
%%CITATION = HEP-PH/0201195;%%

\bibitem{TAUOLA}
{S. Jadach, J. H. Kuhn and Z. Was}, {\it {TAUOLA - a library of Monte Carlo
  programs to simulate decays of polarized $\tau$ leptons}},
  \href{http://dx.doi.org/10.1016/0010-4655(91)90038-M}{{\em Comput. Phys.
  Commun.}} \href{http://dx.doi.org/10.1016/0010-4655(91)90038-M}{{\bf 64}}
  \href{http://dx.doi.org/10.1016/0010-4655(91)90038-M}{(1990)}
  \href{http://dx.doi.org/10.1016/0010-4655(91)90038-M}{275--299}.
%%CITATION = CPHCB,64,275;%%

\bibitem{PHOTOS}
{E. Barberio, B. V. Eijk and Z. Was}, {\it {PHOTOS - a universal Monte Carlo
  for QED radiative corrections in decays}},
  \href{http://dx.doi.org/10.1016/0010-4655(91)90012-A}{{\em Comput. Phys.
  Commun.}} \href{http://dx.doi.org/10.1016/0010-4655(91)90012-A}{{\bf 66}}
  \href{http://dx.doi.org/10.1016/0010-4655(91)90012-A}{(1991)}
  \href{http://dx.doi.org/10.1016/0010-4655(91)90012-A}{115--128}.
%%CITATION = CPHCB,66,115;%%

\bibitem{Czyczula:2012ny}
Z.~Czyczula, T.~Przedzinski, and Z.~Was, {\it {TauSpinner Program for Studies
  on Spin Effect in tau Production at the LHC}},
  \href{http://dx.doi.org/10.1140/epjc/s10052-012-1988-z}{{\em Eur. Phys. J.}}
  \href{http://dx.doi.org/10.1140/epjc/s10052-012-1988-z}{{\bf C 72}}
  \href{http://dx.doi.org/10.1140/epjc/s10052-012-1988-z}{(2012)}
  \href{http://dx.doi.org/10.1140/epjc/s10052-012-1988-z}{1988},
  [\href{http://xxx.lanl.gov/abs/1201.0117}{{\tt arXiv:1201.0117}}].
%%CITATION = ARXIV:1201.0117;%%

\bibitem{ATLAS-CONF-2014-032}
{ATLAS Collaboration}, {\it {Electron efficiency measurements with the ATLAS
  detector using the 2012 LHC proton-proton collision data}},
  \href{http://cdsweb.cern.ch/record/1706245}{ATLAS-CONF-2014-032}, available
  at http://cdsweb.cern.ch/record/1706245.

\bibitem{Aad:2014rra}
{ATLAS Collaboration}, {\it {Measurement of the muon reconstruction performance
  of the ATLAS detector using 2011 and 2012 LHC proton-proton collision data}},
   \href{http://xxx.lanl.gov/abs/1407.3935}{{\tt arXiv:1407.3935}}.
%%CITATION = ARXIV:1407.3935;%%

\bibitem{AntiKT}
{M. Cacciari, G. P. Salam and G. Soyez}, {\it {The anti-$k_t$ jet clustering
  algorithm}},  \href{http://dx.doi.org/10.1088/1126-6708/2008/04/063}{{\em
  JHEP}} \href{http://dx.doi.org/10.1088/1126-6708/2008/04/063}{{\bf 04}}
  \href{http://dx.doi.org/10.1088/1126-6708/2008/04/063}{(2008)}
  \href{http://dx.doi.org/10.1088/1126-6708/2008/04/063}{063},
  [\href{http://xxx.lanl.gov/abs/0802.1189}{{\tt arXiv:0802.1189}}].
%%CITATION = ARXIV:0802.1189;%%

\bibitem{TopoClusterAlgo}
{W. Lampl et al.}, {\it Calorimeter clustering algorithms: Description and
  performance},
  \href{http://cdsweb.cern.ch/record/1099735}{ATL-LARG-PUB-2008-002}, CERN,
  Geneva Switzerland (2008).

\bibitem{ATLASJETEnergyScale}
{ATLAS Collaboration}, {\it {Jet energy measurement and its systematic
  uncertainty in proton-proton collisions at $\sqrt{s}=7$ TeV with the ATLAS
  detector}},  \href{http://xxx.lanl.gov/abs/1406.0076}{{\tt arXiv:1406.0076}}.
%%CITATION = ARXIV:1406.0076;%%

\bibitem{ATLAS-CONF-2014-004}
{ATLAS Collaboration}, {\it {Calibration of b-tagging using dileptonic top pair
  events in a combinatorial likelihood approach with the ATLAS experiment}},
  \href{http://cds.cern.ch/record/1664335}{ATLAS-CONF-2014-004}, available at
  http://cds.cern.ch/record/1664335.

\bibitem{ATLAS-CONF-2014-046}
{ATLAS Collaboration}, {\it {Calibration of the performance of b-tagging for c
  and light-flavour jets in the 2012 ATLAS data}},
  \href{http://cds.cern.ch/record/1741020}{ATLAS-CONF-2014-046}, available at
  http://cds.cern.ch/record/1741020.

\bibitem{ATLASTauIDNew}
{ATLAS Collaboration}, {\it {Identification of the Hadronic Decays of Tau
  Leptons in 2012 Data with the ATLAS Detector}},
  \href{http://cds.cern.ch/record/1562839}{ATLAS-CONF-2013-064}, available at
  http://cds.cern.ch/record/1562839.

\bibitem{ATLASmetNEW}
{ATLAS Collaboration}, {\it {Performance of missing transverse momentum
  reconstruction in proton--proton collisions at 7 TeV with ATLAS}},
  \href{http://dx.doi.org/10.1140/epjc/s10052-011-1844-6}{{\em Eur. Phys. J.}}
  \href{http://dx.doi.org/10.1140/epjc/s10052-011-1844-6}{{\bf C 72}}
  \href{http://dx.doi.org/10.1140/epjc/s10052-011-1844-6}{(2012)}
  \href{http://dx.doi.org/10.1140/epjc/s10052-011-1844-6}{1844},
  [\href{http://xxx.lanl.gov/abs/1108.5602}{{\tt arXiv:1108.5602}}].
%%CITATION = ARXIV:1108.5602;%%

\bibitem{MMCpaper}
{A. Elagin, P. Murat, A. Pranko and A. Safonov}, {\it {A new mass
  reconstruction technique for resonances decaying to di-$\tau$}},
  \href{http://dx.doi.org/10.1016/j.nima.2011.07.009}{{\em Nucl. Instrum.
  Meth.}} \href{http://dx.doi.org/10.1016/j.nima.2011.07.009}{{\bf A 654}}
  \href{http://dx.doi.org/10.1016/j.nima.2011.07.009}{(2011)}
  \href{http://dx.doi.org/10.1016/j.nima.2011.07.009}{481--489},
  [\href{http://xxx.lanl.gov/abs/1012.4686}{{\tt arXiv:1012.4686}}].
%%CITATION = ARXIV:1012.4686;%%

\bibitem{Anastasiou:2003ds}
C.~Anastasiou, L.~J. Dixon, K.~Melnikov, and F.~Petriello, {\it {High precision
  QCD at hadron colliders: Electroweak gauge boson rapidity distributions at
  NNLO}},  \href{http://dx.doi.org/10.1103/PhysRevD.69.094008}{{\em Phys.
  Rev.}} \href{http://dx.doi.org/10.1103/PhysRevD.69.094008}{{\bf D 69}}
  \href{http://dx.doi.org/10.1103/PhysRevD.69.094008}{(2004)}
  \href{http://dx.doi.org/10.1103/PhysRevD.69.094008}{094008},
  [\href{http://xxx.lanl.gov/abs/hep-ph/0312266}{{\tt hep-ph/0312266}}].
%%CITATION = HEP-PH/0312266;%%

\bibitem{LHCHiggsCrossSectionWorkingGroup:2011}
{LHC Higgs Cross Section Working Group}, {\it {Handbook of LHC Higgs Cross
  Sections: 1. Inclusive Observables}},
  \href{http://xxx.lanl.gov/abs/1101.0593}{{\tt arXiv:1101.0593}}.
%%CITATION = ARXIV:1101.0593;%%

\bibitem{lumi}
{ATLAS Collaboration}, {\it {Improved luminosity determination in $pp$
  collisions at $\sqrt{s}$ = 7 TeV using the ATLAS detector at the LHC}},
  \href{http://dx.doi.org/10.1140/epjc/s10052-013-2518-3}{{\em Eur. Phys. J.}}
  \href{http://dx.doi.org/10.1140/epjc/s10052-013-2518-3}{{\bf C 73}}
  \href{http://dx.doi.org/10.1140/epjc/s10052-013-2518-3}{(2013)}
  \href{http://dx.doi.org/10.1140/epjc/s10052-013-2518-3}{2518},
  [\href{http://xxx.lanl.gov/abs/1302.4393}{{\tt arXiv:1302.4393}}].
%%CITATION = ARXIV:1302.4393;%%

\bibitem{ATLAS-CONF-2013-006}
{ATLAS Collaboration}, {\it {Performance of the ATLAS tau trigger in 2011}},
  \href{http://cds.cern.ch/record/1510157}{ATLAS-CONF-2013-006}, available at
  http://cds.cern.ch/record/1510157.

\bibitem{ATLASTauES}
{ATLAS Collaboration}, {\it {Determination of the tau energy scale and the
  associated systematic uncertainty in proton-proton collisions at
  $\sqrt{s}=8\;\textrm{TeV}$ with the ATLAS detector at the LHC in 2012}},
  \href{http://cds.cern.ch/record/1544036}{ATLAS-CONF-2013-044}, available at
  http://cds.cern.ch/record/1544036.
%%CITATION = ATLAS-CONF-2013-044 ETC.;%%

\bibitem{ATLASJETEnergyResolution}
{ATLAS Collaboration}, {\it {Jet energy resolution in proton-proton collisions
  at $\sqrt{s}=7$ TeV recorded in 2010 with the ATLAS detector}},
  \href{http://dx.doi.org/10.1140/epjc/s10052-013-2306-0}{{\em Eur. Phys. J.}}
  \href{http://dx.doi.org/10.1140/epjc/s10052-013-2306-0}{{\bf C 73}}
  \href{http://dx.doi.org/10.1140/epjc/s10052-013-2306-0}{(2013)}
  \href{http://dx.doi.org/10.1140/epjc/s10052-013-2306-0}{2306},
  [\href{http://xxx.lanl.gov/abs/1210.6210}{{\tt arXiv:1210.6210}}].
%%CITATION = ARXIV:1210.6210;%%

\bibitem{ATLASMET}
{ATLAS Collaboration}, {\it {Performance of Missing Transverse Momentum
  Reconstruction in ATLAS studied in Proton-Proton Collisions recorded in 2012
  at 8 TeV}},
  \href{http://cdsweb.cern.ch/record/1570993}{ATLAS-CONF-2013-082}, available
  at http://cdsweb.cern.ch/record/1570993.

\bibitem{egammapaper}
{ATLAS Collaboration}, {\it {Electron reconstruction and identification
  efficiency measurements with the ATLAS detector using the 2011 LHC
  proton-proton collision data}},
  \href{http://xxx.lanl.gov/abs/1404.2240}{{\tt arXiv:1404.2240}}.
%%CITATION = ARXIV:1404.2240;%%

\bibitem{ATLAS_egamma_scale}
{ATLAS Collaboration}, {\it {Electron and photon energy calibration with the
  ATLAS detector using LHC Run 1 data}},
  \href{http://xxx.lanl.gov/abs/1407.5063}{{\tt arXiv:1407.5063}}.

\bibitem{CLs_2002}
A.~L. Read, {\it {Presentation of search results: the CL$_{s}$ technique}},
  \href{http://dx.doi.org/10.1088/0954-3899/28/10/313}{{\em J. Phys.}}
  \href{http://dx.doi.org/10.1088/0954-3899/28/10/313}{{\bf G 28}}
  \href{http://dx.doi.org/10.1088/0954-3899/28/10/313}{(2002)}
  \href{http://dx.doi.org/10.1088/0954-3899/28/10/313}{2693--2704}.
%%CITATION = INSPIRE-599622;%%

\bibitem{CCGV}
{G. Cowan, K. Cranmer, E. Gross and O. Vitells}, {\it {Asymptotic formulae for
  likelihood-based tests of new physics}},
  \href{http://dx.doi.org/10.1140/epjc/s10052-011-1554-0}{{\em Eur. Phys. J.}}
  \href{http://dx.doi.org/10.1140/epjc/s10052-011-1554-0}{{\bf C 71}}
  \href{http://dx.doi.org/10.1140/epjc/s10052-011-1554-0}{(2011)}
  \href{http://dx.doi.org/10.1140/epjc/s10052-011-1554-0}{1554},
  [\href{http://xxx.lanl.gov/abs/1007.1727}{{\tt arXiv:1007.1727}}].
%%CITATION = ARXIV:1007.1727;%%

\end{thebibliography}\endgroup
